\RequirePackage[hyphens]{url}
\documentclass[%
 aip,
 amsmath,amssymb,
 reprint,%
]{revtex4-1}
\usepackage{array}
\usepackage{graphicx}
\usepackage{tabularx, tabularray}
\usepackage{soul}
\usepackage[table,xcdraw]{xcolor}
\renewcommand{\arraystretch}{1.4}
\usepackage{dcolumn}
\usepackage{bm}
\usepackage{hyperref}
\hypersetup{colorlinks=true, linkbordercolor=white, linkcolor=blue, citecolor=blue}
\usepackage[utf8]{inputenc}
\usepackage[T1]{fontenc}
\usepackage{mathptmx}
\usepackage{etoolbox}
\usepackage{xcolor}
\usepackage{braket}
\usepackage{comment}
\usepackage{tikz}
\usetikzlibrary{quantikz2}
\usepackage{booktabs}
\usepackage{adjustbox}
\usepackage[font=small,labelfont=bf,justification=justified,singlelinecheck=true]{caption}
\usepackage{caption,subcaption}
\usepackage{ifthen}
\usepackage{amsmath}

\usepackage{enumitem}
\makeatletter
\def\@email#1#2{%
 \endgroup
 \patchcmd{\titleblock@produce}
  {\frontmatter@RRAPformat}
  {\frontmatter@RRAPformat{\produce@RRAP{*#1\href{mailto:#2}{#2}}}\frontmatter@RRAPformat}
  {}{}
}%
\usepackage{placeins}
\usepackage{float}

\newcommand{\LinearGraph}{
\begin{tikzpicture}[every node/.style={circle,draw,fill=red!20,minimum size=2.5mm,inner sep=0pt,text centered}]
  \foreach \i in {0,...,3} {
    \node (n\i) at (\i,0) {};
  }
  \foreach \i in {0,...,2} {
    \draw (n\i) -- (n\the\numexpr\i+1\relax);
  }
\end{tikzpicture}
}

\newcommand{\twoDGraph}{
\begin{tikzpicture}[every node/.style={circle,draw,fill=red!20,minimum size=2.5mm, inner sep=0pt,text centered}]
  \foreach \x in {0,1,2} {
    \foreach \y in {0,1,2} {
      \node (n\x\y) at (\x,-\y) {};
    }
  }
  \foreach \x in {0,1,2} {
    \foreach \y in {0,1} {
      \draw (n\x\y) -- (n\x\the\numexpr\y+1\relax);
    }
  }
  \foreach \x in {0,1} {
    \foreach \y in {0,1,2} {
      \draw (n\x\y) -- (n\the\numexpr\x+1\relax\y);
    }
  }
\end{tikzpicture}
}
\newcommand{\threeDGraph}{
\begin{tikzpicture}[ every node/.style={circle,draw,fill=red!20,minimum size=2.5mm, inner sep=0pt,text centered}]
  \foreach \x in {0,1,2} {
    \foreach \y in {0,1,2} {
      \foreach \z in {0,1,2} {
        \node (n\x\y\z) at (\x,\y,\z) {};
      }
    }
  }
  \foreach \x in {0,1} {
    \foreach \y in {0,1,2} {
      \foreach \z in {0,1,2} {
        \draw (n\x\y\z) -- (n\the\numexpr\x+1\relax\y\z);
      }
    }
  }
  \foreach \x in {0,1,2} {
    \foreach \y in {0,1} {
      \foreach \z in {0,1,2} {
        \draw (n\x\y\z) -- (n\x\the\numexpr\y+1\relax\z);
      }
    }
  }
  \foreach \x in {0,1,2} {
    \foreach \y in {0,1,2} {
      \foreach \z in {0,1} {
        \draw (n\x\y\z) -- (n\x\y\the\numexpr\z+1\relax);
      }
    }
  }
\end{tikzpicture}
}

\newcommand{\removeshorteringoriginal}{
\begin{tikzpicture}[
    node/.style={circle, draw, minimum size=5mm, font=\small},
    measured/.style={circle, draw, fill=lightgray, minimum size=5mm, font=\small}
]

\foreach \x in {0,...,3} {
    \foreach \y in {0,...,3} {
        \pgfmathtruncatemacro{\isMeasured}{(\x==0 && \y==1) || (\x==1 && \y==1) || (\x==2 && \y==1) || (\x==3 && \y==1)}
        \ifnum\isMeasured=1
            \ifnum\x=0
                \node[measured] (a\x\y) at (\x, -\y) {$\hat{q}$};
            \else
                \ifnum\x=1
                    \node[measured] (a\x\y) at (\x, -\y) {$\hat{q}$};
                \else
                    \ifnum\x=3
                        \node[measured] (a\x\y) at (\x, -\y) {$\hat{q}$};
                    \else
                        \node[measured] (a\x\y) at (\x, -\y) {$\hat{p}$};
                    \fi
                \fi    
            \fi
        \else
            \node[node] (a\x\y) at (\x, -\y) {};
        \fi
    }
}
\foreach \x in {0,...,3} {
    \foreach \y in {0,...,3} {
        \ifnum\x<3
            \draw (a\x\y) -- (a\the\numexpr\x+1\relax\y);
        \fi
        \ifnum\y<3
            \draw (a\x\y) -- (a\x\the\numexpr\y+1\relax);
        \fi
    }
}

\end{tikzpicture}
}

\newcommand{\removeshorteringafter}{
\begin{tikzpicture}[every node/.style={circle, draw, minimum size=5mm}, node distance=2cm]

\node (1) at (0,0) {};
\node (2) at (1,0) {};
\node (3) at (2,0) {};
\node (4) at (3,0) {};
\node (5) at (0,1) {};
\node (6) at (1,1) {};
\node (7) at (2,1) {};
\node (8) at (3,1) {};
\node (9) at (0,3) {};
\node (10) at (1,3) {};
\node (11) at (2,3) {};
\node (12) at (3,3) {};

\draw(1) -- (2);
\draw(2) -- (3);
\draw(3) -- (4);
\draw(5) -- (6);
\draw(6) -- (7);
\draw(7) -- (8);
\draw(7) -- (11);
\draw(9) -- (10);
\draw(10) -- (11);
\draw(11) -- (12);
\draw(1) -- (5);
\draw(2) -- (6);
\draw(3) -- (7);
\draw(4) -- (8);
\end{tikzpicture}
}

\newcommand{\Circuit}[1]{
\ifthenelse{\equal{#1}{CX}}{%
\begin{quantikz}
\lstick{} & \ctrl{1} & \qw \\
\lstick{} & \targ{} & \qw
\end{quantikz}
}{%
\ifthenelse{\equal{#1}{CZ}}{
\begin{quantikz}
\lstick{} & \ctrl{1} & \qw \\
\lstick{} & \gate{Z} & \qw
\end{quantikz}
}{%
\ifthenelse{\equal{#1}{SWAP}}{
\begin{quantikz}
\lstick{} & \swap{1} & \qw \\
\lstick{} & \swap{-1} & \qw
\end{quantikz}
}{%
\ifthenelse{\equal{#1}{TOFFOLI}}{
\begin{quantikz}
\lstick{} & \ctrl{1} & \qw  \\
\lstick{} & \ctrl{1} & \qw  \\
\lstick{} & \targ{}  & \qw  
\end{quantikz}
}{%
\begin{quantikz}
\lstick{} & \gate{#1} & \qw
\end{quantikz}
}}}}%
}

\newcommand{\Matrix}[1]{%
\ifthenelse{\equal{#1}{I}}{%
    $\displaystyle\begin{pmatrix} 1 & 0 \\ 0 & 1 \end{pmatrix}$%
}{%
\ifthenelse{\equal{#1}{X}}{%
    $\displaystyle\begin{pmatrix} 0 & 1 \\ 1 & 0 \end{pmatrix}$%
}{%
\ifthenelse{\equal{#1}{Y}}{%
    $\displaystyle\begin{pmatrix} 0 & -i \\ i & 0 \end{pmatrix}$%
}{%
\ifthenelse{\equal{#1}{Z}}{%
    $\displaystyle\begin{pmatrix} 1 & 0 \\ 0 & -1 \end{pmatrix}$%
}{%
\ifthenelse{\equal{#1}{H}}{%
    $\displaystyle\frac{1}{\sqrt{2}}\begin{pmatrix} 1 & 1 \\ 1 & -1 \end{pmatrix}$%
}{%
\ifthenelse{\equal{#1}{S}}{%
    $\displaystyle\begin{pmatrix} 1 & 0 \\ 0 & i \end{pmatrix}$%
}{%
\ifthenelse{\equal{#1}{T}}{%
    $\displaystyle\begin{pmatrix} 1 & 0 \\ 0 & e^{i\pi/4} \end{pmatrix}$%
}{%
\ifthenelse{\equal{#1}{CX}}{%
    $\displaystyle
        \begin{pmatrix}
            1 & 0 & 0 & 0 \\
            0 & 1 & 0 & 0 \\
            0 & 0 & 0 & 1 \\
            0 & 0 & 1 & 0
        \end{pmatrix}$
}{%
\ifthenelse{\equal{#1}{CZ}}{%
    $\displaystyle
        \begin{pmatrix}
            1 & 0 & 0 & 0 \\
            0 & 1 & 0 & 0 \\
            0 & 0 & 1 & 0 \\
            0 & 0 & 0 & -1
        \end{pmatrix}$
}{%
\ifthenelse{\equal{#1}{SWAP}}{%
    $\displaystyle
        \begin{pmatrix}
            1 & 0 & 0 & 0 \\
            0 & 0 & 1 & 0 \\
            0 & 1 & 0 & 0 \\
            0 & 0 & 0 & 1
        \end{pmatrix}$
}{%
\ifthenelse{\equal{#1}{TOFFOLI}}{%
    $\displaystyle
        \begin{pmatrix}
        1 & 0 & 0 & 0 & 0 & 0 & 0 & 0 \\
        0 & 1 & 0 & 0 & 0 & 0 & 0 & 0 \\
        0 & 0 & 1 & 0 & 0 & 0 & 0 & 0 \\
        0 & 0 & 0 & 1 & 0 & 0 & 0 & 0 \\
        0 & 0 & 0 & 0 & 1 & 0 & 0 & 0 \\
        0 & 0 & 0 & 0 & 0 & 1 & 0 & 0 \\
        0 & 0 & 0 & 0 & 0 & 0 & 0 & 1 \\
        0 & 0 & 0 & 0 & 0 & 0 & 1 & 0 \\
        \end{pmatrix}
    $
}{%
    \textbf{Error: Unknown Matrix!}%
}}}}}}}}}}}%
}

\def\FlagComparisonPDF{1} 

\if\FlagComparisonPDF1

\else

\fi

\makeatother
\begin{document}

\preprint{AIP/123-QED}

\title[Optical Quantum Computing]{Optical Quantum Computing}
\author{Hamza Hasnaoui}
\affiliation{Department of Industrial Engineering, University of Trento, Via Sommarive 9, Povo, 38123, Trento, Italy}
\affiliation{INFN TIFPA, Trento Institute for Fundamental Physics and Applications, Via Sommarive 14, Povo, 38123, Trento, Italy}
\author{Leonardo Limongi}
\affiliation{Department of Industrial Engineering, University of Trento, Via Sommarive 9, Povo, 38123, Trento, Italy}
\affiliation{Center for Sensors and Devices, Bruno Kessler Foundation, Via Sommarive 18, Povo, 38123, Trento, Italy}
\author{Taira Giordani}
\affiliation{Dipartimento di Fisica - Sapienza Universit\`a di Roma, P.le Aldo Moro 5, I-00185 Roma, Italy}
\author{Beatrice Polacchi}
\affiliation{Dipartimento di Fisica - Sapienza Universit\`a di Roma, P.le Aldo Moro 5, I-00185 Roma, Italy}
\author{Alberto Quaranta}
\affiliation{Department of Industrial Engineering, University of Trento, Via Sommarive 9, Povo, 38123, Trento, Italy}
\affiliation{INFN TIFPA, Trento Institute for Fundamental Physics and Applications, Via Sommarive 14, Povo, 38123, Trento, Italy}
\author{Martino Bernard}
\affiliation{Center for Sensors and Devices, Bruno Kessler Foundation, Via Sommarive 18, Povo, 38123, Trento, Italy}
\author{Fabio Sciarrino}
\affiliation{Dipartimento di Fisica - Sapienza Universit\`a di Roma, P.le Aldo Moro 5, I-00185 Roma, Italy}
\author{Mirko Lobino}
\affiliation{Department of Industrial Engineering, University of Trento, Via Sommarive 9, Povo, 38123, Trento, Italy}
\affiliation{INFN TIFPA, Trento Institute for Fundamental Physics and Applications, Via Sommarive 14, Povo, 38123, Trento, Italy}
\email{mirko.lobino@unitn.it}
\date{\today}

\begin{abstract}
Under the label of optical quantum computing, there are a variety of protocols and experiments that use the quantum properties of light to achieve a computational advantage over classical computing machines. In this review, we describe some of the main implementations, which differ in the type of encoding and in how the computation is performed, whether using gates or cluster states. For each protocol, we describe advantages and challenges with an overview of the experimental results obtained, summarized in tables. Because of the great relevance achieved in this field, there is a section dedicated to non-universal quantum computation with photons, where boson sampling, variational quantum eigensolvers, and quantum machine learning applications are described. The aim is to give the reader the broadest overview of the applications where photons and their quantum properties play a key role in computation.
\end{abstract}

\maketitle

\section{\label{sec:intro}Introduction}

A universal quantum computer will represent a paradigm shift in information processing \cite{Usman2025}. Using the principles of quantum mechanics, it will solve problems that are hard to address for classical computers, including prime number factorization and simulation of quantum systems \cite{dalzell2023,Feynman1982}. Impressive achievements towards this goal have been demonstrated on several platforms, including experimental claims of quantum advantage over classical computers\cite{madsen2022,zhong2020quantum,TensorNetworkGBS}. 

Among the various physical platforms for quantum computing—such as superconducting qubits, trapped ions, and silicon-based systems—photons are one of the leading approaches due to their unique advantages, including room-temperature operation, low-noise environments, and scalability \cite{Slussarenko2019}. In particular, it was shown that with linear optical quantum computing\cite{knill2001scheme,kok2007}, computation required only the generation, manipulation in linear optical circuits, and detection of single photons, avoiding the need for nonlinear photon-photon interaction which is experimentally challenging. Two types of encoding are generally associated with photons: discrete-variable encoding, based on degrees of freedom such as polarization or path \cite{Barenco1995}, and continuous-variable encoding, which uses the quadrature of the electromagnetic field \cite{lloyd1999quantum}. Both types of encoding can be used for universal quantum computation, and both have advantages and challenges that will be detailed in sections \ref{sec:Gm QC} and \ref{sec:1way QC} of this review. 

Following the work of Aaronson and Arkhipov on boson sampling \cite{aaronson2011computational}, the scope of photonic quantum computation expanded to include non-universal quantum devices designed to solve specific problems that are believed to be intractable for classical computers. This proposal stimulated a wide range of experiments with progressively increasing complexity. In addition, several protocols inspired by the original boson sampling framework were introduced, ultimately leading to experimental claims of quantum advantage using photonic quantum processors \cite{madsen2022, zhong2020quantum, TensorNetworkGBS}. These results, together with their limitations, are discussed in Sec.~\ref{BSmachines}.

The technological and theoretical progress reached so far demonstrates the maturity of the field and the opinion shared by some researchers that we will see the first quantum computer in one or a few decades. This is certainly the vision of many companies that are now investing considerable resources in optical quantum computing and publishing important results both in theory and experiment \cite{Alexander2025, AghaeeRad2025, maring2024versatile}.

\begin{figure*}[ht!]
    \centering
    \includegraphics[width=\linewidth]{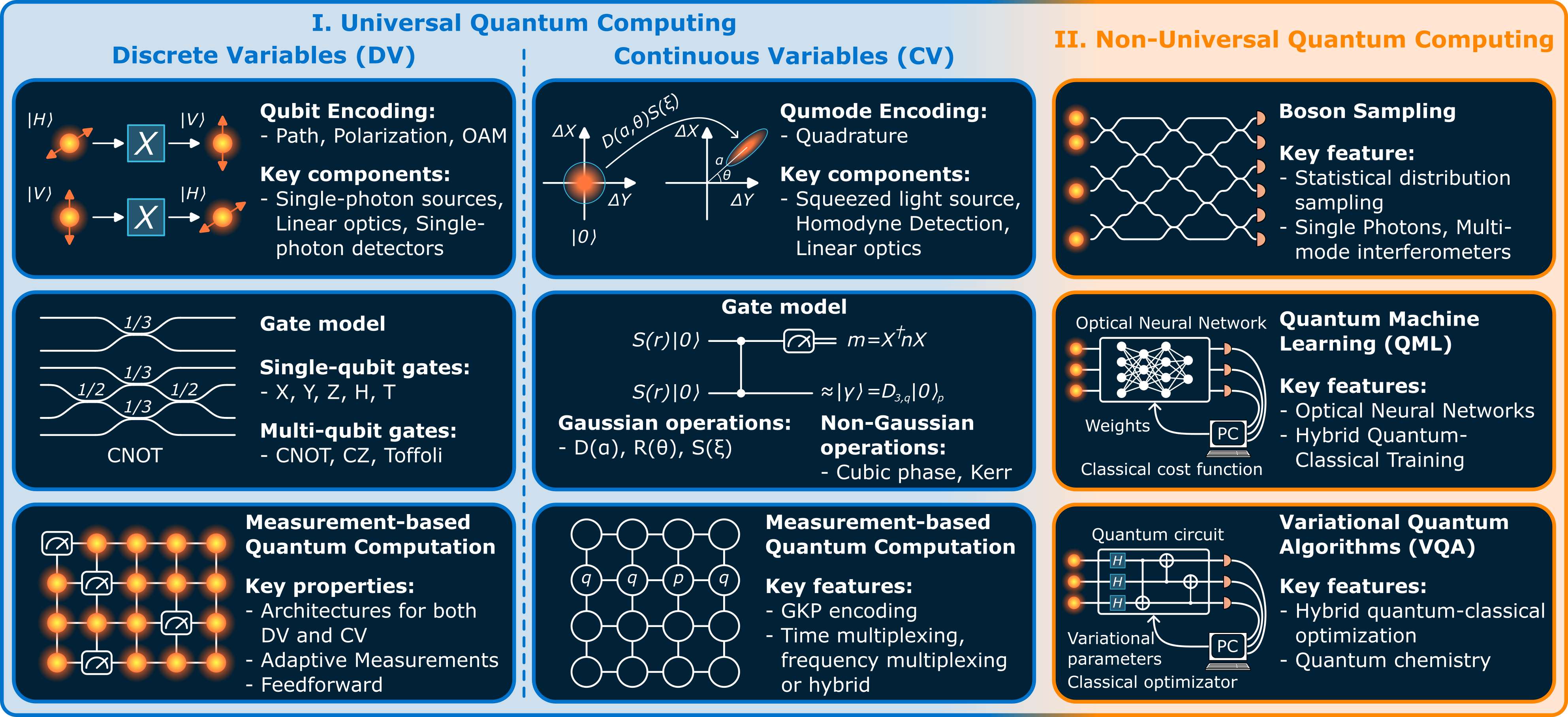}
    \caption{Overview of the optical quantum computing approaches covered in this review. This document is organized into two primary sections: approaches to universal quantum computation (discrete-variable, continuous-variable, gate-based and measurement-based quantum computation), and non-universal approaches (boson sampling, quantum machine learning, and variational algorithms) that provide computational advantages for specific problems. For each topic, we provide tables that summarize the present state of the art.}
    \label{fig:intro}
\end{figure*}

This review surveys the various approaches through which the quantum properties of photons can yield a computational advantage, as illustrated in Fig.~\ref{fig:intro}. These efforts encompass proposals for both universal photonic quantum computing and non-universal photonic protocols, including contributions from the many companies that are now active in this research field. For each approach, we outline the underlying theoretical principles and highlight selected experimental demonstrations, though space limitations preclude an exhaustive coverage of all relevant literature. The review is organized into three primary sections: discrete-variable quantum computation, continuous-variable quantum computation, and non-universal photonic approaches. In the first two sections, we discuss the basic principles of the gate-based model, which was chronologically the first to be developed. However, we dedicate more attention to the measurement-based (cluster-state) model, 
since the field has pivoted heavily toward topological cluster states and measurement-based architectures. The section on non-universal protocols focuses on boson sampling, variational quantum eigensolvers (VQEs), and quantum machine learning. Finally, we conclude by discussing the future directions and challenges of this research field, with particular emphasis on integration and scalability.

\section{\label{sec:Gm QC}Discrete-variable quantum computation}
Discrete-variable (DV) quantum computing encodes quantum information in eigenstates of operators with a discrete spectrum, typically denoted as $\ket{0}$ and $\ket{1}$, in analogy with the 0 and 1 states of a classical bit. In photonic implementations, common encoding degrees of freedom include polarization ($\ket{H}$,$\ket{V}$), orbital angular momentum (OAM), and—particularly in integrated photonics—spatial or path encoding ($\ket{P_0}$,$\ket{P_1}$). 

A key milestone in the development of a photonic quantum computer was the proposal of linear optical quantum computing by Knill et al.,\cite{knill2001scheme} which demonstrated that scalable quantum computation could in principle be achieved using only single photons, linear optical elements, and measurement-induced nonlinearities\cite{Scheel2003} at the price of a polynomial overhead in resources. This result stimulated intense experimental and theoretical activity, leading to the development of photonic qubits encoded in degrees of freedom such as polarization, path, and time bins, as well as techniques for generating and detecting single photons. 

Subsequent work introduced measurement-based models of photonic quantum computation \cite{Raussendorf2001,Briegel2009,Wei2021Measurement}, including cluster-state approaches \cite{raussendorf2003} that rely on the generation of large entangled resource states. This paradigm is particularly promising because it circumvents one of the main limitations of photonic platforms—the difficulty of realizing nonlinear interactions at the single-photon level necessary for the implementation of two qubit gates—by shifting the computational complexity to the preparation of highly entangled states and adaptive measurements.

Over the past two decades, advances in integrated photonics, deterministic photon sources, and high-efficiency detectors have significantly improved the scalability and stability of photonic platforms, establishing optical quantum computing as a major approach within the broader quantum information landscape. Photons offer several advantages compared with other physical systems: their high propagation speed, low decoherence at room temperature, and ability to be transmitted over long distances make them excellent carriers of quantum information. Nevertheless, the performance of photonic quantum devices has not yet reached the level required for large-scale deployment. Key challenges include the realization of near-deterministic single-photon sources and, particularly in integrated photonic systems, the reduction of propagation losses arising from scattering and absorption.

\subsection{\label{DV gate model}Discrete-Variable Gates}

\begin{table*}
\centering
\begin{minipage}[b]{0.45\textwidth}
    \vspace{0pt}
    \centering
    \begin{tblr}{
        baseline = t,
        colspec = {X[15,c,m] X[25,c,b] X[40,c,m] X[20,c,m]},
        rowsep = 2pt,
        row{1} = {font=\bfseries, belowsep = 1pt},
        hline{2} = {solid},
    }
        Symbol  & Circuit       & Matrix        & Name          \\
        $I$     & \Circuit{I}   & \Matrix{I}    & Identity      \\
        $X$     & \Circuit{X}   & \Matrix{X}    & Pauli-X       \\
        $Y$     & \Circuit{Y}   & \Matrix{Y}    & Pauli-Y       \\
        $Z$     & \Circuit{Z}   & \Matrix{Z}    & Pauli-Z       \\
        $H$     & \Circuit{H}   & \Matrix{H}    & Hadamard      \\
        $S$     & \Circuit{S}   & \Matrix{S}    & Phase         \\
        $T$     & \Circuit{T}   & \Matrix{T}    & $\pi/8$       \\
    \end{tblr}
\end{minipage}%
\hfill
\begin{minipage}[b]{0.50\textwidth}
    \vspace{0pt}
    \centering
    \begin{tblr}{
        baseline = t,
        colspec = {X[15,c,m] X[25,c,b] X[50,c,m] X[20,c,m]},
        rowsep = 4pt,
        row{1} = {font=\bfseries, belowsep = 1pt},
        hline{2} = {solid},
    }
        Symbol      & Circuit           & Matrix            & Name                          \\
        \textit{CX}        & \Circuit{CX}      & \Matrix{CX}       & {Control-X \\ or \textit{CNOT}}       \\
        \textit{CZ}        & \Circuit{CZ}      & \Matrix{CZ}       & Control-Z                     \\
        Toffoli     & \Circuit{TOFFOLI}  & \Matrix{TOFFOLI}  & Toffoli                       \\
    \end{tblr}
\end{minipage}

\caption{Schematic representation of most common single- and two-qubit gates and their matrix representation in the computational basis.}
\label{tab:SingleQubit}
\end{table*}

The fundamental concept underlying the discrete-variable quantum gate model (DVGM) is the establishment of a quantum equivalent to the classical universal Turing machine. Often referred to as a universal quantum computer, this model is designed to execute any unitary operator at a desired level of precision, mirroring the classical machine's ability to perform any classical algorithm. The computation is performed by a sequence of quantum logic gates, taken from a set that guarantees universality\cite{NielsenChuang, gottesman_heisenberg_1998}, and that can implement an arbitrary unitary transformation on a defined number of qubits.

Table~\ref{tab:SingleQubit} presents some of the most common single- and multi-qubit gates alongside their matrix representations in the computational basis. Universal quantum computation can be achieved using various non-unique sets of gates acting on the $N$-qubit Hilbert space $(\mathbb{C}^2)^{\otimes N}$. Notable examples of universal gate sets include: (1) the Toffoli, Hadamard, and $\pi/4$ gates \cite{Kitaev1997}; (2) the \textit{CNOT}, Hadamard, and $\pi/8$ gates \cite{Boykin2000}; (3) the \textit{CNOT} gate combined with the set of all single-qubit gates \cite{Barenco1995}; and (4) the Toffoli gate paired with any basis-changing real single-qubit gate \cite{Shi2003}. The Solovay-Kitaev theorem \cite{Kitaev1997,Dawson2006} establishes the functional equivalence of these discrete sets, proving that any universal set can simulate another with only a polylogarithmic overhead in the number of gates. However, while translating between gate sets is efficient, synthesizing an arbitrary unitary transformation to a desired level of precision generally requires a sequence of gates that scales exponentially with the number of qubits \cite{NielsenChuang}. This synthesis can be executed using various methods, such as cosine-sine matrix decomposition \cite{Mottonen2004}, two-level unitary matrices \cite{Li2013}, or customized algorithms tailored to specific problems \cite{Daskin2011,Guseynov2025}.
 
\begin{figure*}[ht!]
    \centering
    \includegraphics{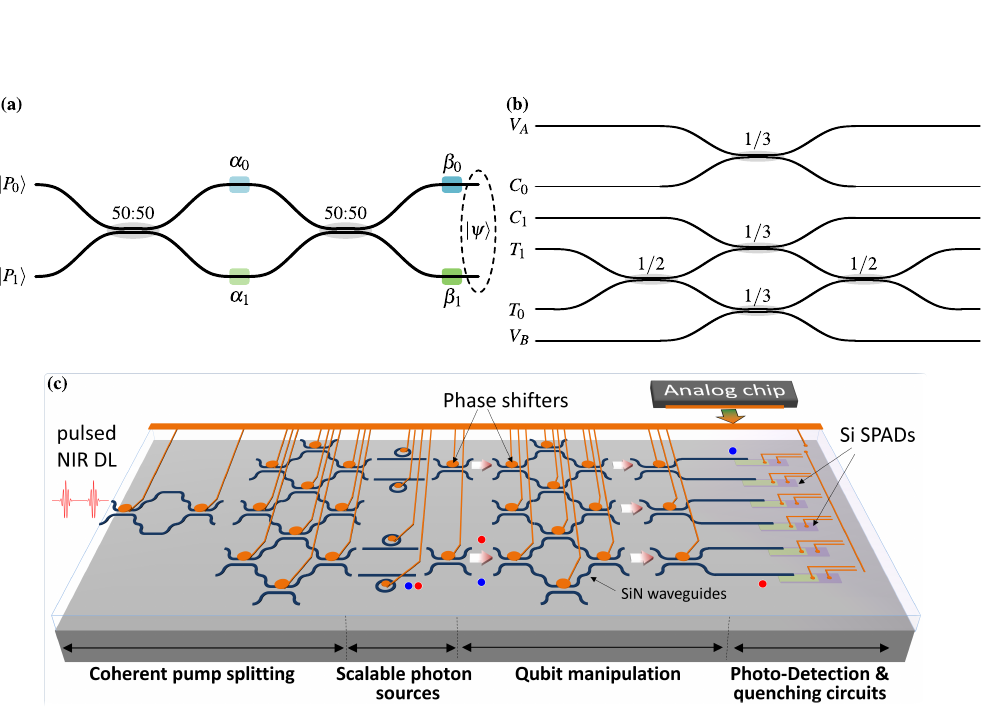}
    \caption{\textbf{(a)} Schematic of a Mach-Zehnder Interferometer (MZI) with phase shifters ($\alpha_0$, $\alpha_1$, $\beta_0$, $\beta_1$). This is a fundamental building block for the photonic implementation of DV gates, allowing the manipulation of path-encoded qubits. The MZI performs arbitrary single qubit operations such as those of Table~\ref{tab:SingleQubit}. \textbf{(b)} Schematic of the probabilistic integrated \textit{CNOT} gate implementation proposed by Ref.~[\onlinecite{Ralph2002Linear}], implemented in bulk optics in Ref.~[\onlinecite{OBrien2003Demonstration}] and integrated into a waveguide device in Ref.~[\onlinecite{politi_silica_CNOT_2008}]. \textbf{(c)} Authors' impression of a Photonic Integrated Circuit for discrete-variable quantum computing where the three key stages of generation, manipulation and detection of photons are realized on the same chip. From H2020 Epiqus GA. No. 899368, courtesy of Dr. Mher Ghulinyan.}
     \label{fig:DV_Gate_Figure1}
\end{figure*}

Among the various degrees of freedom used in discrete-variable (DV) protocols, path encoding is a primary option because it naturally suits waveguide-integrated platforms. However, the theoretical concepts presented for path encoding can be readily generalized to alternative degrees of freedom. In this context, linear gates are primarily implemented using two optical components: beam splitters (BS) and phase shifters (PS). As sketched in Fig.~\ref{fig:DV_Gate_Figure1}a, the beam splitters are typically realized as multimode interferometers (MMIs) or directional couplers (DCs), while the phase shifters exploit thermo-optic, carrier dispersion, or Pockels effect.

Arbitrary single-qubit gates can be fully realized using standard linear optical components. When information is encoded in photon polarization, for instance, a combination of birefringent wave plates can rotate the state to implement any single-qubit rotation. Conversely, in path-encoded systems, universal single-qubit operations are achieved using Mach-Zehnder interferometers (MZIs) and phase shifters, as illustrated in Fig.~\ref{fig:DV_Gate_Figure1}a.

The working principle of a single-qubit gate based on an MZI can be summarized as follows: two spatial modes, $\ket{P_0}$ and $\ket{P_1}$, are combined at a 50:50 beam splitter (BS); a phase shift (PS) $\alpha_j$ is then applied to each mode $\ket{P_j}$ (where $j \in \{0,1\}$); the modes are subsequently recombined at a second 50:50 BS; and finally, a phase shift $\beta_j$ is applied to mode $\ket{P_j}$. The total action of such MZI is represented by the matrix
\begin{equation}
    \text{MZI}= ie^{i\left(\Sigma_\alpha+\Sigma_\beta\right)}
    \begin{pmatrix}
    e^{i\Delta_\beta}\sin\left(\Delta_\alpha\right) & e^{i\Delta_\beta}\cos\left(\Delta_\alpha\right) \\
    e^{-i\Delta_\beta}\cos\left(\Delta_\alpha\right) & -e^{-i\Delta_\beta}\sin\left(\Delta_\alpha\right) 
    \end{pmatrix},
\end{equation}
which is obtained by the direct multiplication of the matrices
\begin{equation}
    \text{BS} = \frac{1}{\sqrt{2}}
    \begin{pmatrix}
        1 & i \\
        i & 1
    \end{pmatrix}
    \quad\text{and}\quad
    \text{PS}\left(\varphi_0,\varphi_1\right) = 
    \begin{pmatrix}
        e^{i\varphi_0} & 0 \\
        0 & e^{i\varphi_1}
    \end{pmatrix},
\end{equation}
and with the definitions
\begin{equation}
    \begin{cases}
        \Sigma_\alpha=\dfrac{\alpha_0+\alpha_1}{2} \\[10pt]
        \Delta_\alpha=\dfrac{\alpha_0-\alpha_1}{2}
    \end{cases},
    \qquad
    \begin{cases}
        \Sigma_\beta=\dfrac{\beta_0+\beta_1}{2} \\[10pt]
        \Delta_\beta=\dfrac{\beta_0-\beta_1}{2}
    \end{cases}.
\end{equation}
Notably, while the global phase is proportional to the sums $\Sigma_\alpha$ and $\Sigma_\beta$, the internal transformation of an MZI is dictated entirely by the phase differences $\Delta_\alpha$ and $\Delta_\beta$ between the two spatial paths (0 and 1). Consequently, the two arms of an MZI are frequently designated as the \textit{push} and \textit{pull} arms.

An arbitrary single-qubit state $\ket{\psi}=\cos\left(\frac{\theta}{2}\right)\ket{P_0}+e^{i\phi}\sin\left(\frac{\theta}{2}\right)\ket{P_1}$ can be synthesized in an MZI. By injecting the input state $\ket{P_0}$, the interferometer applies a transformation yielding:
\begin{equation}
    \text{MZI}\ket{P_0}=
    \sin\left(\Delta_\alpha\right)\ket{P_0}+e^{-2i\Delta_\beta}\cos\left(\Delta_\alpha\right)\ket{P_1},
\end{equation}
with the global phase being omitted. To match this output to the target state $\ket{\psi}$, we map the internal phase variables $(\alpha_0,\alpha_1,\beta_0,\beta_1)$ to the Bloch sphere coordinates $(\phi,\theta)$. Neglecting any global phase factors, one finds that the angles are related by:
\begin{equation}
    \theta=\pm\big(\pi+\alpha_1-\alpha_0\big)\quad\text{and}\quad\phi=\beta_1-\beta_0.
\end{equation}
A similar formalism can be derived for the polarization encoding, where the roles of the BS and phase shifters are played by wave plates~\cite{englert_universal_2001}.

Implementing two-qubit gates in the optical domain is inherently challenging because it necessitates interaction between distinct qubits. The fundamental obstacle—the extremely weak nature of photon-photon interactions—makes the deterministic execution of essential universal gates, such as the \textit{CNOT} or \textit{CZ} (see Table~\ref{tab:SingleQubit} for multi-qubit examples), highly difficult to achieve. Consequently, these operations are primarily realized through probabilistic protocols. The first photonic \textit{CNOT} gate, working in post selection, was realized  by O'Brien et al.\cite{OBrien2003Demonstration} following the probabilistic scheme outlined in Refs.~[\onlinecite{Ralph2002Linear,ralph2003quantum}], while the first heralded \textit{CNOT} gate was realized by Zhao et al.\cite{Zhao2005}.  

Most probabilistic protocols rely on measurement-induced nonlinearity \cite{knill2001scheme, Scheel2003} to achieve the effective photon-photon interaction necessary for two-qubit operations. These schemes typically involve interfering the computational photons on a series of beam splitters alongside auxiliary photons, with the success of the operation determined by specific measurement outcomes. In a heralded scheme \cite{Li2021}, if the measurement of the auxiliary photons yields the desired result, the gate is deemed successful; otherwise, it fails, and the computation must be retried. The inherent success probability of these native gates is generally low—for example, 1/8 for the polarization-encoded \textit{CNOT} in Ref.~[\onlinecite{Li2021}] and 1/9 for the path-encoded implementation in Ref.~[\onlinecite{OBrien2003Demonstration}]—but it can be systematically increased by employing additional auxiliary photons and more complex interferometric networks.

The first integrated photonic \textit{CNOT} gate was demonstrated in Ref.~[\onlinecite{politi_silica_CNOT_2008}] on a silica-on-silicon platform for path encoding, and Ref.~[\onlinecite{Crespi2011}] for polarization encoding. Both experiments employed the same scheme as Ref.~[\onlinecite{OBrien2003Demonstration}]. Figure~\ref{fig:DV_Gate_Figure1}b shows the \textit{CNOT} scheme for path encoding with the target-qubit photon encoded in modes $T_0$ and $T_1$, while modes $C_0$ and $C_1$ are used for the control-qubit photon. The \textit{CNOT} operation is achieved using two 1/2 (50:50) directional couplers on the target modes—effectively acting as two Hadamard gates—flanking a \textit{CZ} gate. This \textit{CZ} gate is implemented probabilistically at the 1/3 directional coupler shared by modes $C_1$ and $T_1$, which applies a $\pi$ phase shift when there is one photon in each output of the coupler. The overall gate is successful if exactly one photon is detected at the output of the control modes and one in the target modes. Finally, two couplers with a 1/3 reflectivity are used to couple modes $C_0$ and $T_0$ with vacuum modes $V_A$ and $V_B$ respectively. This configuration serves to symmetrize the overall success probability to 1/9 across all possible outputs. 

\subsubsection{\label{sec:DCGM_generalwork} Building blocks of a DVGM}
Having established the theoretical framework of the DVGM —including the relevant degrees of freedom and set of logic gates— this section transitions to the hardware domain to detail the physical building blocks studied to realize a photon based discrete-variable quantum computer. A photonic quantum computer operates through three primary functional blocks: the generation, manipulation, and detection of photons. First, single-photon sources\cite{Eisaman2011} generate and initialize the input state—typically a known computational basis state, though the output of a preceding circuit may also serve as the input. Next, a network of quantum logic gates \cite{Carolan2015} manipulates these photonic qubits to execute the target algorithm. Finally, the state is read out using an array of single-photon detectors \cite{natarajan2012superconducting}. Because each detector in the array registers either the presence (1) or absence (0) of a photon, the measurement projects the $N$-qubit quantum state from the Hilbert space $(\mathbb{C}^2)^{\otimes N}$ into a classical output string, denoted as $m \in \{0, 1\}^N$.

Figure~\ref{fig:DV_Gate_Figure1}c illustrates a schematic of a photonic integrated circuit (PIC) that consolidates all three previously described stages into a single monolithic platform. It also depicts the classical computer interface, which enables the dynamic reconfiguration of the quantum gates to execute different algorithms. Ideally, such a device would feature bright, on-demand, high-purity single-photon sources, virtually lossless manipulation circuits, and near-unity efficiency detectors that do not require cryogenic cooling. Achieving these stringent performance metrics is an ongoing effort that requires extensive optimization of various cleanroom fabrication processes.

Generating single photons on demand remains a major challenge in optical quantum computing, yet it is a critical prerequisite for any optical quantum machine. Spontaneous and stimulated parametric processes\cite{Mosley2008,Xiong2011,Eisaman2011} have been the primary approaches for generating photon pairs. These processes can generate time- and energy-correlated photon-pairs via second\cite{Mosley2008} or third order\cite{Xiong2011} non-linear effects, and because this generation is inherently probabilistic, these sources rely on either post-selection schemes or heralding—where the detection of one photon confirms the presence of its partner. Furthermore, pumping the nonlinear medium with strong pulsed lasers enables the parallel operation of multiple photon-pair sources, thereby supplying the quantum circuit with a multitude of heralded, indistinguishable single photons.

To mitigate the probabilistic nature of photon generation, several architectural schemes have been proposed. As early as 2002, researchers suggested multiplexing heralded photon sources and storing heralded SPDC photons in optical cavities for subsequent "on-demand" release\cite{migdall_tailoring_2002,pittman_single_2002}. A parallel approach utilizes optical delay lines to enable interactions between photons generated at different times. For instance, Ref.~[\onlinecite{Humphreys2013Linear}] employed time-shifting to achieve polarization encoding for photons generated sequentially within the same spatial mode. While important proof-of-principle demonstrations, these early multiplexing strategies still had to contend with significant photon losses accumulated during the routing and storage processes.

In recent years, researchers have developed more efficient SPDC sources utilizing compact crystals that produce high-quality photon pairs at high rates \cite{Chekhova2024Spontaneous}. However, miniaturizing these sources without sacrificing yield remains a significant challenge. Furthermore, while the delay and multiplexing techniques originally proposed in Ref.~[\onlinecite{migdall_tailoring_2002}] have been extensively explored and refined\cite{meyer-scott_single-photon_2020}, they can be scaled only with  low-loss and massive, integrated multiplexed sources and circuits\cite{Alexander2025}. Nevertheless, SPDC sources have reached a level of maturity where fiber-coupled photon-pair sources are now commercially available as off-the-shelf components.

Quantum dot sources \cite{Arakawa2020,li_quantum_dots_2023,Senellart2017High} represent a highly promising technology for the deterministic generation of single photons. While the scalable engineering of identical emitters remains a significant challenge, photons emitted sequentially from the same quantum dot have demonstrated near-perfect indistinguishability \cite{somaschi2016near}. Furthermore, the compact physical footprint of these sources makes them well-suited for integration with waveguide circuits, paving the way for fully integrated platforms featuring deterministic photon generation. Recent advancements in this domain include edge-coupling arrays of quantum dots to integrated circuits \cite{ohara2025}, directly fabricating single dots on top of waveguides \cite{Pholsen25}, and utilizing deterministic pick-and-place techniques to position them onto the photonic chip \cite{Zadeh2016}.

Quantum dots typically require cryogenic operating temperatures, and collection losses continue to hinder their large-scale integration. Over the last decade, researchers have continually pushed the boundaries to address these bottlenecks, developing increasingly sophisticated source designs to overcome persistent operational obstacles \cite{Thapa2025}. 
Because quantum-dot-based single-photon sources have been realized across a plethora of material platforms, their operational characteristics—such as emission wavelength, tunability, and compatibility with various fabrication processes—can vary greatly \cite{tomm_bright_2021, li_quantum_dots_2023,ding_high-efficiency_2025}. Nevertheless, this technology has reached a level of maturity where quantum dot sources are now commercially available as stand-alone devices \cite{quandela,sparrow}. 

\begin{figure*}[ht!]
    \centering
    \includegraphics[width=\linewidth]{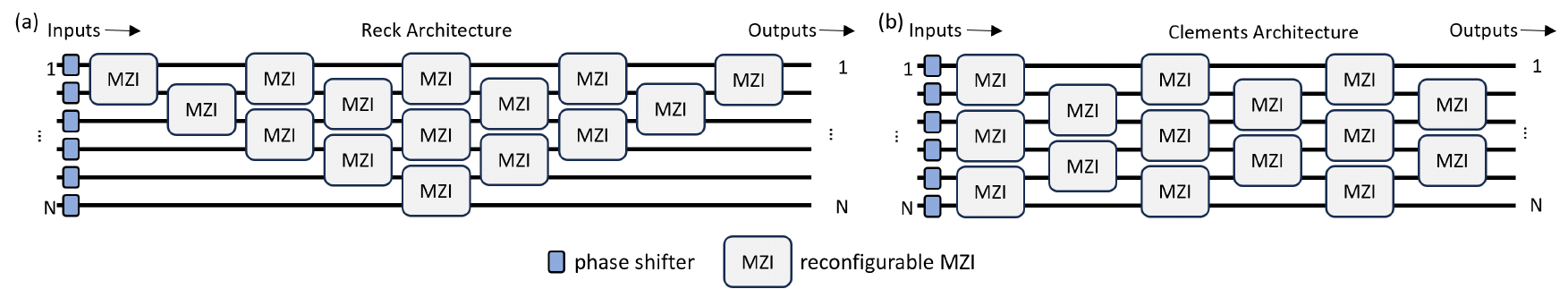}
    \caption{\textbf{(a)} Schematic of the Reck scheme for the implementation of arbitrary unitary transformations using linear optical elements.\cite{Reck_1994} \textbf{(b)} Schematic of the Clements scheme\cite{Clements2016} which reduces optical depth of the interferometer defined  as the longest path through the interferometer, enumerated by counting the number of beam splitters traversed by that path. The Clements scheme has the same number of elements as the Reck scheme but it is more symmetrical and more compact. The reconfigurable MZI is similar to the device shown in Fig.~\ref{fig:DV_Gate_Figure1}b}.
     \label{fig:ReckClements}
\end{figure*}

Beyond quantum dots and parametric processes, optically active point defects in crystalline solids\cite{Esmann2024Solid,Aharonovich2016} have emerged as a robust technology for single-photon generation. Prominent examples include nitrogen- and silicon-vacancy centers in diamond\cite{Beveratos2002}, as well as defects in silicon carbide\cite{Castelletto2021} and two-dimensional host materials\cite{Tran2016}. Because these color centers act as trapped artificial atoms, they can provide stable  emission, with some platforms circumventing the need for cryogenic cooling entirely. Moreover, their ability to entangle emitted photons with long-lived solid-state spin qubits makes them uniquely suited for quantum networking and memory applications. However, to achieve the high indistinguishability and high collection efficiency required for discrete-variable quantum computing, extensive nanofabrication and material optimization are still required to seamlessly integrate these bulk-hosted defects into low-loss waveguide architectures.

Once single photons are successfully generated and coupled into the circuit, they must be manipulated to execute quantum algorithms. Arbitrary unitary transformations can be encoded directly into interferometric networks. The foundational architecture for this manipulation is the Reck scheme\cite{Reck_1994} (see Fig. \ref{fig:ReckClements}a), which demonstrated that any $N \times N$ unitary matrix can be decomposed into a cascaded, triangular arrangement of programmable Mach-Zehnder interferometers (MZIs). While this original topology remains popular for its conceptual simplicity, optimizing the MZI mesh to reduce accumulated errors remains an ongoing challenge. To mitigate the issues of tunability, unequal photon loss, and phase error accumulation, alternative geometries have been developed, including the rectangular Clements scheme\cite{Clements2016} (see Fig. \ref{fig:ReckClements}b), as well as diamond and Bokun structures\cite{Bokun_2023}.

A highly practical realization of the Clements architecture was demonstrated in 2021 over 12 modes using a silicon nitride (SiN) platform equipped with thermo-optical phase shifters. This photonic integrated circuit (PIC) served as the primary component for a commercial $12 \times 12$ universal processor released by QuiX Quantum \cite{taballione_20-mode_2023,QuiX_Quantum}. Two years later, the same authors scaled the system to a $20 \times 20$ Clements matrix, achieving a transformation fidelity of 90\% with approximately 3 dB of optical loss\cite{taballione_20-mode_2023}. Despite these technological advancements allowing for the realization of large, high-fidelity systems\cite{Carolan2015,Wang2016}, purely linear optical implementations of Reck and Clements architectures are not inherently scalable in a computational sense. Because they rely on linear optics, multi-qubit entangling operations are intrinsically probabilistic. For example, the theoretical success probability is limited to $1/9$ for a linear-optical \textit{CNOT} gate\cite{politi_silica_CNOT_2008} and $1/57$ for a three-qubit Toffoli gate\cite{Kieling2010}. Nevertheless, the commercial availability of these programmable meshes, paired with time-multiplexed sources, has catalyzed significant research, enabling versatile, multi-functional quantum processing devices\cite{maring2024versatile}. 

Finally, embedding a given discrete-variable algorithm into an MZI mesh is a complex optimization task in its own right. To bridge the gap between abstract quantum circuits and physical hardware, several dedicated software frameworks have been developed \cite{broughton_tensorflow_2021,bergholm_pennylane_2022,kwon_software_2023,kwon_full_2024}. These compilers facilitate the seamless transfer and manipulation of information from a schematic representation to the precise analog controls required by the actual device.

The final functional block of the discrete-variable quantum architecture consists of single-photon detectors. These devices are crucial for state readout, as they determine the qubit values by measuring the presence or absence of a photon within specific spatial paths. Depending on the operating wavelength and the chosen information encoding scheme, various detector technologies are employed. Historically, single-photon detection began in 1949 with the use of photomultiplier tubes (PMTs)\cite{Morton_PMT_1949}. Since then, detection technologies have advanced enormously in terms of quantum efficiency, count rates, spectral bandwidth, and device compactness.

The direct solid-state successors to PMTs are Single-Photon Avalanche Diodes (SPADs), which exploit similar electron-multiplication principles but operate within the junction of a bulk semiconductor\cite{Cova_1996}. Because they are fabricated using standard semiconductor technologies, SPADs can be seamlessly merged with photonics to create Electronic-Photonic Integrated Circuits (EPICs)\cite{Kimerling_ASEPIC_2006} through both heterogeneous\cite{incoronato_multi-channel_2022} and homogeneous\cite{acerbi_monolithically_2023} integration techniques. A major advantage of silicon SPADs is their relaxed thermal requirements; they can maintain dark count rates (false-positive detections) below 100 Hz at room temperature, making them highly attractive for scalable quantum systems. Alongside SPADs, standard Avalanche Photodiodes (APDs)\cite{shi_avalanche_2024} are also garnering increased interest, particularly due to their high electrical bandwidths, which now exceed 50 GHz.

Currently, the state of the art in single-photon detection is defined by Superconducting Nanowire Single-Photon Detectors (SNSPDs) \cite{peacock1996single,natarajan2012superconducting}. In these devices, a thin superconducting nanowire is biased just below its critical current density. When a photon is absorbed, the localized temperature increase creates a resistive "hot spot" that forces the local current density above the superconducting threshold, generating a measurable voltage pulse. Although SNSPDs require cryogenic cooling to operate, their near-unity quantum efficiency\cite{Qin_SPSCD_2024} and ultra-wide spectral sensitivity \cite{Taylor_2023} make them the premier choice for demanding quantum experiments. Furthermore, SNSPDs have being integrated in waveguides\cite{Sprengers2011,Pernice2012}, demonstrating their compatibility with reconfigurable circuits\cite{Lomonte2021} and multiplexing scheme for photon number resolution\cite{Cheng2023}. Obviously integrating these detectors on an optical chip requires the cryogenic operation\cite{Shadbolt2022} of all the device to avoid  coupling losses. Finally, another class of superconducting devices known as Transition-Edge Sensors (TESs) \cite{Lita08} offers highly efficient photon-number-resolving capabilities, albeit at the expense of slower detection rates. 

\subsubsection{Quantum error correction} 
Quantum error correction (QEC) is an essential component of a DVGM architecture, required to perform fault-tolerant computations with a quantifiable level of precision. In this context, correcting quantum errors differs from classical digital computing in two fundamental ways. First, classical bits are strictly discrete, which inherently prevents a continuous drift between the 0 and 1 states. Qubits, conversely, exist in a continuous state space, meaning they are susceptible to infinitesimal, continuous rotations between the $|0\rangle$ and $|1\rangle$ states. Second, classical logic gates inherently regenerate signal levels at each output, effectively discarding small amounts of analog noise or voltage dissipation between operations. In contrast, quantum algorithms preclude the measurement of qubit states mid-computation to avoid wave-function collapse. Because quantum states cannot be actively "refreshed," noise and decoherence effects accumulate unmitigated throughout the circuit, necessitating specialized error correction protocols.

Quantum error correction protocols mitigate dephasing and bit-flip errors, also known as Pauli errors, by encoding a single logical qubit into multiple entangled physical qubits. Crucially, errors are identified through syndrome measurements—typically parity checks—that extract information about the error without collapsing the underlying state-of-the-logical qubit. The first such scheme, a nine-qubit code, was proposed by Shor in 1995\cite{shor_scheme_1995}. This overhead was subsequently reduced to seven physical qubits by Steane\cite{steane_error_1996}, and ultimately minimized to the five-qubit code detailed in Refs.~[\onlinecite{Bennett1996,Laflamme1996}]. These foundational schemes have since been experimentally realized across various physical platforms, including photonics for the nine-qubit code\cite{zhang_loss-tolerant_2022}, trapped ions for the seven-qubit code\cite{Muller2024}, and nuclear magnetic resonance for the five-qubit code\cite{knill_five2001}.

In optical circuits, the predominant failure mode is not dephasing or qubit flip but the physical loss of photons propagating through the gates. Because the absence of a photon can be heralded by single-photon detectors at the measurement stage, this physical loss is modeled mathematically as an erasure error rather than an arbitrary Pauli error. Consequently, optical architectures benefit from significantly higher theoretical fault-tolerance thresholds, frequently surpassing 30\% for erasure compared to standard depolarizing noise models \cite{stace_thresholds_2009}. While there are proposals to mitigate this loss at the logical level utilizing parity encoding techniques\cite{ralph_loss_2005}, the majority of modern fault-tolerant architectures for photons are based on cluster-state quantum computation, which is the focus of section \ref{sec:DV Cluster model}.

Theoretical efforts to design error correction protocols with minimal overhead go hand-in-hand with the technological development of hardware capable of reducing errors at the physical level. While integrated platforms provide a clear pathway to scalability, they are highly susceptible to coupling losses at the chip interfaces, particularly when relying on off-chip photon sources or detectors. Additionally, ubiquitous propagation losses—arising predominantly from material absorption and surface scattering—continuously degrade the quantum state as it traverses the waveguide network. Mitigating these effects through the fabrication of ultra-low-loss waveguides, fabricating low loss coupling structures, and integrating as many functionalities on chip as possible, is therefore paramount for the viability of large-scale photonic integrated circuits (PICs). Significant strides in this physical scaling have been spearheaded by commercial entities. Notably, PsiQuantum\cite{PsiQuantum,Alexander2025} has demonstrated a platform with propagation losses of 0.5$\pm$0.3 dB/m, splitter losses of 0.5$\pm$0.2 mdB, and fiber-to-chip coupling losses of merely 52$\pm$12 mdB. Concurrently Xanadu\cite{AghaeeRad2025} and Hyperlight  claimed to have yielded integrated systems with 2 dB/m propagation losses and 20 mdB switch losses\cite{Xanadu_news}.
   
While the circuit-based gate model provides a standard framework for fault tolerant quantum computation, its reliance on sequential entangling operations presents significant scaling challenges for photonic platforms. As a powerful alternative, Raussendorf and Briegel introduced the \textit{cluster-state model}—or \textit{one-way quantum computation}—in 2001 \cite{Raussendorf2001}. This paradigm, described in section \ref{sec:DV Cluster model}, represents a fundamental departure from circuit-based architectures, executing quantum logic not through unitary gate sequences, but via adaptive single-qubit measurements. 

\subsection{\label{sec:DV Cluster model}Cluster State Quantum Computation}

\begin{figure*}[ht!] 
\centering
\includegraphics{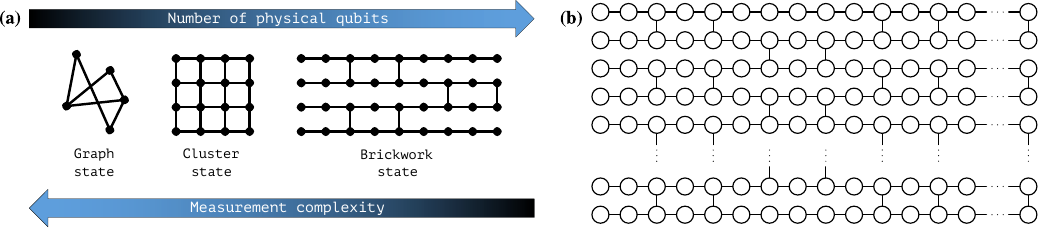}    
\caption{\textbf{Structural comparison of universal resource states in MBQC.} \textbf{(a)}
Schematic representation of the three main classes of graph-based resource states with highlighted trade-off between the number of physical qubits (increasing left to right) and the complexity of required measurement patterns (decreasing left to right). The \texttt{graph state} is an unstructured entanglement pattern, allowing arbitrary topologies at the price of complex qubit measurements. The \texttt{cluster state} is a regularly arranged, two-dimensional square lattice of entangled qubits which increases the number of physical qubits to ease the set of measurements required for universal computing. The \texttt{brickwork state} is a highly regular, layered structure involving an even higher number of physical qubits but with the smallest measurement set. 
\textbf{(b)} The brickwork state as it was presented by Ref.~[\onlinecite{Broadbent2009UniversalBlind}] for universal quantum computation using only single-qubit measurements in the angle set $\left\{0,\frac\pi4,\dots,7\frac\pi4\right\}$.}
\label{fig:DV_Cluster_Figure1}
\end{figure*}

Cluster-state quantum computation (CSQC) \cite{Raussendorf2001} shifts the experimental burden from the active application of gates to the offline preparation of a highly entangled lattice. This lattice, typically referred to as a \textit{resource state}, serves as the substrate upon which the algorithm is executed via adaptive single-qubit measurements. CSQC belongs to the broader framework of measurement-based quantum computing (MBQC), which is characterized by encoding information processing within a specific measurement pattern rather than a sequence of unitary operations. Notably, it has been demonstrated that MBQC can efficiently simulate any standard quantum circuit\cite{Nielsen2006Cluster}, providing a universal model of computation that, when paired with appropriate error correction protocols, can tolerate photon loss rates of up to 50\% \cite{Varnava2006}.

CSQC can be viewed as a natural generalization of quantum teleportation \cite{zhou2000methodology, bennett1993teleporting}. While standard teleportation transfers a quantum state via a single shared entangled pair and a joint measurement, cluster-state computation propagates quantum information across a pre-prepared, multi-qubit entangled resource state. Through a sequence of adaptive single-qubit measurements, the system effectively implements continuous gate teleportation, thereby realizing universal quantum computation. Furthermore, this measurement-driven architecture makes CSQC uniquely suited for delegated settings. In such protocols, a remote server prepares the resource state and executes the measurements, while a client uses classical communication to steer the computation—naturally enabling secure or "blind" quantum computing without revealing the underlying algorithm to the server.

Building on these concepts, this section will introduce different families of resource states within photonic platforms.The photonic platform is inherently aligned with measurement-based protocols; photons naturally facilitate the low-loss routing of quantum information, while integrated and bulk optics supply a robust toolkit for state generation and reconfigurable projection. Formally, these resource states are categorized by their geometrical graph topologies. By increasing graph regularity, one effectively trades a higher physical qubit overhead for vastly simplified measurement sequences, as depicted in Fig.~\ref{fig:DV_Cluster_Figure1}a. A prominent example is the brickwork state \cite{Broadbent2009UniversalBlind} (Fig.~\ref{fig:DV_Cluster_Figure1}b), which pairs a universal architecture with a restricted set of required measurement angles. This feature is highly advantageous for both experimental implementations and the delegated client-server protocols discussed later.

Finally, we note that cluster states are not the only multipartite entangled resources utilized in quantum information. The Greenberger-Horne-Zeilinger (GHZ) state, for instance, plays a prominent role in tests of non-locality, quantum communication \cite{Barz2015Quantum}, and fusion-based quantum computing (FBQC)\cite{Browne2005,Bartolucci2023} which relies on network-wide, two-qubit Bell measurements and can dynamically generate both cluster- and GHZ-type resource states via heralded operations\cite{Paesani2023High}.

\subsubsection{Definition of Cluster States}
Mathematically, cluster states form a subclass of graph states~\cite{hein2004multiparty,Bruss2007Lectures}, in which qubits are arranged on a two-dimensional square lattice. They are generated by preparing each physical qubit in the state $\ket{+}=(\ket{0}+\ket{1})/\sqrt{2}$ and subsequently entangling nearest neighbors through controlled-$Z$ (\textit{CZ}) gates. Denoting by $V$ the set of vertices and by $E$ the set of edges, the resulting cluster state $\ket{C}$ is given by
\begin{equation}
    \ket{C}=\prod_{(i,j)\in E} CZ_{ij} \bigotimes_{i\in V} \ket{+}_i.
\end{equation}

When a two-dimensional lattice serves as the computational substrate, each row corresponds to a logical qubit—often visualized as a distinct \textit{wire}—while the columns represent successive discrete steps in their evolution. Each column thus defines a measurement layer that advances the logical state, playing a role analogous to time in the standard circuit model. Within this framework, the input and output are encoded in the physical qubits located at opposite boundaries of the lattice, and the specific measurement pattern dictates how quantum information propagates across the resource state.

Although MBQC proceeds via inherently probabilistic single-qubit projective measurements, the overarching logical computation must be strictly deterministic. This requirement was formalized in Ref.~[\onlinecite{Danos2006Determinism}] by introducing the \textit{flow condition} for open graph states, thereby defining the families of measurement patterns that guarantee deterministic computation. Under these conditions, the random measurement outcomes induce well-defined Pauli by-products on the logical state. These by-products can be systematically corrected using Pauli operations derived entirely from the underlying graph topology and the input-output geometry. In practice, this necessitates a classical feed-forward mechanism, wherein the basis choices for future measurements are dynamically updated, conditioned upon prior measurement outcomes.

In section \ref{projective}, we detail the formalism of single-qubit projective measurements, characterizing them through Bloch sphere vectors and basis transformations. This framework provides the necessary tools to specify the exact measurement bases required for MBQC, thereby establishing a rigorous link between the physical measurement angles and the resulting logical operations.

\subsubsection{Projective measurements} \label{projective}
Projective measurements offer an intuitive geometric interpretation on the Bloch sphere. Each measurement basis corresponds to a unit vector that defines the measurement axis, conventionally parameterized by the spherical angles $\theta$ and $\phi$. Operationally, however, it is often more convenient to implement these arbitrary projections as a specific unitary rotation followed by a standard measurement in the computational basis.

To this end, recall that single-qubit rotations are generated by Pauli operators. A rotation of angle $\varphi$ around a Cartesian axis $\vec{k}\in\left\{\vec{x},\vec{y},\vec{z}\right\}$ is described by the unitary\cite{Bruss2007Lectures}
\begin{equation}
    \hat{U}_{\vec{k}}\left(\varphi\right)=e^{-\frac{i\varphi}{2}\hat{\sigma}_{\vec{k}}},
\end{equation}
where $\hat{\sigma}_{\vec{k}}$ denotes the Pauli operator along the chosen direction.

Since measures are typically performed in the computational basis, i.e. in the eigenbasis of $\hat{\sigma}_{\vec{z}}$, a projective measurement along a generic axis $(\theta,\phi)$ can be realized by applying a suitable single-qubit basis transformation and then performing a $\hat{\sigma}_{\vec{z}}$ measurement. In the notation adopted here, an explicit sequence achieving this change of measurement axis is
\begin{equation}
\hat{U}_{\vec{z}}\left(\phi+\frac\pi2\right)\hat{U}_{\vec{x}}\left(\theta\right)\hat{\sigma}_{\vec{z}}\hat{U}_{\vec{x}}\left(-\theta\right)\hat{U}_{\vec{z}}\left(-\phi-\frac\pi2\right).
\end{equation}
After this basis change, a standard $\hat{\sigma}_{\vec{z}}$ measurement yields the desired projective measurement along the direction identified by $\theta$ and $\phi$.

This single-qubit measurement procedure is a fundamental ingredient of MBQC: once qubits are entangled into the resource state, local measurements steer the logical evolution. In subsection \ref{Paulibyproduct}, we show explicitly with an example how measurements can both implement a logical transformation and generate measurement-dependent Pauli by-products, thereby motivating the need for classical feedforward.

\begin{figure*}[ht!] 
\centering
\includegraphics{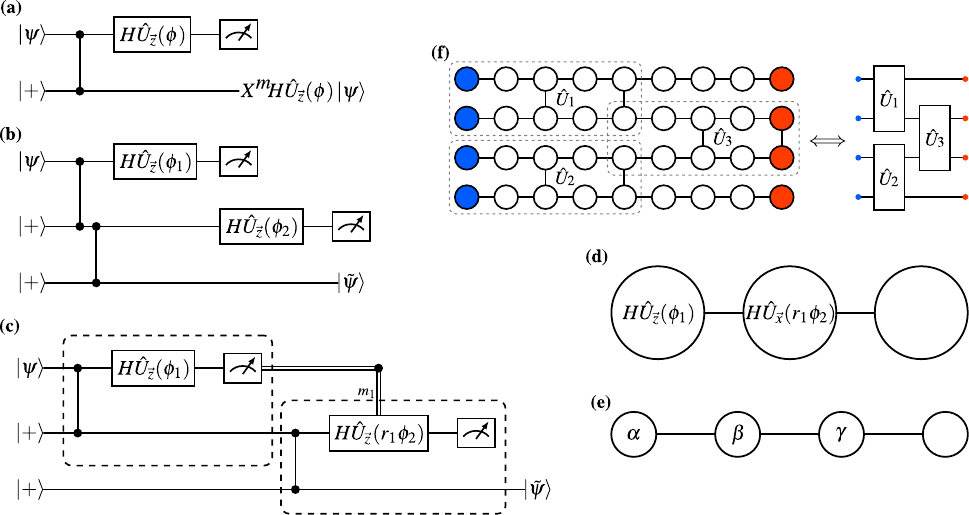}    
\caption{\textbf{From teleportation circuits to cluster state computation in MBQC.} \textbf{(a)}  Schematic depicting the transfer of an arbitrary qubit state $\ket{\psi}$ by entangling it with an ancilla prepared in the $\ket{+}$ state, followed by  a sequence of rotations and projective measurement. \textbf{(b)} Extension of \textbf{(a)} to two chained teleportation protocols enabling the sequential transfer and manipulation of the quantum state through independent measurement stages. \textbf{(c)} A variant of \textbf{(b)} where teleportation operations are performed in a specific temporal sequence, illustrating the interplay between measurement result $m$ and conditional logic in the protocol. \textbf{(d)} Mapping of the circuit in \textbf{(c)} to a linear cluster, where each node and link represent, respectively, a physical qubit and entanglement. The two representations are equivalent. \textbf{(e)} An example of a four-qubit linear cluster for the implementation of arbitrary single-qubit operations via appropriate measurement pattern, represented by a sequence of measurements along the angles $\alpha,\beta,\gamma$. \textbf{(f)} Example of execution of two-qubit gates in brickwork cluster architecture, as explained in Ref.~[\onlinecite{Broadbent2009UniversalBlind}]. The left panel shows a brickwork state, where blue and red nodes indicate input and output states respectively. Grouped nodes represent unitary gates $\hat{U}_1,\hat{U}_2,\hat{U}_3$ in the brickwork state model and the right panel reports its equivalent quantum circuit representation, showing the equivalence between cluster state and circuit computations.}
\label{fig:DV_Cluster_Figure3}
\end{figure*}

\subsubsection{Pauli by-products and Feed forward}\label{Paulibyproduct}
To illustrate the emergence of Pauli by-products in MBQC, it is instructive to analyze a simple protocol, as discussed by Refs.~[\onlinecite{Nielsen2006Cluster},\onlinecite{Bruss2007Lectures}]. Consider a first qubit prepared in the generic superposition $\ket{\psi}=\alpha\ket{0}+\beta\ket{1}$, with the usual normalization, and a second qubit initialized in the $\ket{+}$ state. By applying a \textit{CZ} gate between them, the two-qubit state becomes
\begin{equation}
\ket{\varphi}=CZ\left(\ket{\psi}\ket{+}\right)=\left(\alpha\ket{0}\ket{+}+\beta\ket{1}\ket{-}\right).
\end{equation}
Now, perform a projective measurement of the first qubit in the basis $\left\{\ket{+_\phi},\ket{-_\phi}\right\}$, where $\ket{\pm_\phi}=\left(\ket{0}\pm e^{i\phi}\ket{1}\right)/\sqrt{2}$ which corresponds to measuring in the direction $(\frac{\pi}{2},\phi)$ on the Bloch sphere. By labeling the measurement outcome as $r$, the post-measurement state of the second qubit is either
\begin{equation}
    \begin{gathered}
        \alpha\ket{+}+e^{i\phi}\beta\ket{-}\qquad\text{if}\quad r=+1,\\
        \text{or}\\
        \alpha\ket{+}-e^{i\phi}\beta\ket{-}\qquad\text{if}\quad r=-1.
    \end{gathered}
\end{equation}
Define a binary variable $m\in\{0,1\}$ such that $r=(-1)^m$. The state of the second qubit can be rewritten compactly as
\begin{equation}
    X^mH\hat{U}_{\vec{z}}(\phi)\ket{\psi}.
\end{equation}
This protocol is shown in Fig.~\ref{fig:DV_Cluster_Figure3}a, where both the teleportation of the input state $\ket{\psi}$ and the logical action associated with the measurement angle $\phi$ become explicit.

The $X^m$ factor is referred to as a Pauli by-product and arises directly from the probabilistic nature of quantum measurements. Its effect can be compensated by applying an $X$ correction conditioned on $m$ immediately after the measurement. Alternatively, in specific cases, the same correction can be carried out in post-processing by appropriately bit-flipping the measurement outcome of the second qubit.

When two such protocols are concatenated, as shown in Fig.~\ref{fig:DV_Cluster_Figure3}b, the output state $\ket{\tilde{\psi}}$ is given by
\begin{equation}
     \ket{\tilde{\psi}}=X^{m_2}Z^{m_1}H\hat{U}_{\vec{x}}(r_1\phi_2)H\hat{U}_{\vec{z}}(\phi_1)\ket{\psi}.
     \label{eq:TwoConcatenatedTeleport}
\end{equation}
Due to the commutation relations among $H$, Pauli operators, and the rotation operators $\hat{U}_{\vec{k}}$, the by-products combine into $X^{m_2}Z^{m_1}$, while the intended rotation on the final qubit is affected also by the earlier measurement result through the sign of $\phi_2$, which change $\hat{U}_{\vec{z}}(\phi_2)$ into $\hat{U}_{\vec{x}}(r_1\phi_2)$.

Determinism is recovered by classical feedforward: by forwarding the measurement outcome $m_1$ and updating the subsequent measurement instruction according to the map $\phi_2 \rightarrow (-1)^{m_1}\phi_2$, one can deterministically realize the desired algorithm on the initial state $\ket{\psi}$. This mechanism is reflected in the time-ordered circuit and in the corresponding cluster representation in Figs.~\ref{fig:DV_Cluster_Figure3}c-d.
The same sequence of operation can be expressed within the MBQC framework using a linear three-qubit cluster, as depicted in Fig.~\ref{fig:DV_Cluster_Figure3}d, where each node represents a physical qubit initialized in $\ket{+}$ and logical information flows from left to right. For compactness, the operators applied at each node (or equivalently the measurement angles) can be annotated directly on the corresponding circles.

More generally, concatenating further teleportation-like steps yields an alternating product of $\hat{U}_{\vec{z}}(\cdot)$ and $\hat{U}_{\vec{x}}(\cdot)$ rotations (with signs determined by previous outcomes and corrected via feedforward), as already emerged from the transition $\hat{U}_{\vec{z}}(\phi_2)\rightarrow \hat{U}_{\vec{x}}(r_1\phi_2)$ in Eq.~\ref{eq:TwoConcatenatedTeleport}. Euler's theorem states that any single-qubit SU(2) operation can be decomposed as $\hat{U}_{\vec{z}}(\gamma)\hat{U}_{\vec{x}}(\beta)\hat{U}_{\vec{z}}(\alpha)$, implying that a four-qubit linear cluster, as shown in Fig.~\ref{fig:DV_Cluster_Figure3}e, is sufficient to implement an arbitrary single-qubit unitary transformation through an appropriate measurement pattern. Note that, although the initial resource state involves four physical qubits, the computation acts on a single logical qubit which is progressively delocalized and finally encoded in the last, unmeasured qubit (up to Pauli by-products).

At this stage, it is useful to recall the distinction between physical and logical qubits. Physical qubits correspond to the actual states (like photons) forming the cluster state, whereas the logical qubit is the encoded quantum information that propagates through the entanglement structure as measurements are performed. In this picture, the logical qubit is effectively delocalized over several physical qubits, each associated with successive discrete steps of its evolution.
Finally, it is important to stress that by-products are not a mere technicality but an intrinsic feature of MBQC. Indeed, Ref.~[\onlinecite{Morimae2014Measurement}] showed that, under the no-signaling principle, no universal MBQC resource can be entirely by-product free, thereby formalizing the necessity of feedforward during the computation.

The elementary examples shown in this subsection illustrates how to steer the computation on a single logical qubit and why adaptive updates of measurement bases are essential in practice. In subsection \ref{sec:TheBrickworkState}, we turn to a particularly structured universal resources, the brickwork state and fusion based quantum computing.

\subsubsection{The Brickwork State}\label{sec:TheBrickworkState}
Among the various graph topologies suited for measurement-based protocols, an example capable of universal quantum computation is the \textit{brickwork state}. This resource state enable universal quantum computation via measurements limited only to the $X-Y$ plane of the Bloch sphere\cite{Broadbent2009UniversalBlind}, specifically with measurements along the basis $\{\ket{+_\phi},\ket{-_\phi}\}$ with $\ket{\pm_\phi}=\left(\ket{0}\pm e^{i\phi}\ket{1}\right)/\sqrt{2}$, corresponding to the direction $(\frac{\pi}{2},\phi)$ on the sphere. A key advantage of this approach is that universality can be obtained using only a set of eight measurement angles $\phi\in\{0,\frac\pi4,\dots,7\frac\pi4\}$, at the cost of requiring a larger number of physical qubits in the underlying lattice.

The brickwork state owes its name to the visual similarity of its entanglement graph to a brickwork wall. As formally constructed by Broadbent \textit{et al.}\cite{Broadbent2009UniversalBlind}, a brickwork state $G_{n\times m}$, where $m=5\mod{8}$, is an entangled state of $n\times m$ qubits constructed as follows:
\begin{enumerate}
    \item Prepare all qubits in the state $\ket{+}$ and assign to each qubit an index $(i,j)$, $i$ being a row ($i\in[n]$) and $j$ being a column ($j\in[m]$).
    \item For each row, apply the \textit{CZ} gate between qubits $(i,j)$ and $(i,j+1)$ where $1\leq j \leq m-1$.
    \item For each column $j\equiv3\mod8$ and each odd row $i$, apply the \textit{CZ} gate between qubits $(i,j)$ and $(i+1,j)$ and also on qubits $(i,j+2)$ and $(i+1,j+2)$.
    \item For each column $j\equiv7\mod8$ and each even row $i$, apply the \textit{CZ} gate between qubits $(i,j)$ and $(i+1,j)$ and also on qubits $(i,j+2)$ and $(i+1,j+2)$.
\end{enumerate}
The resulting brickwork state is shown in Fig.~\ref{fig:DV_Cluster_Figure1}b.

Once the resource state is prepared, the computation proceeds via a sequence of single-qubit measurements, each chosen from the finite angle set $\{0, \pi/4, \dots, 7\pi/4\}$. As established previously, classical feed-forward is essential: measurement bases must be adaptively updated based on prior outcomes to ensure deterministic logical evolution. Figure~\ref{fig:DV_Cluster_Figure3}f illustrates the application of two-qubit gates across four logical qubits within this architecture, with red and blue indicating the input and output states, respectively. Crucially, synthesizing an arbitrary two-qubit unitary $U \in \mathrm{SU}(4)$ requires a minimum of three entangling gates (e.g., \textit{CX} or \textit{CZ}) interleaved with single-qubit rotations \cite{Vidal2004Universal}. The brickwork lattice naturally accommodates this requirement for universality: appropriate measurement sequences along each horizontal wire execute arbitrary single-qubit rotations, while the vertical entangling links facilitate two-qubit interactions. By cascading these fundamental primitives across multiple columns, any arbitrary two-qubit operation can be deterministically realized.

The MBQC protocol based on brickwork states can be divided into the following conceptual steps:
\begin{enumerate}
    \item \textit{Initialization:} Prepare all the qubits in the $\ket{+}$ state and entangle them according to the brickwork state structure.
    \item \textit{Measurement sequence:} Measure each qubit $(i,j)$, $i$ for the row and $j$ for the column, along the corresponding angle $\phi_{i,j}\in\{0,\frac\pi4,\dots,7\frac\pi4\}$, according to the algorithm to be executed.
    \item \textit{Feedforward correction:} Owing to the random nature of quantum measurement, after each measurement update the subsequent measurement angle accordingly.
    \item \textit{Computation outcome:} The results are either classical, if all qubits are measured, or quantum, if a subset remains unmeasured as the output register.
\end{enumerate}

Implementing the Brickwork state experimentally is extremely challenging because both photon source and $CZ$ gates are probabilistic, making the generation of a large state with the desired graph structure hard to realize. For this reason attempts to build a large scale photonic quantum computer by private companies use continuous-variable cluster states\cite{xanadu}, which have been generated deterministically over millions of modes\cite{yoshikawa2016invited}, or protocols based on probabilistic fusion gates and percolation\cite{PsiQuantum,Rudoplh2017}.


\subsubsection{Fusion based quantum computation}
While the brickwork state offers an ideal, universal resource state with a minimized measurement requirements, its physical realization in photonic platforms encounters severe hardware limitations. The generation of such a highly entangled, two-dimensional lattice requires deterministic single-photon sources and deterministic two-qubit entangling operations, such as the \textit{CZ} gate. In linear optical quantum computing, however, two-qubit gates are inherently probabilistic. Relying on non-deterministic \textit{CZ} gates to construct a state with a defined connectivity would results in a non-optimal resource state generation, making it practically infeasible for large-scale computation.

To overcome this issue, an alternative, highly resource-efficient approach relies on the concept of \textit{fusion gates}, as introduced by Browne and Rudolph \cite{Browne2005}. Rather than attempting to construct a large-scale cluster state through a sequence of probabilistic gates, the fusion-based approach builds the resource state incrementally from smaller, readily available entangled resources, such as Bell pairs. The core of this architecture relies on two types of non-deterministic, yet heralded, fusion operations. Type-I fusion operates by mixing two spatial modes on a polarizing beam splitter (PBS) and performing polarization-discriminating photon counting after a $45^\circ$ rotation on one output of the PBS. Upon a successful detection signature, which occurs with a 50\% probability, the two distinct cluster states are joined into a single.  If the Type-I fusion is applied to the end-qubits of linear (i.e. one-dimensional) clusters of lengths $n$ and $m$, successful outcomes generate a linear cluster of length $(n+m-1)$ larger cluster. However, a failure outcome measures the input qubits in the computational ($\sigma_z$) basis, effectively severing their bonds. While Type-I fusion is efficient for generating one-dimensional linear clusters, it is insufficient for creating two-dimensional lattices, as failed fusions would continuously split the generated clusters.

To address this limitation and construct 2D topologies, Type-II fusion is employed in conjunction with a redundant encoding strategy. Type-II fusion rotates the polarization of the two photons entering the ports of a PBS by 45$^\circ$ and the operation succeed if one photon is measured from each output port. By encoding a single logical qubit across multiple physical photons, a Type-II fusion attempt that fails merely acts as a $\sigma_x$ measurement on the involved photons. This failure reduces the photon redundancy but crucially does not destroy the existing cluster bonds. Consequently, fusion can be reattempted as long as sufficient redundant photons remain, enabling the dynamic construction of complex two-dimensional networks. 

Quantum computation schemes based on probabilistic Type-II fusion gates that are tolerant to photon losses—even without active error correction—have been proposed \cite{Segovia2015} (tolerating 1.6\% loss per fusion gate), subsequently improved \cite{Bartolucci2023} (up to 10.4\% loss), and recently optimized for specific noise models in quantum emitters \cite{Chan2025}. The fundamental potential of this approach lies in its ability to achieve fault-tolerant quantum computation using inherently probabilistic gates \cite{Auger2018}. It accomplishes this by exploiting entanglement percolation within a macroscopic cluster state, where the links between nodes are established probabilistically. As detailed in Ref. [\onlinecite{Rudoplh2017}], realizing this architecture requires efficient multiplexing schemes to transform arrays of heralded single-photon sources into near-deterministic ones. Furthermore, scaling this approach to the massive number of physical qubits necessary for fault tolerance can only be achieved through integrated optics. This is precisely the strategy adopted by PsiQuantum \cite{PsiQuantum, Alexander2025}, which is developing a scalable technological platform capable of manufacturing the massive volume of integrated devices required for this endeavor.

\subsubsection{Experimental Realizations of Resource States for MBQC}\label{sec:ExperimentalRealizationResourceStates}
\begin{figure*}[ht!]
    \centering
    \includegraphics{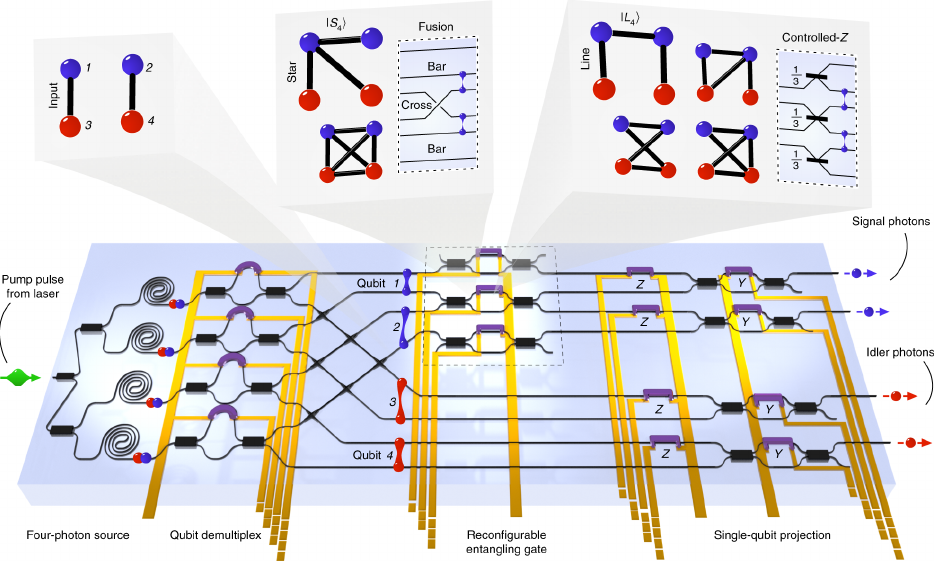}
    \caption{\textbf{Example of photonic integrated circuit for MBQC.} An integrated silicon photonics chip featuring four telecom-band photon-pair sources. A qubit demultiplexer organizes these into two Bell pairs, which are then routed through a reconfigurable entangling gate. Each qubit undergoes individual manipulation and analysis via Mach–Zehnder interferometers enabling Y and Z rotations. Above each schematic, the relevant graph states are displayed, illustrating how the system transitions from two input Bell pairs to either a star or line graph state, depending on the configuration of the entangling gate. Reproduced from Adcock et al., Nat. Commun. 10, 3528 (2019); licensed under a Creative Commons Attribution (CC BY) license.\cite{Adcock2019Programmable}}
    \label{fig:DV_Cluster_Figure4}
\end{figure*}
\begin{table*}[ht!]
\centering
\begin{tabularx}{\textwidth}{ 
   >{\centering\arraybackslash\hsize=.4\hsize\linewidth=\hsize}X 
   >{\centering\arraybackslash\hsize=.8\hsize\linewidth=\hsize}X 
   >{\centering\arraybackslash\hsize=.6\hsize\linewidth=\hsize}X 
   >{\centering\arraybackslash\hsize=1\hsize\linewidth=\hsize}X 
   >{\centering\arraybackslash\hsize=2.2\hsize\linewidth=\hsize}X} 
\hline
\rowcolor[HTML]{FD6864}[\dimexpr\tabcolsep+0.1pt\relax]
\textcolor{white}{\textbf{Year}} & \textcolor{white}{\textbf{Reference}} &  \textcolor{white}{\textbf{Platform}} &\textcolor{white}{\textbf{Resource state}} & \textcolor{white}{\textbf{Experimental highlights}}\\ \hline
2007 & Lu \textit{et al.}\cite{Lu2007Experimental} & Bulk & 6-photon cluster and GHZ & First demonstration of 6-photon cluster and GHZ\\
\rowcolor[HTML]{FFEDEC}[\dimexpr\tabcolsep+0.1pt\relax]  
2016 & Wang \textit{et al.}\cite{Wang2016Experimental} & | & 10-photon cluster & Scalable, complex interferometer stabilization\\   
| & Ciampini \textit{et al.}\cite{Ciampini2016Path} & Integrated & 4-photon linear cluster & On-chip hyperentanglement on path and polarization\\
\rowcolor[HTML]{FFEDEC}[\dimexpr\tabcolsep+0.1pt\relax]  
2019 & Adcock \textit{et al.}\cite{Adcock2019Programmable} & | & 4-photon graph & On-chip programmable graph topologies\\   
2020 & Istrati \textit{et al.}\cite{Istrati2020} & Fiber loop & Linear cluster & Continuous, sequential generation\\   
\rowcolor[HTML]{FFEDEC}[\dimexpr\tabcolsep+0.1pt\relax]  
2021 & Vigliar \textit{et al.}\cite{Vigliar2021Error} & Integrated & Cluster & Error correction via phase estimation MBQC\\   
2023 & Coste \textit{et al.}\cite{Coste2023High} & Bulk & Linear cluster & Spin-photon sequential generation\\
\rowcolor[HTML]{FFEDEC}[\dimexpr\tabcolsep+0.1pt\relax]  
| & Chan \textit{et al.}\cite{Chan2023OnChip} & Integrated & | & Coherent spin-photon interface, on-demand generation\\
| & Cogan \textit{et al.}\cite{Cogan2023} & | & | & GHz rate, scalable, high indistinguishability\\
\rowcolor[HTML]{FFEDEC}[\dimexpr\tabcolsep+0.1pt\relax]  
2024 & Chen \textit{et al.}\cite{Chen2024Heralded} & | & 3-photon GHZ & On-chip heralded states for FBQC\\   
| & Maring \textit{et al.}\cite{maring2024versatile} & | & GHZ and cluster & QD source and programmable photonic chip\\   
\rowcolor[HTML]{FFEDEC}[\dimexpr\tabcolsep+0.1pt\relax]  
| & Pont \textit{et al.}\cite{Pont2024High} & | & | & Integration of QDs and photonic circuits\\   
| & Meng \textit{et al.}\cite{Meng2024} & | & 3-qubit cluster & On-chip, deterministic single-emitters\\   
\rowcolor[HTML]{FFEDEC}[\dimexpr\tabcolsep+0.1pt\relax]  
| & Su \textit{et al.}\cite{Su2024Continuous} & | & Cluster & Continuous, deterministic, high-rate production\\   
| & Cao \textit{et al.}\cite{Cao2024Photonic} & Bulk & 3-photon GHZ & Heralded, high-purity 3-photon GHZ states\\   
\rowcolor[HTML]{FFEDEC}[\dimexpr\tabcolsep+0.1pt\relax]  
| & Lib \textit{et al.}\cite{Lib2024Resource} & | & Cluster & High-dimensional, spatial encoding, effective scaling\\   
2025 & Huet \textit{et al.}\cite{Huet2025Deterministic} & Bulk 
& | & Real-time reconfigurable, 
topology control\\   
\rowcolor[HTML]{FFEDEC}[\dimexpr\tabcolsep+0.1pt\relax]  
\hline
\end{tabularx}
\caption{\textbf{Experimental demonstration of photonic cluster and GHZ state generation.} This table summarizes the advances in the experimental generation of cluster and GHZ states using both bulk and integrated photonic platforms. Moreover, the works listed highlight the evolution from foundational demonstrations of multi-photon entanglement to scalable, programmable, and deterministic on-chip resource generation for MBQC. Symbol "|" means that the entry is the same as above.}
\label{tab:ResourceStates}
\end{table*}

The experimental generation of photonic cluster states for MBQC has been demonstrated across several platforms, including bulk free-space optics, integrated waveguide chips, and deterministic solid-state emitters. In 2007, Ref.~[\onlinecite{Lu2007Experimental}] reported the realization of six-photon graph and GHZ states using spontaneous parametric down-conversion (SPDC) sources in a bulk-optic regime, providing a crucial early testbed for multipartite entanglement. By 2016, improvements in active phase stabilization and source brightness allowed the entanglement complexity to be scaled up to a ten-photon state \cite{Wang2016Experimental}, further advancing the limits of the probabilistic approach based on parametric photon generation.

Recognizing the inherent scalability limits of bulk free-space optics, researchers investigate the potential of integrated photonic architectures. Leveraging the enhanced stability and reduced physical footprint of these platforms, early on-chip experiments successfully generated four-photon linear cluster states using hyperentanglement—simultaneously encoding quantum information in both path and polarization degrees of freedom\cite{Ciampini2016Path}. This capability was expanded in 2019 with the demonstration of a fully programmable four-photon graph state on a silicon waveguide device. By networking on-chip photon sources through tunable interferometric circuits \cite{Adcock2019Programmable}, the system dynamically accessed a diverse array of entanglement topologies (see Fig.~\ref{fig:DV_Cluster_Figure4}). Building upon this integrated paradigm, Ref.~[\onlinecite{Vigliar2021Error}] successfully implemented error-protected logical qubits on a silicon chip in 2021. By executing a phase-estimation algorithm, the study established a critical milestone for on-chip fault tolerance in MBQC, yielding success probabilities in excess of 95\%—a stark improvement over the 62\% baseline achieved without error correction using the same photonic overhead.

Parallel to the advancement of platform performance, accelerated progress in photon source quality has heavily driven the scaling of resource states. In 2020, Ref.~[\onlinecite{Istrati2020}] leveraged a single quantum dot within a fiber-loop setup to sequentially generate linear cluster states. More recently, in 2024, Refs.~[\onlinecite{Chen2024Heralded}] and [\onlinecite{Cao2024Photonic}]—utilizing integrated and bulk-optics approaches with engineered photon sources, respectively—demonstrated heralded three-photon GHZ states. These states act as critical building blocks for large-scale fusion-based quantum computing (FBQC)\cite{Bartolucci2023}. Furthermore, Refs.~[\onlinecite{maring2024versatile}] and [\onlinecite{Pont2024High}] demonstrated that combining deterministic quantum dot emitters with reprogrammable photonic circuits allows for the direct on-chip generation of high-fidelity entangled states, effectively merging generation efficiency with architectural scalability. Finally, Ref.~[\onlinecite{Lib2024Resource}] opened the door to parallel approaches using high-dimensional spatial encoding, proving that scalable resource generation is not strictly limited to increasing the absolute number of photons.

At the time of this review, the deterministic generation regime is largely investigated by employing spin-photon interfaces and 
resource engineering \cite{deGliniasty2024,Chan2025practical}.
Recently, the works in Refs.~[\onlinecite{Coste2023High},\onlinecite{Chan2023OnChip},\onlinecite{Cogan2023},\onlinecite{Meng2024}] used spin-photon entanglement in quantum emitters to generate multi-photon cluster states on-demand, thus establishing  solid, state-of-the-art realizations of cluster state strings generation according to the original proposal by Ref.~[\onlinecite{Lindner2009Proposal}] in 2009. 
In 2024, Ref.~[\onlinecite{Su2024Continuous}] proposed a strategy toward continuous, high-rate deterministic generation of large cluster states, with photon indistinguishability suitable for integration into large architectures, thus improving cluster state generation scalability.
Finally, in 2025, Ref.~[\onlinecite{Huet2025Deterministic}] introduced methods for 
reconfigurability of cluster states topology and tested in an hybrid spin-photon platform.
As aforementioned, the recently introduced FBQC technique \cite{Bartolucci2023} also constitutes a promising route toward the generation of large photonic resource states for fault-tolerant quantum computing \cite{Paesani2023High}. Temporal fusion of entangled states from a quantum emitter was demonstrated in Ref.~[\onlinecite{Meng2025}].
To improve the scalability of fusion operations enabled by a hybrid spin-photon hardware platform, Ref. [\onlinecite{Chan2025}] introduced a noise-resilient architecture specifically designed to mitigate the physical noise sources inherent in such devices. Concurrently, Ref. [\onlinecite{Wein2024minimizing}] proposed schemes to minimize the associated resource overhead.

Table~\ref{tab:ResourceStates} provides a concise summary of the experimental resource state implementations, which pave the way for practical applications like delegated and blind quantum computing discussed below.

\subsubsection{Rapidly growing applications of MBQC: Delegated and Blind Quantum Computing}
\begin{figure*}[htb]
    \centering
    \includegraphics[width=0.9\textwidth]{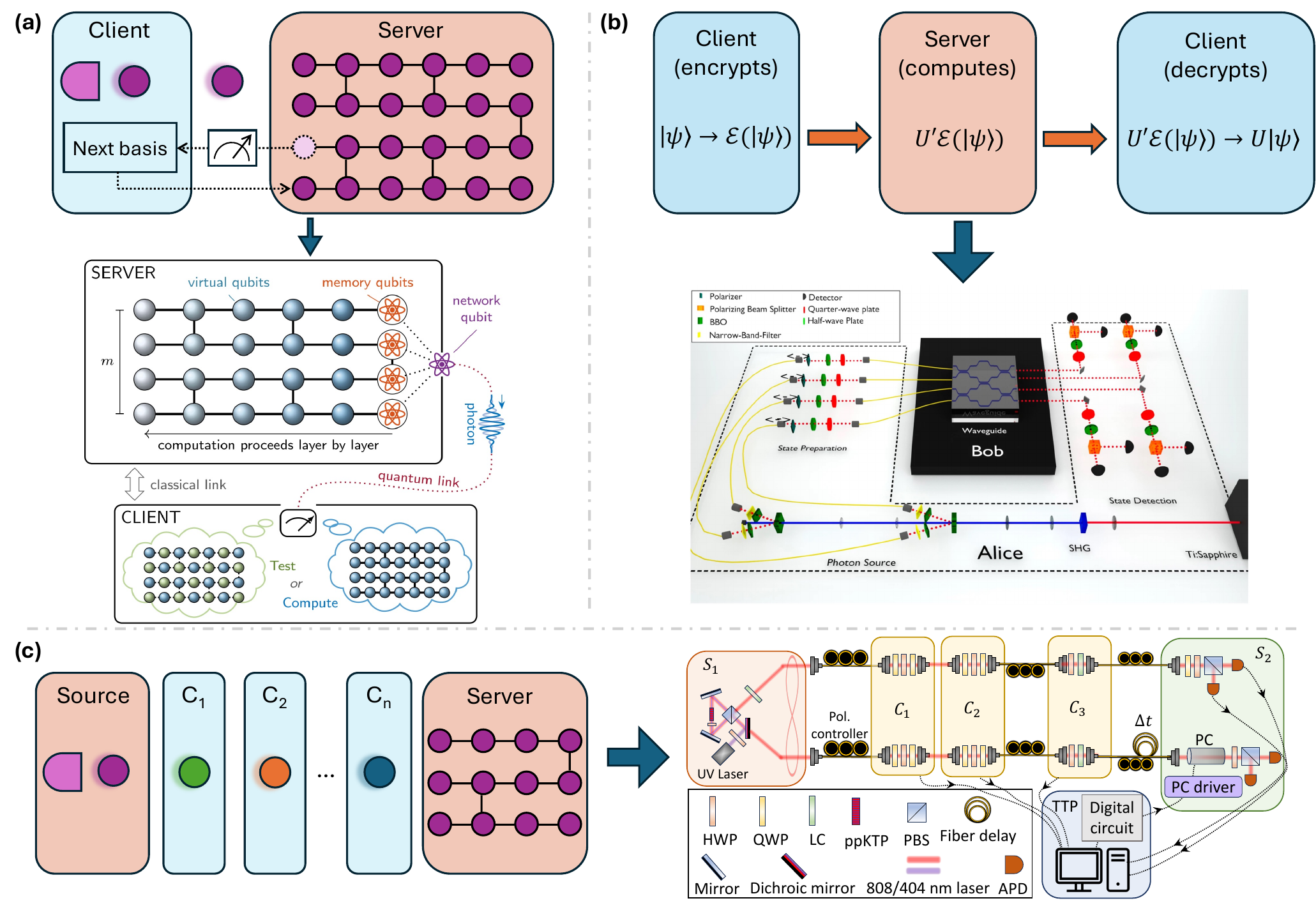}
    \caption{\textbf{Blind quantum computing and related protocols. (a)} In the original BQC protocol, a client prepares random qubits and sends them to a remote quantum server which produces a cluster state. At the end of this round of quantum communication, the client and the server interacts only classically. At each classical round, the client instructs the server with a single-qubit measurement basis, the server performs the measurement, and then returns the outcome to the client for the computation of the next measurement basis. The idea is that the server contains $m$ memory qubits (orange) and one network qubit (violet), which can be entangled with each other. The network qubit can also be entangled with a photon, enabling the client to control the server’s qubit without revealing its state, via photon measurement. Below in the panel, figure reproduced from Drmota et al., Phys. Rev. Lett. 132 (2024); licensed under a Creative Commons Attribution (CC BY) license.\cite{ Drmota2024} \textbf{(b)} In quantum homomorphic encryption (QHE), the client applies an encryption protocol to their private data, i.e., a quantum state $\ket{\psi}$, and sends the encrypted data, $\mathcal{E}(\ket{\psi})$, to a server. The server performs an evaluation on the encrypted data, obtaining an encrypted outcome $U'\mathcal{E}(\ket{\psi})$. This is returned to the client, who applies a decryption protocol to the outcome to retrieve the plain computation outcome $U\ket{\psi}$. Below in the panel, the experimental set-up reproduced from Zeuner et al., npj Quantum Information 7 (2021); licensed under a Creative Commons Attribution (CC BY) license. \cite{Zeuner2021}. This experiment realizes a QHE protocol based on the Boson Sampling and the quantum walk paradigms. \textbf{(c)} In Refs.~[\onlinecite{polacchi2023multi},\onlinecite{Polacchi2025}], the authors consider a multi-client setting by employing the Qline architecture \cite{Doosti2024}. A source of single qubits sequentially sends qubits to a chain of $n$ clients who apply random single-qubit rotations to each qubit before sending them to the remote quantum server. This is implemented through the setup shown on the right side of the panel, reproduced from Polacchi et al., Phys. Rev. Lett. 134 (2025); licensed under a Creative Commons Attribution (CC BY) license.\cite{Polacchi2025} Here the source produces pairs of entangled photons through SPDC, clients use bulk polarizing optics to apply polarization rotations, while the server performs adaptive polarization measurements on the received qubits, through an active feed-forward system enabled by an electro-optical modulator. PC: Pockels Cell; HWP/QWP: Half/Quarter Wave-Plate; PBS: Polarizing Beam Splitter; LC: Liquid Crystal; APD: Avalanche Photo-Diode.}
    \label{fig:immagine_BQC}
\end{figure*}

Despite significant progress toward the realization of large-scale quantum computers, these machines will inherently require sophisticated hardware, specialized maintenance, and stringent environmental conditions to operate reliably. Consequently, in the near term, quantum processors will likely be housed in a limited number of centralized facilities worldwide and accessed remotely by users via cloud services \cite{Caleffi2024}. This scenario closely mirrors the current deployment of artificial intelligence chatbots and large language models, which execute their complex workloads on massive data-center infrastructures while interfacing with end-users through standard personal computers.

A natural consequence of this cloud-based scenario is the emergence of significant privacy concerns. Specifically, transmitting sensitive user data to a remote quantum server inherently introduces the risk of information leakage to potentially malicious parties, thereby necessitating robust protocols to guarantee user privacy. In this context, MBQC provides a framework for the realization of secure delegated quantum computing protocols. As ideal quantum information carriers, photons will play a major role in these applications, serving not only as the secure communication link but also as a computational resource itself. Examples of such protocols are blind quantum computing\cite{Broadbent2009UniversalBlind} (BQC), and quantum homomorphic encryption\cite{Tan2016} (QHE). Here, we briefly revise them and present some photonic experimental demonstrations. Existing experimental implementations of BQC, QHE, and related private delegated computation protocols are summarized in Table~\ref{tab:delegated_computing}.

\begin{table*}[htb]
\centering
\begin{tabularx}{\textwidth}{
   >{\centering\arraybackslash\hsize=.5\hsize\linewidth=\hsize}X 
   >{\centering\arraybackslash}X 
   >{\centering\arraybackslash\hsize=1.5\hsize\linewidth=\hsize}X   
   >{\centering\arraybackslash}X 
   >{\centering\arraybackslash}X 
   }
\hline
\rowcolor[HTML]{54C556}[\dimexpr\tabcolsep+0.1pt\relax]
\textcolor{white}{\textbf{Year}} & \textcolor{white}{\textbf{Reference}}  & \textcolor{white}{\textbf{Task}}  & \textcolor{white}{\textbf{Encoding}} & \textcolor{white}{\textbf{Number of nodes}}  \\ \hline
2012 & Barz et al. \cite{Barz2012}                     & Delegated BQC                                             & Bulk optics, polarization                                  & 1 C, 1 S      \\ 
\rowcolor[HTML]{D5F9D4}[\dimexpr\tabcolsep+0.05pt\relax]
2013 & Barz et al. \cite{Barz2013}                     & Verifiable delegated BQC                                  & Bulk optics, polarization                                   & 1 C, 1 S      \\ 
2014 & Fisher et al. \cite{Fisher2014}                 & Quantum computing on encrypted data                       & Bulk optics, polarization                                   & 1 C, 1 S      \\ 
\rowcolor[HTML]{D5F9D4}[\dimexpr\tabcolsep+0.05pt\relax]
2016 & Greganti et al. \cite{Greganti2016}             & Verifiable measurement-only BQC                           & Bulk optics, polarization                                   & 1 C, 1 S      \\
2016     & Marshall et al. \cite{Marshall2016}             & Continuous-variable QHE                                   & Field quadratures                              & 1 C, 1 S      \\ 
\rowcolor[HTML]{D5F9D4}[\dimexpr\tabcolsep+0.05pt\relax]
2017 & Huang et al. \cite{Huang2017}                   & Verifiable classical client BQC                           & Bulk optics, polarization                                   & 1 C, 2 S      \\ 
2019 & Jiang et al. \cite{Jiang2019}                   & Remote blind state preparation with weak coherent pulses  & Bulk optics, polarization                                   & 1 C, 1 S      \\ 
\rowcolor[HTML]{D5F9D4}[\dimexpr\tabcolsep+0.05pt\relax]
2020 & Tham et al. \cite{Tham2020}                     & Quantum fully homomorphic encryption                      & Bulk optics, polarization                                                     & 1 C, 1 S \\
2021 & Zeuner et al. \cite{Zeuner2021}                 & Quantum homomorphic encryption                            & Integrated optics, polarization + path                                   & 1 C, 1 S      \\
     
\rowcolor[HTML]{D5F9D4}[\dimexpr\tabcolsep+0.05pt\relax]
2023 & Polacchi et al. \cite{polacchi2023multi}        & Multi-client delegated BQC                                & Bulk optics, polarization                                   & 2 C, 2 S      \\ 
2024 & Polacchi et al. \cite{Polacchi2025} & Multi-client verifiable delegated BQC                     & Bulk optics, polarization                                   & 3 C, 2 S      \\ 
\rowcolor[HTML]{D5F9D4}[\dimexpr\tabcolsep+0.05pt\relax]
2024     & Drmota et al. \cite{Drmota2024}                 & Verifiable delegated BQC                                  & Trapped ions, photon polarization              & 1 C, 1 S      \\ 
2024     & Li et al. \cite{Li2024}                         & Quantum homomorphic encryption                            & Integrated optics, path                & 1 C, 1 S \\ 
\rowcolor[HTML]{D5F9D4}[\dimexpr\tabcolsep+0.05pt\relax]
2025 & Wei et al. \cite{Wei2025}                       & Verifiable delegated BQC                                  & SiV centers in diamond, photonic time-bin        & 1 C, 2 S      \\ 
2025 & Delle Donne et al. \cite{DelleDonne2025}             & Delegated QC                                  & NV centers in diamond, photonic time-bin        & 1 C, 1 S      \\ 

\hline

\end{tabularx}
\caption{\textbf{Photonic demonstrations of delegated quantum computing.} Chronology of experiments that demonstrated privacy-preserving delegated quantum computing protocols on a quantum-optical platform. For each reference, we report the focus of the protocol, the chosen encoding, and the number of nodes involved in the network, i.e., C = number of clients; S = number of servers. QHE = quantum homomorphic encryption; BQC = blind quantum computing. }
\label{tab:delegated_computing}
\end{table*}

\textit{Blind quantum computing} was originally proposed in Ref.~[\onlinecite{Broadbent2009UniversalBlind}] as a delegated protocol for universal quantum computation; a systematic overview of its evolution can be found in Ref.~[\onlinecite{Fitzsimons2017}]. Here, we review its basic working principles and present recent experimental progress on the topic. 

Operating within the MBQC paradigm, a standard BQC architecture encompasses one or more clients and servers. The protocol is defined by two fundamental security properties: blindness and verifiability. Blindness guarantees that the algorithm, along with all input and output data, is perfectly concealed from the server and any eavesdropping adversaries. Verifiability provides the client with a mechanism to detect malicious or faulty server operations without compromising the blindness of the computation.

A schematic of the BQC protocol is shown in Fig.~\ref{fig:immagine_BQC}a and proceeds as follows: during an initial quantum communication round, the client prepares single qubits in random states and transmits them to a remote quantum server, which entangles them to produce a brickwork state (see Sec.~\ref{sec:TheBrickworkState}). Following this, the protocol shifts to $O(n)$ classical communication rounds between the two parties. These classical rounds are necessary because, in the MBQC framework, an algorithm is executed via a specific pattern of adaptive single-qubit measurements across the resource state. During these rounds, the client provides the server with an encrypted measurement basis for each targeted qubit. After the server measures a qubit, it returns the outcome to the client, who uses this classical data to adapt the measurement basis for the subsequent qubit. Crucially, this iterative structure naturally accommodates the insertion of hidden traps, allowing for the unconditional verification of the computation\cite{Fitzsimons2017Unconditionally}.

A measurement-only variant of this protocol\cite{Morimae2013} reverses the standard preparation phase: the server generates the initial resource state and sequentially sends single qubits to the client for measurement. Driven by the need to accommodate clients with increasingly limited hardware, further progress has led to ancilla-driven BQC protocols. In these schemes, clients only need the capacity to measure single qubits or implement rudimentary single-qubit gates \cite{Dai2023Ancilla}, yet they retain the ability to verify the server's honesty with high probability. Pushing this minimization to its absolute limit, fully classical clients can delegate quantum computation to multiple entangled, remote servers \cite{Reichardt2013}. While this completely eliminates the client's quantum hardware requirements, it introduces the stringent operational assumption that the servers cannot communicate with one another. Finally, from a practical hardware perspective, alternative schemes have been developed wherein the client prepares and transmits weak coherent pulses rather than demanding ideal single-photon sources \cite{Dunjko2012}.

On the experimental side, most BQC implementations have been carried out on fully photonic platforms. This is a natural consequence of the delegated nature of such protocols, as photons are the ideal candidates for establishing quantum communication channels between a client and a server. Furthermore, as discussed in Sec.~\ref{sec:ExperimentalRealizationResourceStates}, recent technological advancements have vastly improved the generation of photonic cluster states for MBQC. Beyond purely photonic systems, BQC has also been demonstrated on hybrid light-matter platforms, such as trapped ions and point defects in diamond, where a solid-state server interfaces with photon-based clients \cite{DelleDonne2025,Wei2025,Drmota2024}. These results significantly broaden the landscape of potential implementations, proving that BQC can be tailored to a variety of physical architectures.

Historically, the first verifiable BQC implementation \cite{Barz2012,Barz2013} utilized a fully photonic bulk-optics setup to generate four-qubit cluster states encoded in the polarization of single photons. Using this platform, the authors successfully demonstrated blind single- and two-qubit gates, alongside Grover's and Deutsch's algorithms. Verifiable measurement-only BQC was demonstrated several years later on a similar platform \cite{Greganti2016}, while a classical-client implementation leveraging two entangled servers was reported in Ref.~[\onlinecite{Huang2017}]. Pursuing the ultimate goal of minimizing client hardware requirements, BQC based on weak coherent pulses was also successfully demonstrated \cite{Jiang2019}.

More recently, the single-client framework was extended by Refs.~[\onlinecite{polacchi2023multi}] and [\onlinecite{Polacchi2025}], which demonstrated two and three clients performing verifiable federated blind quantum computing on a remote server. The protocol and corresponding experimental setup are depicted in Fig.~\ref{fig:immagine_BQC}c. This multi-client architecture was realized by introducing a classical trusted third party (TTP) to secure all communications between the clients and the server. Additionally, it employed a modular linear quantum network architecture known as Qline \cite{Doosti2024}, which reduces the clients' necessary quantum capabilities to simple single-qubit rotations.

Finally, research within the BQC framework has gained traction in the industry sector. Recently, a protocol for secure delegated quantum computation based on semi-classical communication \cite{Bourdoncle2025} was proposed to dramatically lower the technological requirements for both parties. In this highly practical approach, the client only needs an attenuated laser source with a random phase and a small set of waveplates, while the server is not required to perform complex non-demolition photon-number measurements or deterministic photon gates.

\textit{Homomorphic encryption} (HE)\cite{Gentry2009} refers to a class of protocols developed with the goal of securely delegating quantum computation from a client to a server, while keeping the client's input and output data private. The general idea behind homomorphic encryption is depicted in Fig.~\ref{fig:immagine_BQC}b. The client has two efficient and secure algorithms ($\mathcal{E}$, $\mathcal{D}$) to perform respectively data encryption and decryption, that satisfy the condition $\mathcal{D}(Eval(\mathcal{E}(x))) = f(x)$, for any function $f$ in a given set $C$.
Therefore, the server evaluates the function $f$ on the encrypted input $\mathcal{E}(x)$ by using the algorithm $Eval$ and returns the outcome to the client, who can decrypt it, obtaining $f(x)$.
Such a procedure should be performed without the server being able to seize information about $x$ from $\mathcal{E}(x)$.
If the set $C$ contains all polynomial-sized circuits, the scheme becomes \textit{fully} homomorphic\cite{Gentry2009,Aaronson2019}.
Such schemes allow for computing on encrypted data without requiring any additional communication between the client and the server, which is a desirable feature in the context of delegated computation.
However, this comes at the cost of some computational assumptions that lower the level of security achieved by the protocol\cite{Newman2018,Yu2014,Aaronson2019}.
Quantum versions of the HE (QHE) have been derived\cite{Broadbent2015,Tan2016,Mahadev2020,Rohde2012} with the goal of equipping HE with the possibility of processing quantum data with quantum circuits. The first demonstration of quantum computing with encrypted data was provided by Ref.~[\onlinecite{Fisher2014}]. 
Later on, QHE was demonstrated on bulk platforms where qubits were encoded in the polarization of single photons \cite{Tham2020} and within the continuous-variable framework\cite{Marshall2016}. More recently, QHE schemes were also demonstrated on compact photonic integrated circuits\cite{Zeuner2021,Li2024}.

\section{\label{sec:1way QC}Continuous-variable quantum computation}

As a fundamental alternative to discrete-variable paradigms, Lloyd and Braunstein introduced continuous-variable (CV) quantum computation in 1999\cite{lloyd1999quantum,weedbrook_gaussian_2012}. This encoding architecture relies on observables with continuous eigenspectra, most notably the position quadrature $\hat{q}=(\hat{a}+\hat{a}^{\dagger})/\sqrt{2}$ and the momentum quadrature $\hat{p}=(\hat{a}-\hat{a}^{\dagger})/(i\sqrt{2})$ of the quantized electromagnetic field. These operators are related to the in-phase and in-quadrature amplitudes of the electromagnetic field, adhere to the canonical commutation relation $[\hat{q},\hat{p}]=i$ (where $\hbar=1$), and exhibit a continuous spectrum spanning from $-\infty$ to $+\infty$.

An advantage of the CV approach is the ability to deterministically generate macroscopic quadrature entanglement through parametric processes in nonlinear media \cite{Lenzini2018, Polkinghorne1999}. However, it also presents distinct challenges, particularly regarding the implementation of non-Gaussian transformations\cite{weedbrook_gaussian_2012}, which are necessary for universal quantum computation and will be detailed in subsections \ref{sec:CV Gate model} and \ref{sec:CV cluster state}.

CV encoding is exceptionally well-suited for cluster-state quantum quantum computation \cite{Nielsen2006Cluster,gu2009quantum} because massive, two-dimensional cluster states can be generated deterministically using squeezed vacuum sources and linear optical components \cite{menicucci2006universal, larsen2019deterministic, asavanant2019generation}. Furthermore, by embedding a discrete logical encoding within the continuous quadrature space using Gottesman-Kitaev-Preskill (GKP) states \cite{gottesman2001encoding}, CV systems become fully compatible with quantum error correction, exhibiting a quantifiable fault-tolerance threshold \cite{menicucci2014fault}.

This section provides an overview of the foundational concepts driving CV quantum computation. We first present a concise review of the CV approach to quantum information processing, followed by the specific formalism of CV one-way quantum computation. Finally, we review recent theoretical descriptions and experimental progress, placing a particular emphasis on scalability, advanced quantum operations, and the pathway to fault-tolerant computation via GKP encoding.

\subsection{Gate based quantum computation}\label{sec:CV Gate model}
\subsubsection{Quantum states and the phase space representation}
The continuous-variable equivalent of a \textit{qubit} is known as a \textit{qumode}, which resides within the infinite-dimensional Hilbert space of the quadrature operators. By utilizing the continuous basis states $\ket{s}_q$ and $\ket{s}_p$, defined by the eigenvalue equations $\hat{q}\ket{s}_q=s\ket{s}_q$ and $\hat{p}\ket{s}_p=s\ket{s}_p$, an arbitrary state $\ket{\psi}$ of the electromagnetic field can be expressed as
\begin{equation}
    \ket{\psi}=\int_{-\infty}^{+\infty} \psi_q(s)\ket{s}_q ds=\int_{-\infty}^{+\infty} \psi_p(s)\ket{s}_p ds.
\end{equation}
Here, $\psi_q(s)={}_{q}\langle s|\psi\rangle$ and $\psi_p(s)={}_{p}\langle s|\psi\rangle$ denote the state's wavefunctions in the position and momentum representations, respectively. This continuous integral formalism maps directly onto the discrete Fock-basis expansion, $\ket{\psi}=\sum_{n=0}^{\infty}a_n\ket{n}$, where $\ket{n}$ represents the $n$-photon eigenstate of the number operator such that $\hat{n}\ket{n}=n\ket{n}$.

Beyond state vectors and density matrices, CV quantum states are frequently represented using continuous phase-space distributions\cite{LEONHARDT199589}. The most widely used of these is the Wigner function\cite{wigner1932quantum}, defined as
\begin{equation}
    W(q,p)=\frac{1}{2\pi}\int_{-\infty}^{+\infty} {}_{q}\langle q+\frac{1}{2}x|\hat{\rho}|q-\frac{1}{2}x\rangle_{q} e^{ixp}dx,
\end{equation}
where $\hat{\rho}=\sum_{m,n=0}^{\infty}a_{m,n}\ket{m}\bra{n}$ is the density matrix operator of the quantum state. The Wigner function is classified as a quasiprobability distribution because it relaxes certain classical probability constraints, for example allowing for regions of negativity that serve as a signature of non-classicality. Despite this, its marginals reconstruct the physical probability distributions for the respective quadratures:
\begin{equation}
\begin{aligned}
    P(q)=\int_{-\infty}^{+\infty} W(q,p)dp,\\
    P(p)=\int_{-\infty}^{+\infty} W(q,p)dq.
\end{aligned}
\end{equation}

\begin{figure}
    \begin{subfigure}[b]{0.45\textwidth}
        \centering
        \includegraphics[width=\textwidth]{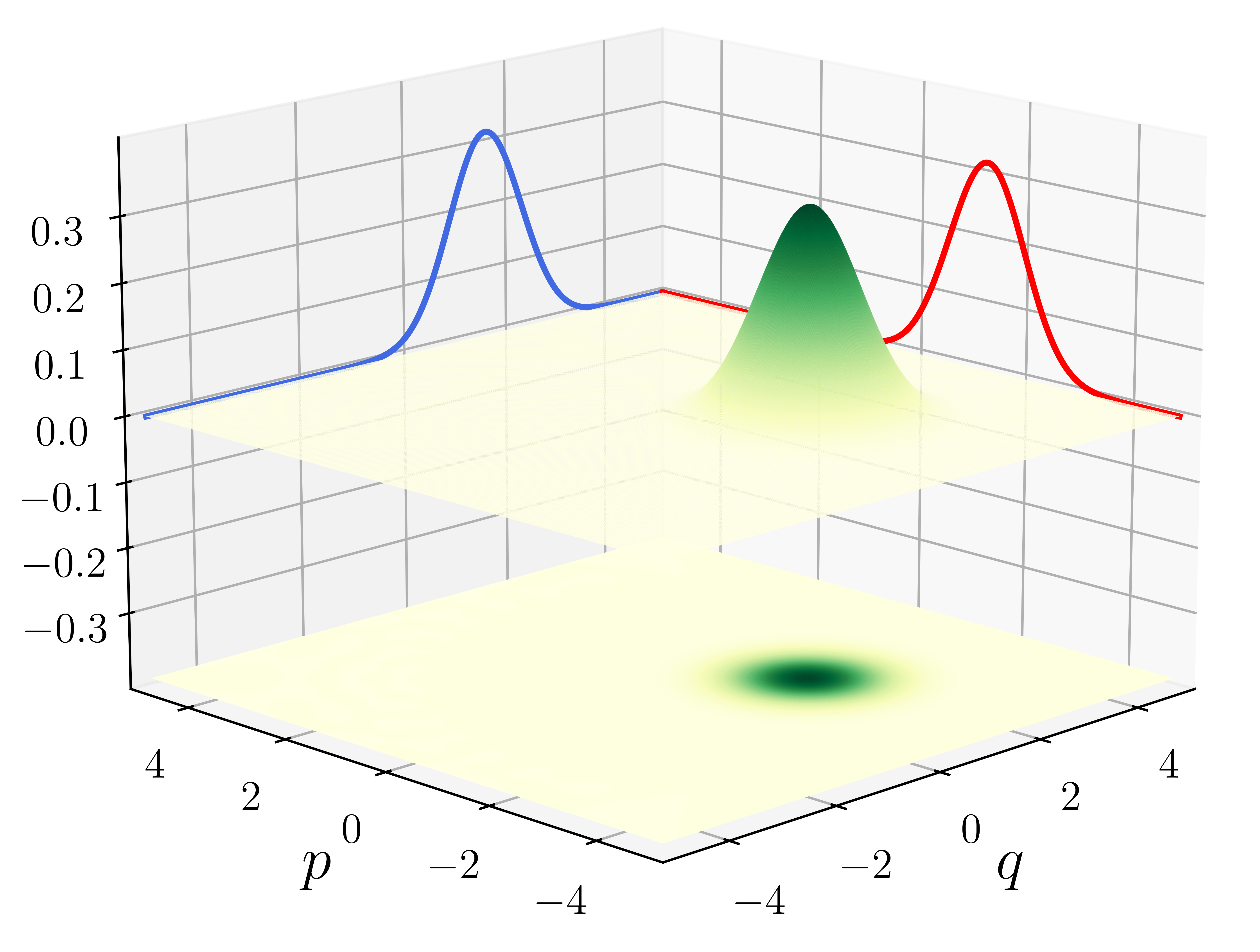}
        \caption{}
        \label{Wigner coherent}
    \end{subfigure}
\hfill
    \begin{subfigure}[b]{0.45\textwidth}
        \centering
        \includegraphics[width=\textwidth]{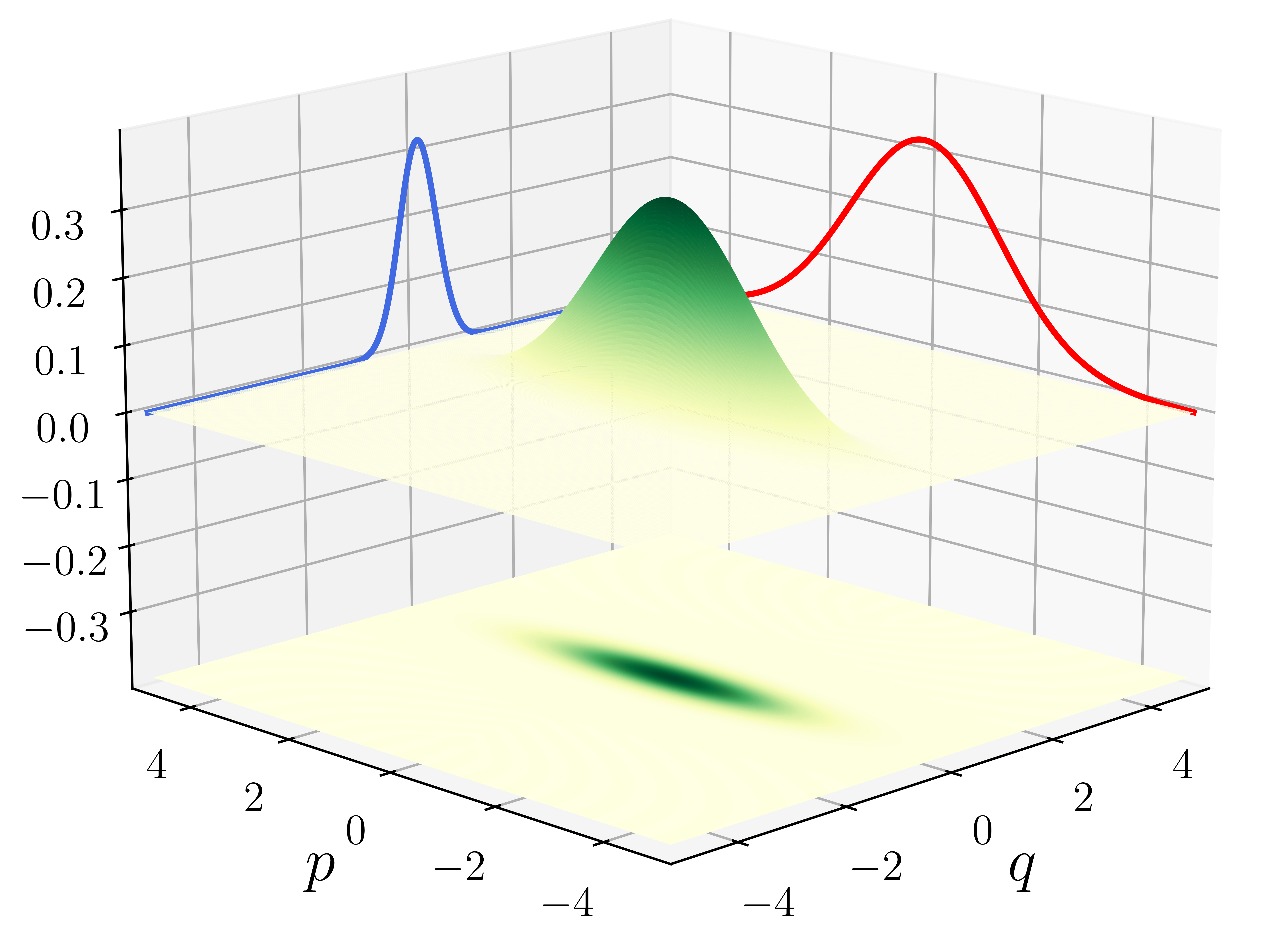}
        \caption{}
        \label{Wigner Squeezed}
    \end{subfigure}
    \hfill
    \begin{subfigure}[b]{0.45\textwidth}
        \centering
        \includegraphics[width=\textwidth]{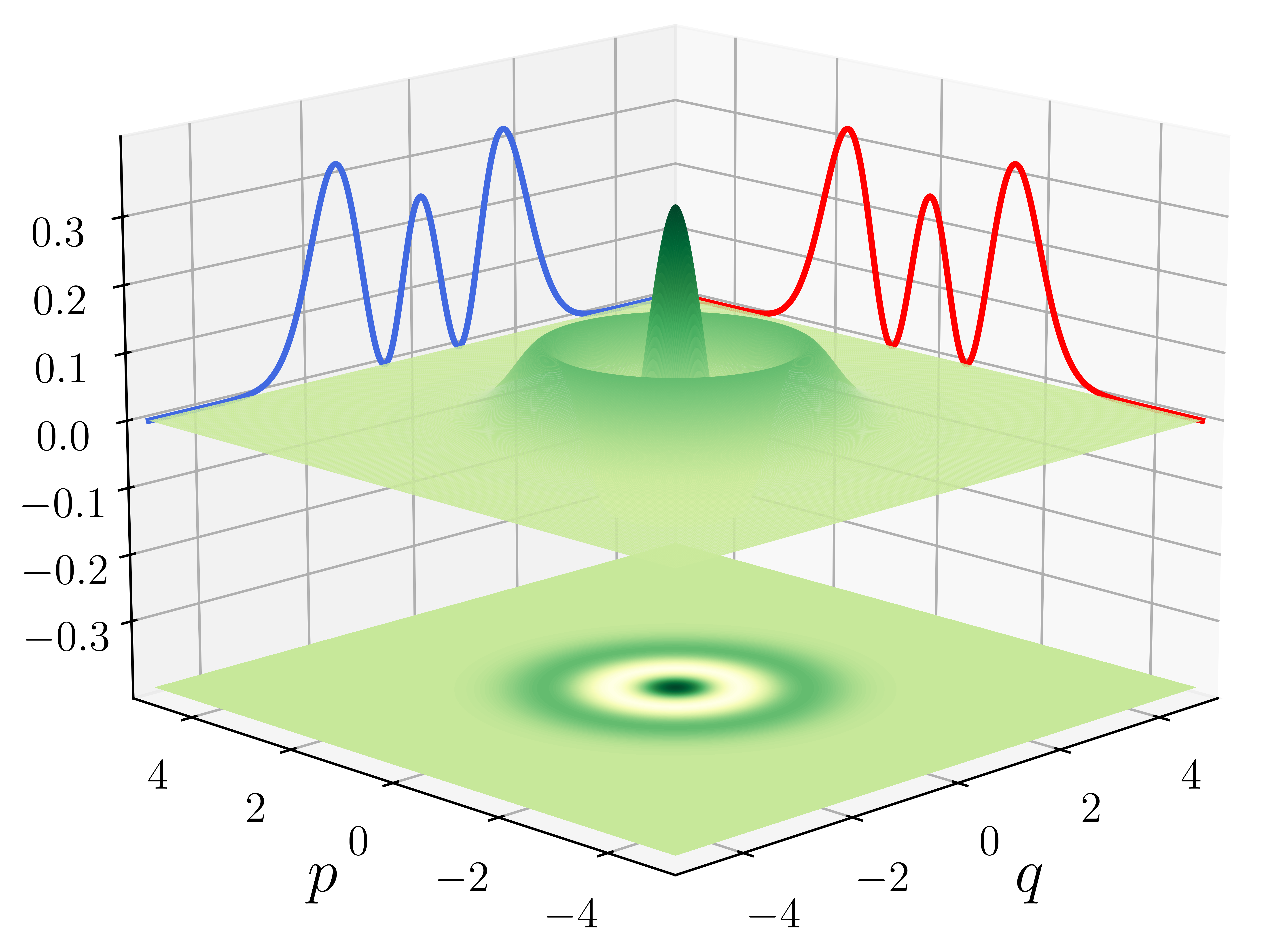}
        \caption{}
        \label{Wigner Fock}
    \end{subfigure}

    \caption{Wigner function, and the quadrature distributions P(q) (blue) and P(p) (red) of : (a) Coherent state $\ket{\alpha}$, (b) Squeezed vacuum state $\ket{\xi}$ with a phase of squeezing $\phi=0$, and (c) Fock state $\ket{n=2}$.}
    \label{Wigner function}
\end{figure}

From both an experimental and theoretical standpoint, the simplest state to prepare and characterize is the vacuum state, $\ket{n=0}$. Its Wigner function is a perfectly symmetric Gaussian centered at the origin of phase space, characterized by the minimal uncertainty variances $\langle\Delta\hat{q}\rangle^2 = \langle\Delta\hat{p}\rangle^2 = 1/2$. A closely related Gaussian state is the coherent state, $\ket{\alpha}$, which maintains the same minimal quadrature variances as the vacuum but is displaced to the phase-space coordinates $(q_0, p_0)$. As illustrated in Fig.~\ref{Wigner function}a, this displacement point directly defines the mean values of the respective quadrature operators. Defined as the eigenstate of the annihilation operator ($\hat{a}\ket{\alpha} = \alpha\ket{\alpha}$), the coherent state serves as the quantum mechanical description of an ideal laser, possessing a mean photon number $\langle\hat{n}\rangle = |\alpha|^2$.

A fundamental resource for CV quantum computation is the squeezed vacuum state. This non-classical state has a Gaussian Wigner function centered at the origin of phase space, shown in Fig.~\ref{Wigner function}b, with different variances for the different quadrature operators. To strictly satisfy the Heisenberg uncertainty principle, the variance of one quadrature is reduced below $1/2$, while the variance of the conjugate quadrature is proportionally increased above $1/2$. A single-mode squeezed vacuum state is generated by applying the squeezing operator $\hat{S}(z)=\exp[\frac{1}{2}(z^*\hat{a}^2-z\hat{a}^{\dagger 2})]$ to the vacuum, where $z=re^{i\phi}$ is the complex squeezing parameter. For a real squeezing parameter ($\phi=0$), the quadrature variances are $\langle\Delta\hat{q}\rangle^2=\frac{1}{2}e^{-2r}$ and $\langle\Delta\hat{p}\rangle^2=\frac{1}{2}e^{2r}$. In the theoretical limit of infinite $q$-squeezing ($r \to \infty$), the state corresponds to the eigenstate of the position operator $\ket{0}_q$. However, because such a state requires infinite energy, it is physically unrealizable.

The finite degree of squeezing achievable in physical systems degrades the operational fidelity of the continuous-variable encoding, a practical limitation strictly bounding protocols such as quantum teleportation \cite{BraunsteinTeleport1998}. Despite this constraint, squeezed states are a resource across different areas of quantum technologies, driving critical enhancements in quantum metrology\cite{SCHNABEL20171} and broad-scale quantum communications\cite{braunstein2005quantum, Pirandola2020}. The most prevalent approach for generating squeezed vacuum states relies on second-order $\chi^{(2)}$ nonlinear interactions, including degenerate optical parametric amplification (OPA) \cite{Peace2022, Kashiwazaki2023} and optical parametric oscillation (OPO) \cite{Wu:87}. Alternatively, third-order $\chi^{(3)}$ nonlinear processes, such as four-wave mixing (FWM) \cite{Slusher1985, Chembo2016}, are employed in the generation of squeezed states directly within optical fibers and integrated photonic platforms \cite{Zhao2020}.

States possessing a non-Gaussian Wigner function are necessary for universal CV quantum computation, as the evolution of purely Gaussian states can be efficiently simulated on a classical computer \cite{Mari_2012, Veitch_2012}. Fock states provide a fundamental example of non-Gaussian states. While they possess a fixed number of photons, they are highly non-classical, exhibiting Wigner functions that take on negative values in certain regions of phase space—with the vacuum state being the sole exception. For an arbitrary $n$-photon state $\ket{n}$, the Wigner function is given by \cite{LEONHARDT199589}
\begin{equation}
    W_n(q,p)=\frac{1}{\pi}(-1)^n e^{-(q^2+p^2)} L_n (2(q^2+p^2)),
\end{equation}
where $L_n$ is the Laguerre polynomial of order $n$. Figure~\ref{Wigner function}c shows the Wigner function of the $\ket{n=2}$ state alongside its associated marginal distributions. Unlike coherent and squeezed states, which feature smooth Gaussian profiles, Fock states display distinct oscillatory structures and localized negativity due to their fixed photon number. Figure~\ref{Wigner function} also shows how the marginal probability distributions of the quadratures are always non-negative.

\subsubsection{Gates and transformations for qumodes}\label{CVgates}
The first theoretical proposal for universal CV quantum computation using the gate model was introduced by Lloyd and Braunstein\cite{lloyd1999quantum}. They demonstrated that a CV quantum computing system is universal if it can simulate any unitary transformation of the form $\hat{U}=\exp[iH(\hat{q}_i, \hat{p}_i)]$, where the Hamiltonian $H(\hat{q}_i, \hat{p}_i)$ is a polynomial of the quadratures $\hat{q}_i$ and $\hat{p}_i$ over a set of $N$ qumodes ($i=1, 2, \dots, N$). Similar to the discrete-variable gate model (see Sec.~\ref{DV gate model}), a finite set of elementary gates can approximate these arbitrary unitary transformations. These operations are broadly divided into Gaussian and non-Gaussian operations. Gaussian operations are generated by a Hamiltonian with a polynomial degree of at most two. They naturally transform states with Gaussian Wigner functions—such as coherent or squeezed states—into other Gaussian states. Because these states are completely characterized by a finite set of parameters (specifically, their mean displacement vector and covariance matrix), computations relying exclusively on Gaussian operations can be simulated efficiently on a classical computer, as shown in [\onlinecite{bartlett2002}], which is the CV equivalent of the Gottesman-Knill theorem for DV protocols\cite{gottesman_heisenberg_1998}. 
To approximate a polynomial transformation of arbitrary degree, and thereby achieve true universality, the inclusion of a non-Gaussian operation is strictly necessary\cite{sefi2011decompose}. These operations, typically defined by the operator $\hat{D}_{k,\hat{q}}(s)=\exp(is\hat{q}^{k}/k)$ for integer degrees $k \geq3$, introduce the required non-linearity (for some interaction strength $s$). The lowest-order non-Gaussian operation is the cubic phase gate, generated by the operator $\hat{D}_{3,\hat{q}}(s)=\exp(is\hat{q}^{3}/3)$.

\begin{figure}
    \centering
    \includegraphics[width=0.75\linewidth]{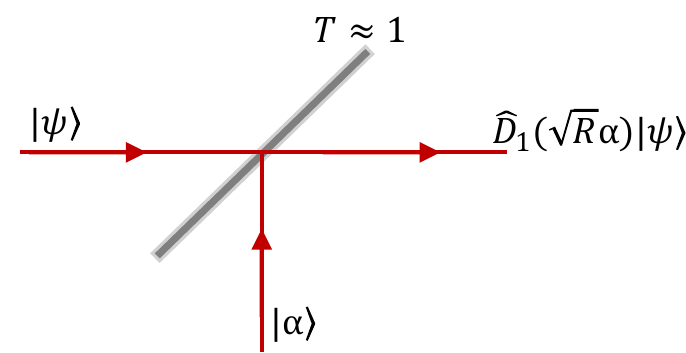}
    \caption{Optical implementation of displacement operation, $R=1-T$.}
    \label{Displacement gate}
\end{figure}

Among the operations detailed below, the first five (a)--(e) are strictly Gaussian, meaning they preserve the Gaussian character of the quantum state, whereas the final operation (f) is inherently non-Gaussian. This fundamental continuous-variable gate set comprises:\\
\noindent\textit{(a) The displacement operation}, $\hat{D}(\alpha) = \exp(\alpha\hat{a}^{\dagger}-\alpha^*\hat{a}) = \exp[-i\sqrt{2}(Re(\alpha)\hat{p} - Im(\alpha)\hat{q})]$, shifts the quantum state’s position in phase space by a complex amplitude $\alpha$. Specifically, the operator $\hat{D}_{1,\hat{q}}(s) = \exp(is\hat{q})$ (also denoted as $\hat{Z}(s)$) shifts the momentum quadrature $\hat{p}$ of a qumode by $s$, while $\hat{D}_{1,\hat{p}}(s) = \hat{X}(s) = \exp(-is\hat{p})$ shifts the position quadrature $\hat{q}$ by $s$:
\begin{align}
    \hat{Z}(s)\ket{r}_p&=\ket{r+s}_p,\\
    \hat{X}(s)\ket{r}_q&=\ket{r+s}_q .    
\end{align}
Experimentally, a small displacement operation is implemented using a highly asymmetric beam splitter with a transmissivity of $T \approx 1$. As illustrated in Fig.~\ref{Displacement gate}, the state of interest $\ket{\psi}$ interferes with a strong coherent field $\ket{\alpha}$, producing a displacement $s = \sqrt{R}\alpha$, where $R = 1-T$.
\\

\noindent\textit{(b) The rotation gate}, $\hat{R}(\theta) = \exp(i\theta\hat{a}^{\dagger}\hat{a})$, transforms the phase-space quadratures according to:
\begin{align}
    \hat{q}&\rightarrow\hat{q}_{\theta}=\hat{q}cos\theta-\hat{p}sin\theta,\\    \hat{p}&\rightarrow\hat{p}_{\theta}=\hat{q}sin\theta+\hat{p}cos\theta.
\end{align}
In particular, a rotation by an angle $\theta = \pi/2$ corresponds to the continuous variable Fourier transform, mapping the quadratures such that $\hat{F}\ket{s}_q = \ket{s}_p$. Conversely, for $\theta = -\pi/2$, the inverse transform yields $\hat{F}^{\dagger}\ket{s}_p = \ket{s}_q$. From an experimental standpoint, the rotation gate is straightforwardly implemented by introducing an optical phase shift, or delay, to the mode of interest.\\

\noindent\textit{(c) The squeezing gate}, represented by the unitary operator $\hat{S}(z) = \exp[\frac{1}{2}(z^*\hat{a}^2 - z\hat{a}^{\dagger 2})]$ with a complex squeezing parameter $z = re^{i\phi}$, applies a phase-dependent scaling to the quantum state's quadratures. For a real squeezing parameter ($\phi=0$), the gate exponentially attenuates the position quadrature while amplifying the momentum quadrature, transforming them according to:
\begin{align} 
    \hat{q} &\rightarrow \hat{q}e^{-r},\\ 
    \hat{p} &\rightarrow \hat{p}e^{r}. 
\end{align}
Unlike displacements and rotations, which are easily realized using linear optics, a deterministic squeezing gate acting on an arbitrary input state—known as inline squeezing—is experimentally challenging. It requires high-efficiency optical parametric amplifiers (OPAs) and perfect mode matching between the input mode and the squeezer. For these reasons, initial experiments utilized an offline approach\cite{Filip2005} based on teleportation, which involved interfering a quantum state with a squeezed vacuum at a beam splitter (BS), followed by a feedforward displacement operation\cite{yoshikawa2007demonstration}. Nevertheless, despite these stringent technical requirements, deterministic inline squeezing has recently been successfully demonstrated \cite{WangInlinesqueezer2022}.\\

\noindent\textit{(d) The shearing gate}, frequently referred to as the quadratic phase gate, is represented by the unitary operator $\hat{D}_{2,\hat{q}}(s)=\exp({is\hat{q}^2/2})$, where $s$ is a real-valued shearing parameter. Geometrically, this Gaussian operation distorts the quantum state's phase space by leaving the position quadrature invariant while shifting the momentum quadrature by an amount strictly proportional to the position $\hat{q}$\cite{weedbrook_gaussian_2012}. The quadratures transform according to:
\begin{align}
\hat{p} &\rightarrow \hat{p} + s\hat{q},\\
\hat{q} &\rightarrow \hat{q}.
\end{align}
Because the shearing gate is a purely Gaussian operation, it can be theoretically decomposed into a sequence of standard rotation and squeezing gates. In particular, it can be expressed as the operator $\hat{D}_{2,\hat{q}}(s)=\hat{R}(\theta)\hat{S}(re^{i\phi})$ with a suitable choice of parameters:
\begin{align}
    \theta &= \arctan(s/2),\\
    r &= \text{arccosh}\left(\sqrt{1+(s/2)^2}\right),\\
    \phi &= -\text{sgn}(s)\frac{\pi}{2}-\theta.
\end{align}
However, rather than cascading complex inline active components, modern continuous-variable architectures typically implement arbitrary shearing transformations using measurement-based techniques\cite{menicucci2006universal}. By utilizing offline squeezed resource states, linear optical interferometry, and classical feedforward, the required phase-space shear can be applied deterministically to the propagating quantum state \cite{ukai2011demonstration}.\\

\noindent\textit{(e) The controlled-Z gate}, also known as the continuous-variable controlled-phase gate, is the primary two-mode entangling operation in continuous-variable architectures. It is represented by the unitary operator $CZ=\exp({is\hat{q}_1\hat{q}_2})$, where $s$ represents a real-valued interaction weight or gain parameter. 
Geometrically, this gate leaves the position quadratures of both optical modes invariant while translating the momentum quadrature of each mode by an amount strictly proportional to the position of the other mode. The joint transformations are:
\begin{align}
    \hat{q}_1& \rightarrow\hat{q}_1, \quad\hat{p}_1\rightarrow\hat{p}_1+s\hat{q}_2,\\
    \hat{q}_2& \rightarrow\hat{q}_2, \quad\hat{p}_2\rightarrow\hat{p}_2+s\hat{q}_1.
\end{align}
While a direct inline nonlinear interaction is experimentally prohibitive, a deterministic two-mode $CZ$ gate can be effectively synthesized using purely Gaussian resources. Specifically, it is realized by interfering the two input modes on a balanced beam splitter, applying opposite single-mode squeezing operations to the separated optical paths, and interfering them again on a second beam splitter\cite{gu2009quantum}. \\


\noindent\textit{(f) The cubic phase gate} is the lowest-order non-Gaussian operation necessary for universal quantum computation. It is generated by the operator $\hat{D}_{3,\hat{q}}(s)=\exp(is\hat{q}^{3}/3)$, where $s$ is a real-valued nonlinear coupling parameter. While the shearing gate applies a linear shift to the momentum, the cubic phase gate distorts phase space by applying a quadratic shift to the momentum quadrature while leaving the position quadrature invariant. The transformations are given by:
\begin{align} 
    \hat{q} &\rightarrow \hat{q}, \\ 
    \hat{p} &\rightarrow \hat{p} + s\hat{q}^2. 
\end{align}
The introduction of the cubic phase gate breaks the efficient simulability of Gaussian operations, enabling universal fault-tolerant quantum algorithms\cite{bartlett2002}. 
Because single-photon optical nonlinearities are vanishingly weak in standard optical media, implementing a deterministic inline cubic phase gate is exceptionally difficult. Alternative architectures propose synthesizing this operation with measurement-based techniques, which involve preparing an auxiliary non-Gaussian resource state (a cubic phase state), teleporting the data mode into this state via a continuous-variable Bell measurement, and applying active non-Clifford feedforward corrections\cite{Miyata2016}. But this approach only shifts the problem from implementing the cubic phase gate to generating a cubic phase state, which is also experimentally challenging and requires heralded protocols with non-Gaussian operations like photon number resolving detection\cite{Ghose2007}.

In the gate set detailed above, Gaussian operations are readily implemented using standard optical setups; conversely, realizing the required non-Gaussian gate deterministically and with high fidelity remains a significant experimental challenge. This disparity is unsurprising, as it directly parallels the fundamental bottleneck in DV architectures: inducing deterministic nonlinear interactions at the single-photon level, which similarly precludes the direct realization of a deterministic \textit{CNOT} gate. 


\subsection{\label{sec:CV cluster state}CV measurement-based quantum computation}
Continuous-variable quantum computation finds its most natural implementation within the measurement-based quantum computation (MBQC) paradigm, largely due to the availability of massively scalable, deterministically generated cluster states\cite{larsen2019deterministic, asavanant2019generation, chen2014experimental}. First introduced in 2006 \cite{menicucci2006universal, zhang2006continuous}, these states extend the concept of discrete-variable cluster states into the continuous-variable regime. Ideally, a CV cluster state is a highly entangled multimode Gaussian state. It is typically synthesized by interfering an array of individually squeezed vacuum states through a linear optical network of beam splitters and phase shifters, defined by a specific graph topology\cite{menicucci2011graphical}. To achieve multiplexing and scale the number of cluster nodes, the two primary degrees of freedom utilized in optics are time\cite{menicucci2010arbitrarily} and frequency\cite{menicucci2007ultracompact}. However, much like the CV gate model, relying solely on Gaussian resources is insufficient. As detailed in subsection \ref{DefCluster}, the integration of non-Gaussian resources is strictly necessary to achieve both universal quantum computation and viable quantum error correction.

\subsubsection{Definition of CV cluster state} \label{DefCluster}
The construction of a CV cluster state follows a procedure closely analogous to the discrete-variable framework described in Sec.~\ref{sec:DV Cluster model}. However, instead of initializing qubits in the $\ket{+}$ state, the constituent qumodes are prepared in the zero-momentum eigenstate $\ket{0}_p$. The vertices of the cluster are then entangled using the $CZ$ gates detailed in  section \ref{CVgates}. Following this analogy, an ideal CV cluster state is defined by a graph\cite{kockum2023lecture} $G = (V, E)$ comprising $N$ vertices $V=\{v_1, \dots, v_N\}$ and a set of edges $E$ containing elements $e_{jk}=\{v_j,v_k\}$. With all vertices initialized in the $\ket{0}_p$ state, the cluster state is generated by applying the entangling operations across all defined edges:
\begin{equation}
    \ket{C} = \prod_{e_{jk}\in E} CZ_{jk} \ket{0}_p^{\otimes N} = \prod_{e_{jk}\in E} \exp\left(\frac{i}{2}\hat{q}_j\hat{q}_k\right) \ket{0}_p^{\otimes N}.
\label{G}
\end{equation}

Arguably, the primary reason why CV-MBQC is considered a highly promising approach for realizing a full-scale universal quantum computer is that generating macroscopic cluster states requires only squeezed vacuum sources and linear optical interferometers\cite{menicucci2010arbitrarily}. Indeed, such states have already been experimentally realized, demonstrating linear clusters of more than $10^6$ qumodes\cite{yoshikawa2016invited} as well as scalable 2D topologies\cite{asavanant2019generation, larsen2019deterministic}. However, a cluster state relying exclusively on squeezed vacua and linear optics is insufficient for fault-tolerant universal quantum computation. To implement robust error correction protocols, it is strictly necessary to encode information into orthogonal non-Gaussian states, such as the GKP states\cite{gottesman2001encoding}, which remain experimentally challenging to produce. By coupling these encoded CV qubits to the cluster via \textit{CZ} gates, universal quantum computation can be driven entirely through local homodyne measurements and classical feedforward. Furthermore, it is possible to correct for errors and achieve fault tolerance \cite{menicucci2014fault}, provided the resource states are generated with a squeezing level that exceeds the threshold of the error correction protocol.

In the CV framework, the stabilizer formalism—commonly referred to as the theory of \textit{nullifiers} \cite{barnes2004stabilizer, van2007building}—can completely specify a CV cluster state \cite{zhang2008graphical} through a set of operators for which the state is an eigenstate with an eigenvalue of 1. An arbitrary $N$-qumode CV cluster state $\ket{C}$, as defined by Eq.~\eqref{G}, is stabilized by the set of operators:
\begin{equation}
    \hat{S}_i(s) = \hat{X}_i(s) \prod_{j\in N(i)} \hat{Z}_j(s) \quad \text{for all } s \in \mathbb{R}, 
\label{S}
\end{equation}
where $N(i) = \{j | (v_j,v_i)\in E\}$ represents the set of indices defining the neighbors of vertex $v_i$, such that $\hat{S}_i(s)\ket{C} = \ket{C}$. Every stabilizer can be expressed as the exponential of a Hermitian operator $\hat{H}_i$ (i.e., $\hat{S}_i(s) = \exp[-is\hat{H}_i]$). Because $\hat{S}_i(s)\ket{C} = \ket{C}$, it follows that $\hat{H}_i\ket{C} = 0$ for all $i$. These Hermitian generators $\hat{H}_i$ are the \textit{nullifiers} of the cluster state. They can be explicitly derived \cite{kockum2023lecture} via $\hat{H}_i = i \left. \frac{d\hat{S}_i(s)}{ds} \right|_{s = 0}$, which yields:
\begin{equation}
    \hat{H}_i = \hat{p}_i - \sum_{j\in N(i)} \hat{q}_j \quad \text{for} \quad i=1,\dots,N.
\end{equation}
A foundational example is the two-mode EPR state \cite{ou1992realization}, which is specified by the nullifier relations:
\begin{equation}
    (\hat{q}_1 - \hat{q}_2)\ket{\text{EPR}} = 0, \quad (\hat{p}_1 + \hat{p}_2)\ket{\text{EPR}} = 0.
\end{equation}
Scaling up, the nullifier set for a linear 4-qumode cluster state is given by:
\begin{equation}
    \{\hat{p}_1 - \hat{q}_2, \quad \hat{p}_2 - \hat{q}_1 - \hat{q}_3, \quad \hat{p}_3 - \hat{q}_2 - \hat{q}_4, \quad \hat{p}_4 - \hat{q}_3\}.
\end{equation}
Nullifier operators are particularly valuable for analyzing realistic Gaussian CV cluster states generated with finite squeezing. In the laboratory, the measured variances of these nullifiers serve as a direct metric of how closely the physical state approximates the ideal cluster state \cite{yokoyama2013ultra}. Furthermore, the entanglement structure of these states—which is central to their utility in quantum computing—is typically verified using the van Loock–Furusawa (vLF) criterion \cite{van2003detecting}. The vLF framework establishes a set of classical bounds for linear combinations of quadrature operators (closely related to the nullifiers). To certify fully inseparable entanglement across the cluster, the measured variances must violate these classical bounds by falling below a specific threshold. Consequently, these criteria serve as practical, quantitative benchmarks for assessing the quality of the state generated experimentally\cite{kashif2023physical, Lenzini2018}.

\subsubsection*{Quantum computation on CV cluster states}
\begin{figure}[h]
\centering
\begin{subfigure}[t]{0.5\textwidth}
\begin{tikzpicture}[every node/.style={circle, draw=black, minimum size=0.8cm}, node distance=2cm]
    \node[fill=white, label={[yshift=0.05cm]above:$\ket{\phi}$}] (1) at (0,0) {1};
     \node[fill=white, label={[yshift=0.0cm]above:$\ket{{0}_p}$}] (2) at (1.5,0) {2};
     \node[fill=white, label={[yshift=0.0cm]above:$\ket{{0}_p}$}] (3) at (3,0) {3};
     \node[fill=white, label={[yshift=0.0cm]above:$\ket{{0}_p}$}] (4) at (4.5,0) {4};
     \draw[-, thick] (1) -- (2) -- (3) -- (4);
\end{tikzpicture}
\caption{}
\end{subfigure}
\begin{subfigure}[t]{0.5\textwidth}
\begin{tikzpicture}[every node/.style={circle, draw=black, minimum size=0.8cm}, node distance=2cm]
     \node[fill=lightgray, label={[yshift=-0.2cm,xshift=0.3cm]above:$M_1\longrightarrow$}] (1) at (0,0) {1};
     \node[fill=lightgray, label={[yshift=-0.2cm,xshift=0.3cm]above:$M_2\longrightarrow$}] (2) at (1.5,0) {2};
     \node[fill=lightgray, label={[yshift=0.05cm]above:$M_3$}] (3) at (3,0) {3};
     \node[fill=white, label={[yshift=-0.08cm]above:$U\ket{\phi}$}] (4) at (4.5,0) {4};
     \draw[-, thick] (1) -- (2) -- (3) -- (4);
\end{tikzpicture}
\caption{}
\end{subfigure}
\caption{Quantum computation with 1D CV cluster state. (a) Initial cluster state, (b) Cluster state after measurement.}
\label{1D_cluster_MBQC}.
\end{figure}
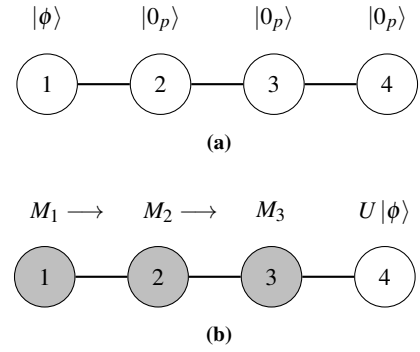

As for DV-MBQC, also in the CV paradigm, a unitary operation $\hat{U}$ can be implemented through a sequence of state preparation and adaptive measurements on a sufficiently large cluster state. Stated more formally, for any unitary $\hat{U}$, there always exists a graph $G = (V, E)$ and a partitioning of vertices into input ($V_{\text{in}}$) and output ($V_{\text{out}}$) sets that allow the implementation of $\hat{U}$ via the following steps:
\begin{enumerate}[label=\roman*]
    \item Designate $k$ vertices of $G$ as input vertices $V_{\text{in}}$, which encode the initial quantum state $\ket{\phi}$, while the remaining qumodes are prepared in the zero-momentum eigenstate $\ket{0}_p$.
    \item Apply continuous-variable $CZ$ gates between the relevant qumodes to establish the target cluster topology defined by the edges $E$.
    \item Perform sequential single-mode measurements on designated qumodes. The outcome of each measurement dictates the choice of basis for subsequent measurements. As the measured modes are projected out, the quantum state effectively propagates through the cluster, ultimately leaving the unmeasured output qumodes $V_{\text{out}}$ in the transformed state $\hat{U}\ket{\phi}$.
\end{enumerate}
Figure~\ref{1D_cluster_MBQC} illustrates this MBQC protocol using a 1D cluster state. Qumode 1 encodes the arbitrary input state $\ket{\phi}$, while the remaining qumodes are initialized in the $\ket{0}_p$ state and entangled via $CZ$ operations. The desired target state $\hat{U}\ket{\phi}$ is achieved by executing adaptive single-qumode measurements ($M_1, M_2, M_3$) on the sequential nodes, where the choice of the measurement basis ($M_i$) strictly depends on the classical outcome of the preceding measurement ($M_{i-1}$).

The operational logic of CV cluster computation is most intuitively understood as a generalized sequence of quantum teleportation steps\cite{zhou2000methodology}. When an input quantum state is coupled to a node of the cluster and a quadrature measurement is subsequently performed on that node, the quantum information is effectively ``teleported'' to the adjacent vertex. The specific choice of measurement basis at each node determines the quantum gate applied during this teleportation step. This measurement-induced teleportation provides a deterministic mechanism for gate execution: by appropriately selecting the quadrature measurement angles, one can imprint an arbitrary Gaussian unitary transformation onto the input state as it systematically propagates through the cluster lattice.

Single mode Gaussian operations—such as displacements $\hat{D}_{1,\hat{q}}(s)$, Fourier transforms $\hat{F}=\hat{R}(\pi/2)$, and shear gates $\hat{D}_{2,\hat{q}}(s)$—are implemented by tuning the homodyne measurement angles on individual qumodes where each measurement corresponds to a projection onto a rotated quadrature $\hat{q}_\theta = \hat{q}\cos(\theta) - \hat{p}\sin(\theta)$. When combined with the intrinsic entanglement of the cluster state, the effect of this projective measurement on the logical input state is equivalent to applying a single-mode Gaussian gate. Squeezing operations, denoted by $\hat{S}(z)$, alongside multimode Gaussian operations such as the $CZ$ gate, can be realized by embedding the appropriate connectivity into the graph structure of the cluster (i.e., specific entangling patterns).

\begin{figure}[b]
\centering
\begin{subfigure}[b]{0.3\textwidth}
    \begin{quantikz} 
   \lstick{$\ket{\phi}$}&\ctrl{1}&&\meterD{\hat{p}}&\setwiretype{c}\quad m\\
    \lstick{$\ket{0}_p$}&\control{}&&& \hat{X}(m)\hat{F}|\phi\rangle
    \end{quantikz}
    \label{Fourier}
    \caption{}
\end{subfigure}\hfill
\begin{subfigure}[b]{0.4\textwidth}
\begin{tikzpicture}[every node/.style={circle, draw=black, minimum size=0.8cm}, node distance=2cm]
    \node[fill=lightgray, label={[yshift=0cm]above:$\hat{p}$}] (1) at (0,0) {1};
     \node[fill=white, label={[yshift=-0.35cm]above:$\hat{X}(m)\hat{F}\ket{\phi}$}] (2) at (1.5,0) {2};
     \draw[-, thick] (1) -- (2);
\end{tikzpicture}
\caption{}
\end{subfigure}
\caption{(a) The basic circuit for CV MBQC, (b) the equivalent graph representation.}
\label{One way}
\end{figure}
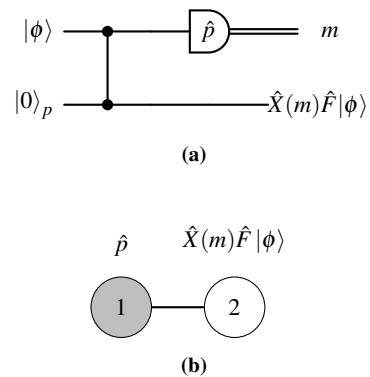

The quantum circuit for gate teleportation, depicted in Fig.~\ref{One way}, shows the most fundamental implementation of a Gaussian operation within the MBQC framework.
In this circuit, the first qumode encodes an arbitrary input state $\ket{\phi}$, while the second auxiliary qumode is prepared in the momentum vacuum $\ket{0}_p$. The two modes are then entangled via a $CZ$ gate. Upon measuring the first qumode in the $\hat{p}$-basis—yielding a classical measurement outcome $m$—the state of the unmeasured second qumode collapses to the output state $\ket{\phi}_{\text{out}} = \hat{X}(m)\hat{F}\ket{\phi}$. This circuit functions as a universal state teleporter. 

\begin{figure}[h]
\begin{center}
\begin{quantikz} 
\lstick{$\ket{\phi}$}&\ctrl{1}&&\meterD{\hat{D}^{\dagger}\hat{p}\hat{D}}&\setwiretype{c}\quad m\\
\lstick{$\ket{0}_p$}&\control{}&&& \hat{X}(m)\hat{F}\hat{D}|\phi\rangle
\end{quantikz}
\end{center}
\caption{Circuit with measurement in a rotated basis and equivalent transformation.}
\label{rotated}
\end{figure}
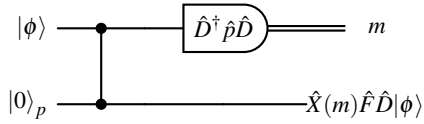

If we instead process a transformed state $\ket{\psi'} = \hat{D}\ket{\phi}$ (where $\hat{D} = \exp(if(\hat{q}))$ is diagonal in the computational basis) and perform a $\hat{p}$ measurement on the first qumode, the second qumode collapses to $\ket{\phi}_{\text{out}} = \hat{X}(m)\hat{F}\hat{D}\ket{\phi}$. Equivalently, as shown in Fig.~\ref{rotated}, this exact transformation can be induced on the original state $\ket{\phi}$ by modifying the measurement on the first qumode to a rotated basis defined by $\hat{D}^{\dagger}\hat{p}\hat{D}$.

This teleportation motif forms the bedrock of CV-MBQC, illustrating that any continuous-variable operation can be realized strictly through projective measurements. The byproduct operator $\hat{X}(m)\hat{F}$ depends inherently on the stochastic measurement result $m$ of the first qumode. This shift can be fully compensated for by adaptively updating the measurement bases in subsequent steps, or by merely tracking it classically as a Pauli frame update for the final state. By concatenating this basic circuit and incorporating classical feedforward, the complete set of Gaussian operations can be executed via the following measurement strategies:
\begin{itemize}
\item The Fourier transform $\hat{F}$ is intrinsically applied at each step of the cluster computation.
\item The momentum displacement $\hat{D}_{1,\hat{q}}(s) = \exp(is\hat{q})$ is achieved by measuring the observable $\hat{p}_{s,1} = \hat{p} + s$. Physically, this is equivalent to measuring $\hat{p}$ and simply adding the classical value $s$ to the result.
\item The shear gate $\hat{D}_{2,\hat{q}}(s) = \exp(is\hat{q}^2/2)$ is realized by measuring the observable $\hat{p}_{s,2} = \hat{p} + s\hat{q}$. This corresponds to homodyne detection in a rotated quadrature basis proportional to $\hat{p}\cos(\theta) - \hat{q}\sin(\theta)$, followed by a classical rescaling of the measurement outcome by a factor of $1/\cos(\theta)$.
\end{itemize}
The optical implementation of all Gaussian operations can be executed using strictly non-adaptive homodyne measurements on a suitably structured CV cluster state. Because the feedforward adaptations required for these measurements commute with Gaussian operations, the adaptation is considered trivial. The measurement results can simply be corrected via classical post-processing, meaning Gaussian measurements can be parallelized or performed in an arbitrary temporal sequence\cite{menicucci2006universal}.

\begin{figure}[h]
\centering
\begin{subfigure}[t]{0.5\textwidth}
    \centering
    \begin{quantikz} 
    \lstick{$\hat{S}(r)\ket{0}$}&\ctrl{1}&\meter{}&\setwiretype{c}\quad m = \hat{X}^\dagger\hat{n}\hat{X}\\
    \lstick{$\hat{S}(r)\ket{0}$}&\control{}&& \quad \approx \ket{\gamma} = \hat{D}_{3,\hat{q}}\ket{0}_p
    \end{quantikz}
    \caption{}
\end{subfigure}\hfill
\begin{subfigure}[t]{0.5\textwidth}
    \centering
    \begin{tikzpicture}[every node/.style={circle, draw=black, minimum size=0.8cm}, node distance=2cm]
    \node[fill=lightgray, label={[yshift=-0.1cm]above:$\hat{X}^\dagger\hat{n}\hat{X}$}] (1) at (0,0) {1};
     \node[fill=white, label={[yshift=0.05cm]above:$\ket{\gamma}$}] (2) at (1.5,0) {2};
     \draw[-, thick] (1) -- (2);
\end{tikzpicture}
    \caption{}
\end{subfigure}
\caption{The preparation of the cubic phase state. (a) Quantum circuit, (b) Cluster state representation.}
\label{Cubic phase state}
\end{figure}
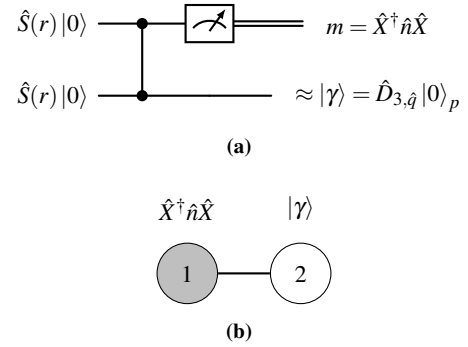

\begin{figure}[h]
\centering
\begin{subfigure}[t]{0.45\textwidth}
    \centering
    \begin{quantikz}    
    \lstick{$\ket{\phi}$}&\ctrl{1}&&\meterD{\hat{p}}&\setwiretype{c}\quad m_1\\
    \lstick{$\ket{\gamma}$}&\control{}&\ctrl{1}&\meterD{\hat{p}}&\setwiretype{c}\quad m_2\\
    \lstick{$\ket{0}_p$}&&\control{}&&\quad \approx \ket{\phi'} = \hat{D}_{3,\hat{q}}\ket{\phi}
    \end{quantikz}
    \caption{}
\end{subfigure}\hfill
\begin{subfigure}[t]{0.45\textwidth}
    \centering
    \begin{tikzpicture}[every node/.style={circle, draw=black, minimum size=0.6cm}, node distance=2cm]
    \begin{scope}
    \node[fill=white, label={[yshift=0.05cm]above:$\ket{\phi}$}] (1) at (0,0) {1};
     \node[fill=white, label={[yshift=0.05cm]above:$\ket{\gamma}$}] (2) at (1,0) {2};
     \node[fill=white, label={[yshift=-0.05cm]above:$\ket{{0}_p}$}] (3) at (2,0) {3};
     \draw[-, thick] (1) -- (2) -- (3);
     \end{scope}
     \draw[->, thick] (3, 0) -- (4, 0);
     \begin{scope}
    \node[fill=lightgray, label={[yshift=0.05cm]above:$\hat{p}$}] (1) at (5,0) {1};
     \node[fill=lightgray, label={[yshift=0.05cm]above:$\hat{p}$}] (2) at (6,0) {2};
     \node[fill=white, label={[yshift=-0.05cm]above:$\ket{\phi'}$}] (3) at (7,0) {3};
     \draw[-, thick] (1) -- (2) -- (3);
     \end{scope}
     \end{tikzpicture}
    \caption{}
\end{subfigure}
\caption{Implementation of cubic phase gate for an arbitrary state $\ket{\phi}$. (a) Quantum circuit, (b) Cluster state representation.}
\label{Cubic phase gate}
\end{figure}
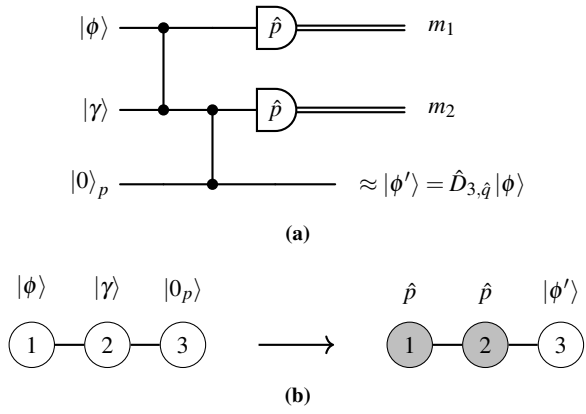

However, this operational triviality breaks down for non-Gaussian operations such as the cubic phase gate, which would theoretically be obtained by measuring the observable $\hat{p}_{s,3} = \hat{p} + s\hat{q}^2$. Because non-Gaussian operations do not commute trivially with Pauli displacements, adaptive measurements become strictly necessary. Consequently, the optical implementation becomes experimentally more challenging than that of Gaussian operations. One practical approach to implementing the cubic phase gate relies on gate teleportation via the offline preparation of a cubic phase state, defined as:
\begin{equation}
\ket{\gamma} = \hat{D}_{3,\hat{q}}\ket{0}_p = \frac{1}{\sqrt{2\pi}} \int dq \, \exp(i\gamma q^3)\ket{q}.
\end{equation}
Generating this state directly is notoriously difficult due to its requirement for cubic nonlinear resources (i.e.,  crystals with very large third-order $\chi^{(3)}$ nonlinearity). Such optical nonlinearities are extremely weak at the single-photon level and difficult to harness. To circumvent this, Gottesman, Kitaev, and Preskill \cite{gottesman2001encoding} demonstrated that the cubic phase state could be closely approximated using highly squeezed vacuum states $\hat{S}(r)\ket{0}$ combined with a displaced photon-counting measurement $\hat{X}^\dagger\hat{n}\hat{X}$, as illustrated in Fig.~\ref{Cubic phase state}.
Once successfully prepared offline, this cubic phase state is injected into the computation by a \textit{CZ} gate with a specific node of the Gaussian cluster state. A sequence of standard Gaussian measurements is then performed, and the outcomes are classically processed to deterministically apply the desired cubic transformation to the input state $\ket{\phi}$. As shown in Fig.~\ref{Cubic phase gate}, this approach deviates from the foundational CV-MBQC scheme by executing Gaussian homodyne measurements ($\hat{p}$) on a hybridized, non-Gaussian cluster state \cite{gu2009quantum}.

An alternative architecture preserves the original macroscopic Gaussian cluster but shifts the non-Gaussianity entirely into the measurement step\cite{gu2009quantum}. In this scenario, the full set of computational measurements must include both Gaussian and non-Gaussian detection schemes. A critical consequence of this approach is the loss of measurement parallelism. Due to the probabilistic nature of non-Gaussian measurements—and the fact that displacement errors are actively distorted by non-Gaussian unitaries—adaptive Gaussian corrections (such as feedforward displacements and squeezing) must be dynamically calculated and applied in real-time. This strict causal ordering ensures that the effective transformation corresponds exactly to the target cubic phase gate\cite{weedbrook_gaussian_2012}.

\begin{figure}[h]
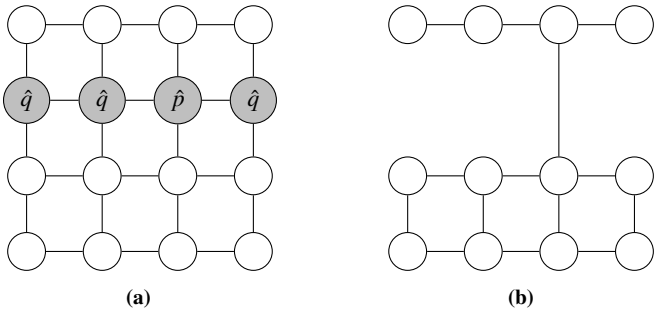

\centering
\begin{subfigure}[t]{0.2\textwidth}
    \removeshorteringoriginal
    \caption{}
\end{subfigure}\hfill
\begin{subfigure}[t]{0.2\textwidth}
    \removeshorteringafter
    \caption{}
\end{subfigure}
\caption{Cluster transformation in a 4$\times$4 cluster state. (a) Initial square lattice. (b) resulting cluster state after measuring selected nodes in $\hat{q}$ and $\hat{p}$ quadratures.}
\label{vertex removal and wire shortening}
\end{figure}

Finally, to dynamically manipulate and reconfigure entanglement graphs for MBQC, operations such as vertex removal and wire shortening \cite{weedbrook_gaussian_2012} are essential. These operations can be intuitively understood through the nullifier formalism and are executed via targeted quadrature measurements. Vertex removal is accomplished by measuring the position quadrature, $\hat{q}$, of a selected qumode; this effectively deletes the corresponding vertex from the graph and severs all of its entanglement links. Conversely, wire shortening involves measuring the momentum quadrature, $\hat{p}$, of an intermediate node. This measurement removes the target vertex while preserving and directly linking the correlations between its adjacent neighbors, effectively ``shortening'' the entanglement wire. Figure~\ref{vertex removal and wire shortening} illustrates the resulting cluster state topology after a sequence of different quadrature measurements is applied to an initial $4 \times 4$ cluster.

\subsubsection{Optical implementation of CV cluster states}
Ideal CV cluster states are constructed from zero-momentum eigenstates $\ket{0}_p$, which optically correspond to infinitely squeezed vacuum states requiring infinite energy. Because these states cannot be normalized, they are strictly unphysical; in practice, only finite levels of squeezing are achievable due to experimental constraints such as optical losses, phase noise, and limited pump power. Consequently, in real-world experiments, the theoretical vertices of the cluster are replaced by finitely squeezed states in the momentum quadrature. Despite this limitation, a key advantage of the continuous-variable optical implementation over its DV counterpart is the deterministic generation of cluster states \cite{menicucci2006universal, yoshikawa2007demonstration} utilizing squeezed light and linear optics \cite{zhang2009quantum, van2007building}. Furthermore, the homodyne detection used for CV measurements is highly efficient and well-developed, enabling high-speed, minimal-noise readout \cite{yuen1983noise} at room temperature.

Optical parametric oscillators (OPOs) have proven to be highly effective tools for generating both squeezed vacua and entangled cluster states. Notably, Vahlbruch \textit{et al.} demonstrated the generation of single-mode squeezed light with a record noise reduction of 15 dB using a monolithic PPKTP-based (Periodically Poled Potassium Titanyl Phosphate) OPO  \cite{vahlbruch2016detection}. Achieving such high levels of squeezing is essential for improving computational fidelity and crossing the fault-tolerance threshold\cite{menicucci2014fault, larsen2021fault}. Depending on the specific configuration, OPOs also possess the capacity to produce multimode squeezing and standard two-mode entangled states\cite{ou1992realization}. Another distinct advantage of this nonlinear process is that high-gain optical parametric amplification (OPA) can be leveraged for ultra-broadband quadrature measurements, beyond the bandwidth limit of the homodyne detector and effectively covering the entire bandwidth of the squeezed light\cite{shaked2018lifting, Li2020} which for OPA can be of several THz.
\begin{table*}[ht!]
\centering
\begin{tabularx}{\textwidth}{
   >{\centering\arraybackslash\hsize=.5\hsize\linewidth=\hsize}X 
   >{\centering\arraybackslash}X 
   >{\centering\arraybackslash\hsize=1\hsize\linewidth=\hsize}X 
   >{\centering\arraybackslash\hsize=0.7\hsize\linewidth=\hsize}X   
   >{\centering\arraybackslash\hsize=1.3\hsize\linewidth=\hsize}X  
   >{\centering\arraybackslash}X
    >{\centering\arraybackslash\hsize=1.5\hsize\linewidth=\hsize}X 
   }
\hline
\rowcolor[HTML]{3166FF}[\dimexpr\tabcolsep+0.1pt\relax]
\textcolor{white}{Year}&\textcolor{white}{Reference}  &  \textcolor{white}{Dimension }& \textcolor{white}{$\#$ of modes }  & \textcolor{white}{Insep Crteria }&\textcolor{white}{Multiplexing time}&\textcolor{white}{Platform} \\ \hline
2013 & Yokoyama et al  \cite{yokoyama2013ultra} & 1D (XEPR) & $10^4$  &$\checkmark$ &  157.5 ns  & Free space + delay line \\
\rowcolor[HTML]{ECF4FF}[\dimexpr\tabcolsep+0.05pt\relax]
2016 & Yoshikawa et al  \cite{yoshikawa2016invited} &1D (XEPR)  & $1.2\times10^6$&$\checkmark$ (first $6\times10^5$ qumode)&160 ns &  Free space + delay line\\ 
2019 &Larsen et al  \cite{larsen2019deterministic}& 2D & $3\times10^4$& $\checkmark$& 271 ns &Free space + delay line   \\ 
\rowcolor[HTML]{ECF4FF}[\dimexpr\tabcolsep+0.05pt\relax]
2019& Asavanant et al  \cite{asavanant2019generation} &2D&$24690$&$\checkmark$&40 ns& Free space + delay line     \\ 
2019&Takeda et al.\cite{takeda2019demand}&1D&up to 1000&$\checkmark$&66 ns& Based loop circuit\\
\rowcolor[HTML]{ECF4FF}[\dimexpr\tabcolsep+0.05pt\relax]
2023& Zhou et al \cite{zhou2023ultra}&1D&$2\times20400$&$\checkmark$&140 ns&Free space + delay line\\ \hline
\end{tabularx}
\caption{\textbf{Optical implementations of TDM CV cluster state}. For each work, we report the dimension, number of modes, the time of multiplexing, and the employed platform.}
\label{TDM}
\end{table*}

Historically, one of the earliest experimental realizations was the generation of a four-mode linear CV cluster state using four independent squeezed sources and a network of beam splitters\cite{su2007experimental}. Building on this, Yukawa \textit{et al.} demonstrated a scalable methodology for synthesizing arbitrary-shaped small cluster states using offline-prepared squeezed states and linear optics\cite{yukawa2008experimental}. The successful generation of spatially separated, eight-mode CV cluster states was subsequently achieved by multiplexing eight squeezed light sources driven by four nondegenerate optical parametric amplifiers (NOPAs)\cite{su2012experimental}. 

Currently, integrated photonics represents a highly promising platform for CV-MBQC, enabling scalable, mechanically stable, and compact quantum optical circuits. Integrating components—such as squeezers, beam splitters, and phase shifters—directly onto a chip\cite{Lenzini2018} can reduce optical propagation losses, and enhance phase stability and mode matching, which are paramount for manipulating large-scale cluster states. Materials exhibiting high optical nonlinearity and low propagation loss, such as lithium niobate (LN), silicon nitride (SiN), and silicon-on-insulator (SOI), are exceptionally well-suited for on-chip quantum operations \cite{Wang2019, elshaari2020hybrid, Peace2022}. Thin-film lithium niobate (TFLN)\cite{hu2025integrated, Milad2022}, in particular, is notable for its strong optical confinement, high $\chi^{(2)}$ nonlinearity, and rapid electro-optic tunability, making it vital for efficient squeezing and fast circuit reconfigurability. While challenges remain—chiefly overcoming coupling inefficiencies and propagation losses in large-scale circuits to preserve high squeezing levels—integrated photonics continues to push the boundaries of scalable continuous-variable quantum technologies.

\subsubsection{Scalability of CV cluster states}
The initial CV cluster state experiments discussed above\cite{su2007experimental,yukawa2008experimental,su2012experimental} are inherently limited in scalability due to their physical footprint, extreme alignment sensitivity, and optical losses that compound with the number of modes. These physical constraints have motivated a necessary transition toward more scalable architectures. Several approaches have been proposed and experimentally demonstrated to generate massive CV cluster states by leveraging different physical degrees of freedom. These primary strategies include spatial, time-domain, and frequency-domain multiplexing, alongside hybrid combinations of these techniques. Each method presents distinct trade-offs regarding resource efficiency, experimental complexity, and compatibility with integrated photonics. The rest of this section explores these approaches and their potential for scalable CV quantum information processing.

\textit{Spatial multiplexing} is one of the earliest approaches explored for scaling CV-MBQC, encoding quantum information into distinct spatial modes of light, such as Hermite-Gaussian (HG) \cite{Wang2022} or Laguerre-Gaussian (LG) modes \cite{Zhang:17}. It has been demonstrated that OPOs natively support spatial entanglement \cite{janousek2009optical, Midgley_2010, zambrini2003polarization} as well as hyperentanglement in orbital angular momentum (OAM) \cite{dosSantos2009}. Several studies have confirmed that spatial multiplexing offers a powerful method for the optical implementation of large-scale CV cluster states \cite{Pooser_2014, Yang:16, Zhang:17}. For example, Armstrong \textit{et al.} \cite{Armstrong_2012} experimentally demonstrated the generation of up to 8 entangled modes, including programmable cluster states (of 2, 3, and 4 modes) via emulated linear optical networks. Later, Zhang \textit{et al.} \cite{Zhang2020} generated a four-mode CV cluster state using spatially separated LG modes from a single squeezed beam, rigorously verifying the inseparability of the modes via homodyne detection. In 2020, Wang \textit{et al.} \cite{Wang2020} demonstrated combined OAM multiplexing with spatial pump shaping to create a CV entangled network. By employing a hot atomic vapor to drive multiple four-wave mixing (FWM) processes, they shaped the pump spatially to produce hexapartite entanglement—specifically, six spatially separated LG beams distributed across 11 orthogonal OAM channels. Because each channel comprised six entangled modes, the system yielded a total of 66 entangled OAM modes. Using homodyne detection on the six beams and verifying over 31 bipartitions, they confirmed full inseparability, successfully demonstrating a multiplexed CV quantum network. A subsequent experiment \cite{Wang2022} demonstrated the simultaneous and deterministic generation of spatial entanglement across 56 distinct HG mode pairs using FWM in a single hot $^{85}\text{Rb}$ atomic vapor cell. Despite these achievements, the deterministic implementation of an arbitrary, large-scale CV cluster state using strictly spatial multiplexing remains inefficient. This is primarily due to the experimental difficulty of individually accessing, manipulating, or separating each spatial mode within a single macroscopic beam. However, hybridizing this approach with time-domain or frequency-domain multiplexing may offer a  promising pathway to reduce the complexity of interferometric setups.
\begin{table*}[th!]
\centering
\begin{tabularx}{\textwidth}{
   >{\centering\arraybackslash\hsize=.5\hsize\linewidth=\hsize}X 
   >{\centering\arraybackslash}X 
   >{\centering\arraybackslash\hsize=0.75\hsize\linewidth=\hsize}X 
   >{\centering\arraybackslash\hsize=0.75\hsize\linewidth=\hsize}X   
   >{\centering\arraybackslash}X
    >{\centering\arraybackslash\hsize=2\hsize\linewidth=\hsize}X 
   }
\hline
\rowcolor[HTML]{08a332}[\dimexpr\tabcolsep+0.1pt\relax]
\textcolor{white}{Year}&\textcolor{white}{Reference}  &  \textcolor{white}{Dimension }& \textcolor{white}{$\#$ of modes }  &\textcolor{white}{Platform} & \textcolor{white}{Key notes}\\ \hline
2011& Pysher et al.\cite{Pysher2011} &2D&60&Bulk optic&Entanglement between cavity modes\\
\rowcolor[HTML]{9ad29c}[\dimexpr\tabcolsep+0.05pt\relax]
2014&Roslund et al.\cite{roslund2014wavelength}&-& 511 bipartitions& Bulk optic& Entanglement between supermodes  \\ 
2014&Chen et al.\cite{chen2014experimental}&1D dual-rail& 60& Bulk optic& Entanglement between cavity modes    \\ 
\rowcolor[HTML]{9ad29c}[\dimexpr\tabcolsep+0.05pt\relax]
2014&Wang et al.\cite{Wang2014}&nD dual-rail&$10^4$&Proposal&Hypercubic cluster\\
2017&Cai et al.\cite{cai2017multimode}&1D,2D&6&Bulk optic &Reconfigurable graph
states\\
\rowcolor[HTML]{9ad29c}[\dimexpr\tabcolsep+0.05pt\relax]
2021&Zhu et al.\cite{zhu2021hypercubic}&nD&$10^4$ &Proposal&Hypercubic cluster\\
2023&Hurvitz et al.\cite{Hurvitz_2023}&2D&-&Proposal&Bright squeezed state as resource\\
\rowcolor[HTML]{9ad29c}[\dimexpr\tabcolsep+0.05pt\relax]
2025&Jia et al.\cite{jia2025continuous}&1D,2D&8&Integrated microcomb&Supermode entanglement\\
2025&Wang et al.\cite{wang2025large}&1D,2D&60&Integrated microcomb&Programmable 1D and 2D cluster states\\\hline
\end{tabularx}
\caption{\textbf{Optical implementations and proposals of FDM CV cluster state}. For each work, we report the dimension, number of modes, and the employed platform. The sign (-) indicates that the specific information is not mentioned in the paper.}
\label{FDM}
\end{table*}

\begin{figure*}[!htb] 
    \centering
    \begin{subfigure}{1\textwidth}
   \centering 
    \includegraphics[width=\textwidth]{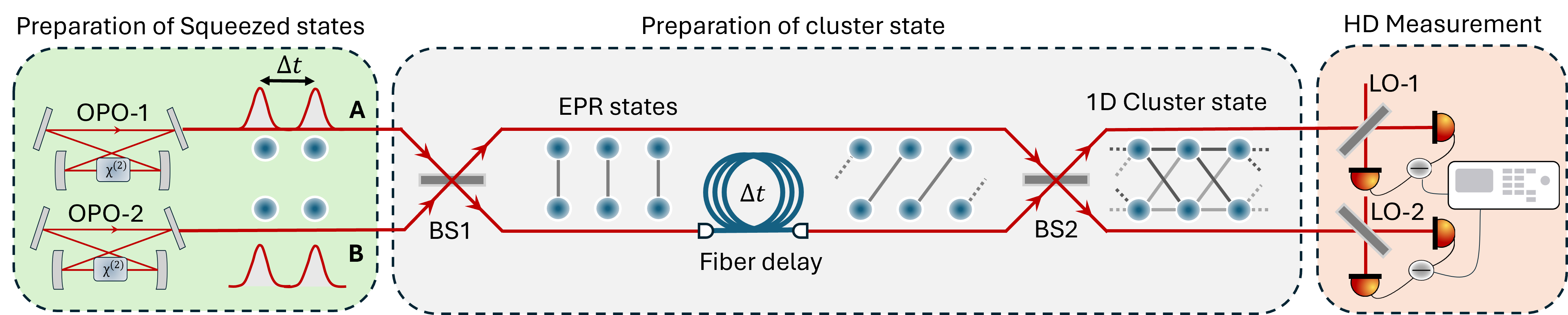}
    \caption{}

\end{subfigure}
    \hfill
\begin{subfigure}{1\textwidth}
\centering
    \includegraphics[width=\textwidth]{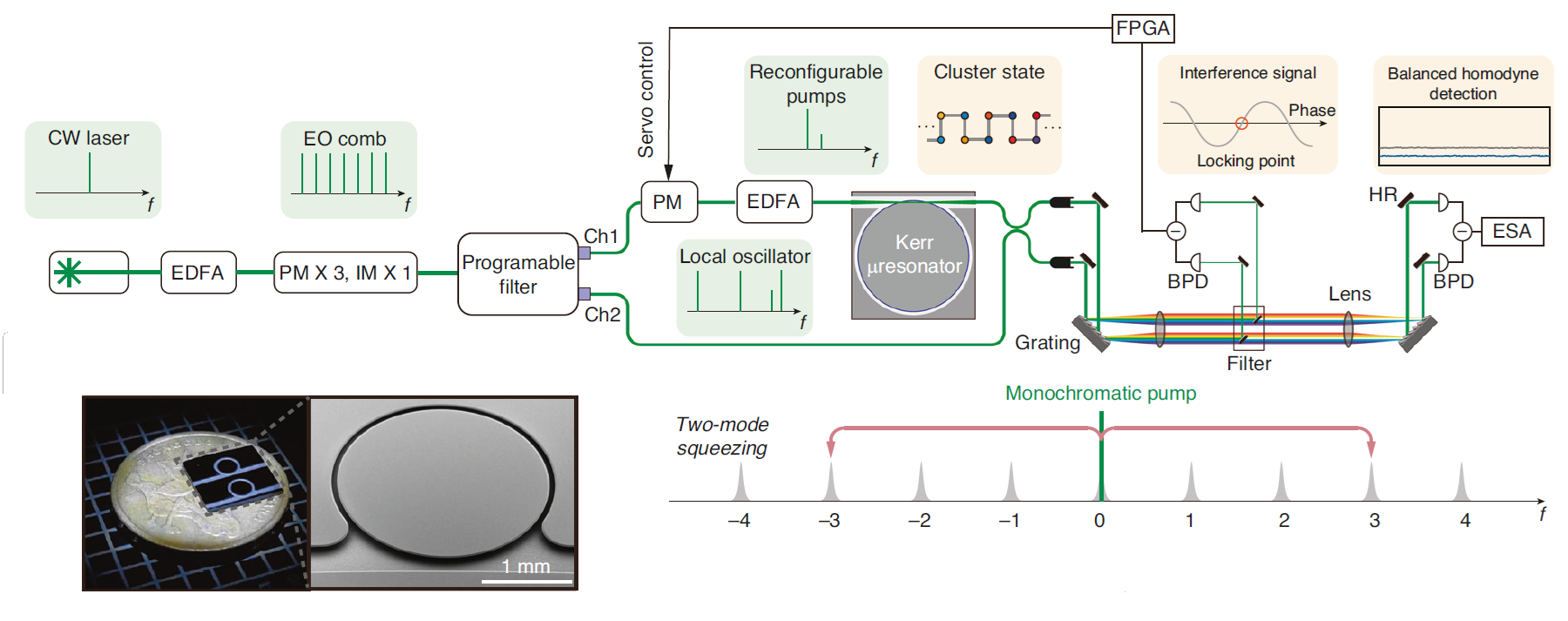}
    \caption{}
\end{subfigure}\hfill
    
\caption{\textbf{Implementation of CV-cluster states}. (a) Schematic of the optical setup to generate a 1D CV cluster state multiplexed in the time domain. Temporal modes of squeezed vacuum are generated in two spatial modes A and B interfered to generate EPR states in BS1. The EPR pairs are connected into a 1D cluster state using a $\Delta t$ delay in mode B and interference in BS2.  (b) Experimental setup for the generation of large-scale cluster quantum microcombs. (CW) continuous-wave laser, (EO comb) electro-optic comb, (EDFA) erbium-doped fiber amplifier, (PM) phase modulator, (IM  intensity modulator, (BPD) balanced photodetector, (ESA) electrical spectrum analyzer, (FPGA) field-programmable gate array. Figure reproduced from Wang et al. Light: Science \& Applications 14, 164 (2025); licensed under a Creative Commons Attribution (CC BY) license.\cite{wang2025large}} 
    \label{CV cluster}
\end{figure*}

\textit{Time-domain multiplexing (TDM)} was first introduced in 2010 \cite{menicucci2010arbitrarily, menicucci2011temporal} with the aim of generating large cluster states with the minimum number of resources in terms of squeezing sources and optical components. This strategy facilitates the propagation of multiple $qumodes$ within a single beam, distinguished and ultimately made orthogonal by their separation in time. In 2013, Yokoyama \textit{et al.} \cite{yokoyama2013ultra} demonstrated the first generation of a 1D ultra-large-scale CV cluster state comprising $10^4$ entangled qumodes, experimentally proving the feasibility of time-domain multiplexing for massive entanglement generation. In this experiment, shown in Fig.~\ref{CV cluster}a, two OPOs were used to generate single-mode squeezed vacuum beams, which were conceptually divided into time bins to define a continuous train of pulsed squeezed states. These states were interfered on a 50:50 beam splitter, creating EPR pairs separated by a time interval $\Delta t$. Subsequently, one optical mode was delayed by a duration of $\Delta t$ via a fiber delay line before being recombined with the second mode on another 50:50 beam splitter. Repeating this process, deterministically entangles successive pulses, leading to the formation of a 1D dual-rail cluster state, commonly referred to as an extended EPR (XEPR) state. Three years later, in 2016, this experimental architecture was refined and extended to generate more than $10^6$ qumodes \cite{yoshikawa2016invited}. In 2019, Asavanant \textit{et al.} \cite{asavanant2019generation} and Larsen \textit{et al.} \cite{larsen2019deterministic} pushed this approach further by realizing 2D cluster states—which are strictly necessary for universal MBQC—by incorporating additional delay lines and squeezing resources. Building on this, Zhou \textit{et al.} \cite{zhou2023ultra} recently demonstrated the generation of ultra-large-scale clusters utilizing a time-delayed quantum interferometer \cite{yurke19862}. Furthermore, a novel methodology for generating 3D CV cluster states \cite{fukui2020temporal} was proposed in 2020. This technique combines time-multiplexing with a divide-and-conquer approach: a massive theoretical CV cluster state is split into smaller, local sub-clusters, and time-domain entangling operations are subsequently used to reconnect (conquer) them step-by-step. This effectively synthesizes 3D cluster states, which are a critical prerequisite for topological quantum error correction in MBQC. Most recently, Segev \textit{et al.} \cite{segev2025continuous} proposed an innovative method to create TDM cluster states utilizing photonic time-crystals (PTCs)—dielectric materials whose refractive indices are rapidly modulated in time. Table~\ref{TDM} summarizes several key experiments that have demonstrated the expanding feasibility of the TDM framework. Despite the experimental challenges in terms of optical loss, synchronization stability, and the execution of real-time active feedforward, this protocol demonstrated entanglement over the largest number of modes and is the main architecture pursued by leading companies like Xanadu\cite{xanadu}. 

\textit{Frequency-domain multiplexing (FDM)} encodes different optical modes into distinct frequency bins or comb lines, usually generated from a single OPO cavity or resonator. This approach enables multiple modes to coexist within a single spatial and temporal channel. By leveraging the broad spectral bandwidth of spontaneous parametric processes alongside the spectral line selectivity of a cavity and its associated frequency comb, FDM can entangle many modes simultaneously, reducing the spatial footprint of the device that generate the cluster state \cite{Pfister2004, Menicucci2008, Pinel2012, Pfister_2020}. Pysher \textit{et al.} \cite{Pysher2011} demonstrated the first generation of cluster states in the optical frequency domain, creating 15 quadripartite cluster states over 60 cavity qumodes. In 2014, Chen \textit{et al.} \cite{chen2014experimental} generated a CV cluster state by pumping a single optical parametric oscillator (OPO) with two frequencies to produce an entangled optical frequency comb comprising 60 distinct modes. These modes, each corresponding to a spectral line in the comb, were pairwise entangled via parametric down-conversion and coherently linked into a one-dimensional cluster state using a passive interferometric network. Homodyne detection allowed them to reconstruct the multimode covariance matrix and verify genuine multipartite entanglement across all 60 modes, marking a significant step toward scalable MBQC in the frequency domain. Concurrently, Roslund \textit{et al.} \cite{roslund2014wavelength} demonstrated the single-step generation of a highly multimode quantum cluster state by parametrically down-converting a femtosecond optical frequency comb within a synchronously pumped OPO. Instead of treating each frequency bin individually, they applied a supermode decomposition approach \cite{Gerke2014supermode, Fabre2020supermode} to identify orthogonal eigenmodes of the squeezing interaction, revealing at least eight significantly squeezed and independently addressable supermodes. Through ultrafast pulse shaping and homodyne detection \cite{Medeiros2014}, they characterized spectral entanglement across 10 discrete frequency bands, verifying entanglement across all 511 possible bipartitions. Table~\ref{FDM} summarizes key experiments and proposals validating the FDM method.
\begin{table*}[t!]
\centering

\begin{tabularx}{\textwidth}{
   >{\centering\arraybackslash\hsize=.5\hsize\linewidth=\hsize}X 
   >{\centering\arraybackslash}X 
   >{\centering\arraybackslash\hsize=0.75\hsize\linewidth=\hsize}X 
   >{\centering\arraybackslash\hsize=0.75\hsize\linewidth=\hsize}X   
   >{\centering\arraybackslash\hsize=1.5\hsize\linewidth=\hsize}X
    >{\centering\arraybackslash\hsize=1.5\hsize\linewidth=\hsize}X 
   }
\hline
\rowcolor[HTML]{d4a50d}[\dimexpr\tabcolsep+0.1pt\relax]
\textcolor{white}{Year}&\textcolor{white}{Reference}  &  \textcolor{white}{Dimension }& \textcolor{white}{$\#$ of modes }  &\textcolor{white}{Platform} & \textcolor{white}{type of multiplexing}\\ \hline
2020 &Yang  et al.\cite{Yang_2020} &2D&8-60&Proposal&Spatiotemporal\\
\rowcolor[HTML]{e8cf7c}[\dimexpr\tabcolsep+0.05pt\relax]
2020&Wu et al.\cite{wu2020quantum}& 1,2,3D & - & Proposal& Time-frequency   \\ 
2022&Guo et al.\cite{Guo:22}&-& 14& Bulk optic& Space-frequency    \\ 
\rowcolor[HTML]{e8cf7c}[\dimexpr\tabcolsep+0.05pt\relax]
2023&Du et al.\cite{Du_2023}&1D,3D&$10^6 -10^{10}$&Proposal&Time-frequency\\
2023&Kouadou   et al.\cite{kouadou2023spectrally}&-&21&Bulk optic &Time-frequency\\
\rowcolor[HTML]{e8cf7c}[\dimexpr\tabcolsep+0.05pt\relax]
2025&Roh  et al.\cite{roh2025generation}&3D&36&Bulk optic&Time-frequency\\
2025&Du et al.\cite{du2025compactonewayfaulttolerantoptical}&3D&$10^3$ &Proposal&Space-frequency\\
\rowcolor[HTML]{e8cf7c}[\dimexpr\tabcolsep+0.05pt\relax]
2025&Du et al.\cite{Du_2025}&2D&-&Proposal&Spatiotemporal\\
\hline
\end{tabularx}
\caption{\textbf{Optical implementations and proposals of CV cluster state via the Hybrid approach}. The sign (-) indicates that the specific information is not mentioned in the paper.}
\label{Hybrid}
\end{table*}  

Recent studies have shown a marked shift in focus toward the implementation of CV cluster states on-chip using the FDM approach. Jia \textit{et al.} \cite{jia2025continuous} demonstrated the first deterministic generation of CV multipartite entanglement on an integrated photonic chip utilizing a silicon-nitride microresonator. By pumping the microcomb with a phase-locked polychromatic pump and employing polychromatic homodyne detection, they generated an eight-mode squeezed vacuum state. Entanglement was validated by measuring nullifier correlation matrices to verify the violation of van vLF inseparability criteria. These inequalities were successfully violated for four-qumode states (linear, box, and star topologies) at sideband frequencies up to 200 MHz, and for a six-qumode linear cluster state at up to 100 MHz. Further advancing the field, Wang \textit{et al.} \cite{wang2025large} generated a CV cluster state spanning 60 qumodes using an on-chip optical microresonator pumped by multiple laser lines (see Fig.~\ref{CV cluster}b). Leveraging Kerr-induced four-wave mixing (FWM) enhanced by high-Q resonator modes, they programmably formed either one- or two-dimensional cluster state graphs simply by tuning the pump configurations. These states exhibited high-quality entanglement with raw two-mode squeezing levels exceeding 3 dB, setting a new record for chip-scale platforms. They confirmed full inseparability via covariance matrix measurements and the positive partial transposition (PPT) criteria \cite{HORODECKI1997333}, demonstrating robust multipartite quantum correlations. This work represents the largest-scale CV cluster state generated from an integrated system to date, suggesting a potential pathway for future on-chip applications via dispersion engineering and advanced photonic integration.

Notably, FDM offers a fundamentally lossless approach to multiplexing compared to the TDM method, which typically relies on lossy optical delay lines. However, the detection part can be more complex than TDM because of mode selectivity issues since all frequencies travel in the same spatial mode. Individually addressing closely spaced frequency modes, and the execution of real-time adaptive measurements are challenges of this approach.

\textit{Hybrid multiplexing} combines multiple optical degrees of freedom—such as time, frequency, and spatial modes—to maximize the scalability and flexibility of CV cluster states for MBQC. By exploiting a variety of multiplexing strategies in parallel, hybrid architectures can increase the number of accessible entangled modes without a concomitant rise in experimental complexity. For instance, creating two-dimensional cluster states by combining time-domain and frequency-domain multiplexing \cite{Humphreys_2014, Alexander2016, wu2020quantum, Du_2023, kouadou2023spectrally, roh2025generation} enables the dense arrangement of a massive number of modes across both temporal and spectral axes. Similarly, incorporating spatial multiplexing facilitates the parallelization of multiple hybrid streams, thereby further augmenting the total entangled resource \cite{Yang_2020, Guo:22, Du_2025, du2025compactonewayfaulttolerantoptical}.

Recent proposals and proof-of-principle demonstrations highlight the potential of hybrid approaches to circumvent the physical limitations inherent to individual multiplexing techniques. Notably, Roh \textit{et al.} \cite{roh2025generation} demonstrated the deterministic generation of a 3D cluster state, which is a resource for fault-tolerant MBQC. Their innovation lies in exploiting the ultrafast time-frequency modes of squeezed light from a synchronously pumped OPO. By selecting complex mode bases via active pulse shaping and homodyne detection, they successfully established programmable connectivity structures in 1D, 2D, and 3D. The resulting multimode Gaussian states were fully characterized using continuous-variable quantum state tomography \cite{Cramer_2010, Lvovsky_2009}. Key entanglement properties, such as nullifier uncertainties and full multipartite inseparability, were rigorously validated, demonstrating high-quality 3D entanglement. Because 3D clusters are the minimal resource required to enable built-in topological error correction, this result represents a step toward fault-tolerant, universal photonic quantum computing. A list of experiments utilizing this approach is presented in Table~\ref{Hybrid}. Ultimately, hybrid multiplexing represents a highly promising avenue for achieving the extremely large, structured entangled states necessary for universal CV quantum computation.

\subsubsection{Optical implementation of CV quantum operation}
Recent research has made significant progress in tackling the challenges of CV-MBQC. Asavanant \textit{et al.} \cite{asavanant2021time} developed a high-speed (25 MHz clock) programmable homodyne detection system integrated with a 1D time-multiplexed CV cluster state, enabling the first experimental realization of quantum gate operations across multiple temporal modes. This demonstration encompassed arbitrary single-mode Gaussian operations, validated the preservation of non-classical features, and successfully executed over 100 sequential operations. Concurrently in 2021, Larsen \textit{et al.} \cite{larsen2021deterministic} implemented a multimode set of measurement-induced quantum gates within a large two-dimensional (2D) cluster state. 

Recent developments have leveraged feedback-based loop architectures \cite{motes2014scalable, takeda2017universal} that operate as programmable photonic processors \cite{takeda2019demand}. This enables high-speed, stable quantum processing using time-multiplexed architectures combined with active feedforward control \cite{enomoto2021programmable}. Yokoyama \textit{et al.} \cite{yokoyama2025fullstackanalogopticalquantum} recently introduced a scalable CV optical quantum computing system that integrates cutting-edge hardware, software, and cloud access. Building upon the TDM generation of large-scale 2D entangled cluster states, their system processes sequential analog inputs across up to 100 channels at a  clock rate of 100 MHz without requiring full error correction. Instead of relying on complex fault-tolerant schemes, they suppress accumulated analog noise and biases through precise calibration and repeated trial averaging. Delivered as a full-stack platform complete with a cloud interface and a Python SDK (software development kit) for streamlined accessibility, this work represents a step forward in analog optical quantum computing, with potential applications to high-speed quantum neural networks and practical quantum information processing. 

Despite these results with Gaussian operations, the implementation of the non-Gaussian cubic phase gate remains the main challenge in the field of CV-MBQC. Numerous experimental proposals \cite{marek2011deterministic, marek2018general, arzani2017polynomial} and architectural arrangements have been formulated to make the original concept \cite{gottesman2001encoding, gu2009quantum} more experimentally feasible. An important result in this context is the work of Miyata \textit{et al.}\cite{Miyata2016}, who demonstrated a proof-of-principle implementation of a cubic phase gate using an offline-prepared non-Gaussian ancillary state, linear optics, and measurement-induced operations within a CV cluster state. Their approach involved generating an approximate cubic phase state via photon subtraction from a squeezed vacuum, subsequently performing gate teleportation via homodyne measurements. Alternatively, a weak cubic phase can be approximated by generating superpositions of Fock states up to $\ket{3}$ \cite{sabapathy2018states, yukawa2013generating}. Most recently, research has increasingly focused on implementing tunable cubic phase gates optimized and enhanced by machine learning protocols \cite{sabapathy2019production}.

\subsubsection{Fault-tolerant CV quantum computation}
In realistic CV-MBQC architectures, errors arise primarily from the degradation of squeezing due to optical losses, which introduces Gaussian noise into the quantum state. This noise degrades the fidelity of gate operations and the overall computation, fundamentally limiting the usable entanglement of the initial resource states. As the computation progresses via sequential measurements, these errors accumulate—particularly in large-scale systems—distorting the underlying gate teleportation mechanism \cite{gu2009quantum}. Implementing robust quantum error correction (QEC) is therefore strictly necessary to maintain computational accuracy within fault-tolerant thresholds. In CV systems, QEC protocols digitize quantum information by defining two discrete logical codewords, $\ket{0}_L = \sum_n c_n^0\ket{n}$ and $\ket{1}_L = \sum_n c_n^1\ket{n}$. This bosonic encoding takes advantage of the infinite-dimensional Hilbert space of a single qumode. Prominent examples include cat codes \cite{ralph2003quantum}, binomial codes \cite{michael2016new}, and Gottesman-Kitaev-Preskill (GKP) codes \cite{gottesman2001encoding}. In contrast to cat and binomial codes, which are principally designed to protect against photon loss, the grid-like phase space structure of the GKP code uniquely enables the correction of small shift errors in both position and momentum quadratures. This feature makes the GKP code a highly promising candidate for realizing fault-tolerant, error-corrected quantum information processing using CV encoding.\\

In their ideal form, GKP states are grid-like superpositions of infinitely squeezed position eigenstates spaced periodically in phase space, forming a lattice structure robust against small displacement errors. Mathematically, they are composed of Dirac delta functions positioned at integer multiples of $\sqrt{\pi}$. In the $q$ basis, the logical codewords are given by
\begin{equation}
    \ket{0}_{GKP} = \sum_n\ket{2n\sqrt{\pi}}_q, \quad  \ket{1}_{GKP} = \sum_n\ket{(2n+1)\sqrt{\pi}}_q.
\end{equation}
Note that, when measuring $\hat{q}$ or $\hat{p}$, we would obtain an outcome $s$ that would be a multiple of $\sqrt{\pi}$, and the GKP code is stabilized by two displacement operators $\{S_q= e^{-2i\sqrt{\pi}\hat{p}}, S_p=e^{2i\sqrt{\pi}\hat{q}}\}$.

Recent advances have shown that by concatenating GKP codes with CV cluster architectures, fault tolerance thresholds can be achieved, despite the challenges posed by finite squeezing and noise\cite{fukui2018high}. The GKP encoding bridges the gap between CV and DV models, allowing hybrid architectures that take advantage of both worlds, CV systems for scalability, and DV systems for error correction and logical operations, providing a pathway to universality. For example, all logical Clifford gates can be implemented using only Gaussian operations.  More precisely, they allow for a direct equivalence between universal gate sets in CV and DV as shown in the table \ref{tab:my_label}.
 
\begin{table}[t!]
    \centering
    \begin{tabular}{c>{\centering\arraybackslash}p{5cm}}\toprule
         Logical Gate (DV) & CV realisation\\\midrule
         $X$ (Pauli-X)& $\hat{D}(\sqrt{\pi})$\\
         $Z$ (Pauli-Z)& $\hat{D}(i\sqrt{\pi})$\\
         $H$ (Hadamard)& $\hat{F} = \hat{R}(\theta/2)$\\
         $S$ (Phase)& $S = e^{i\hat{q}^2/2}$ (Shear gate) \\
         \textit{CZ}& $e^{is\hat{q}_1\hat{q}_2}$ (Controlled-Z)\\
         \textit{CNOT} &  $\hat{U}_{\text{SUM}} = e^{-i\hat{q}_1\hat{p}_2}$\\ \bottomrule
    \end{tabular}
    \caption{Logical qubit operations (DV) and their equivalences in CV systems for ideal GKP encoding.}
    \label{tab:my_label}
\end{table}
To complete a universal gate set, one must introduce a non-Gaussian operation that enables the implementation of a logical non-Clifford gate, such as the T-gate. However, as noted previously, the practical realization of the cubic phase gate presents a major experimental challenge in CV systems. This is particularly true for GKP encoding, as it requires strong optical nonlinearities, photon-number-resolving detectors, or complex ancilla-based gate teleportation methods—all of which introduce significant overhead, noise, and sensitivity to optical loss \cite{marshall2015repeat}. To circumvent these difficulties, an alternative approach based on GKP magic states has been proposed\cite{Bravyi2005, reichardt2005quantum, Konno2021}. These are pre-prepared, non-Gaussian resource states that can be injected directly into a Gaussian circuit. This strategy allows universal quantum computation to be achieved using only Gaussian operations combined with GKP magic state distillation, entirely removing the need to implement the problematic cubic phase gate directly\cite{menicucci2014fault}. Here, the problem is shifted towards the generation of GKP magic states, which is experimentally challenging.

\begin{figure}[b!]
    \centering
    \includegraphics[width=1\linewidth]{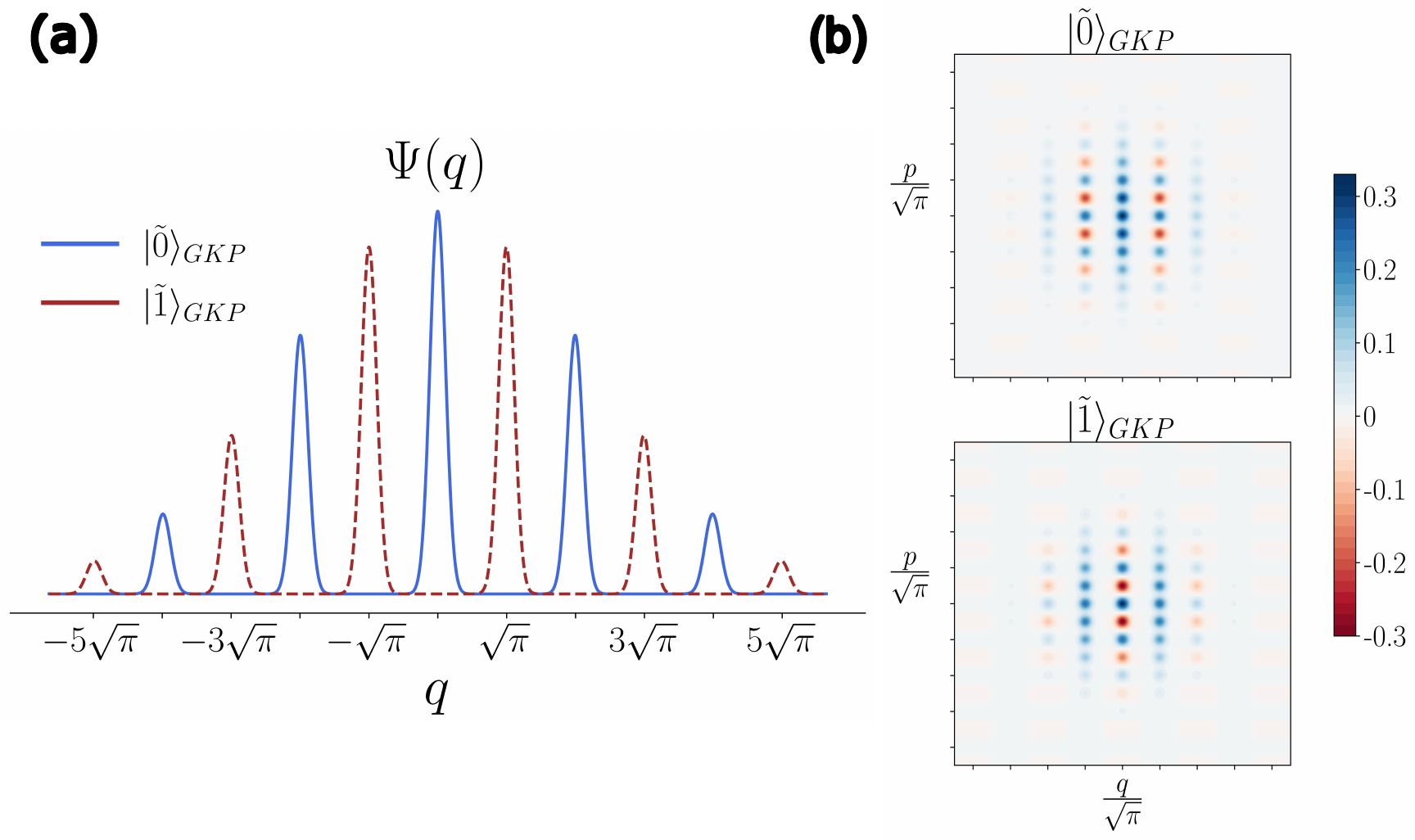}
    \caption{\textbf{Real GKP states}. (a) Wavefunction in position representation, (b) Wigner function representation of $\ket{\tilde{0}}_{GKP}$ and $\ket{\tilde{1}}_{GKP}$ for $\Delta_{dB}= 10$ dB.}
    \label{GKP vis}
\end{figure}

In practice, physical GKP states can be constructed from finitely squeezed Gaussian states of widths $\Delta$ that replace the Dirac peaks and apply an overall Gaussian envelope width $\delta$ to ensure finite energy. In the position basis, the logical codes $\ket{0}_L$ and $\ket{1}_L$ of the approximate GKP states are given by
\begin{align}
    \ket{\tilde{0}}_{GKP} & \propto \sum_n\int e^{-\frac{[2n\sqrt\pi]^2}{2\delta^2}}e^{-\frac{[s-2n\sqrt\pi]^2}{2\Delta^2}}\ket{s}_qds,\\
   \ket{\tilde{1}}_{GKP} & \propto \sum_n\int e^{-\frac{[(2n+1)\sqrt\pi]^2}{2\delta^2}}e^{-\frac{[s-(2n+1)\sqrt\pi]^2}{2\Delta^2}}\ket{s}_qds.
\end{align}
Note that the peak width $\Delta$ is directly related to the squeezing strength. If we consider pure states with symmetric squeezing and anti-squeezing, we find $\Delta = \delta^{-1}$, as the product of the variances for the quadratures $\hat{q}$ and $\hat{p}$ is minimized. A smaller $\Delta$ yields better discrimination between the logical codewords $\ket{\tilde{0}}_{\text{GKP}}$ and $\ket{\tilde{1}}_{\text{GKP}}$ during homodyne measurement. In the asymptotic limit where $\Delta \rightarrow 0$ and $\delta \rightarrow \infty$, these finite-energy states converge into ideal GKP qubits. Figure~\ref{GKP vis} illustrates the wavefunction in position representation, and its Wigner function for the GKP state with $\Delta=\delta^{-1}$ and $\Delta^2 =0.05$ (equivalent to $\Delta^2_{dB}>-10log_{10}(2\Delta^2)$=10 dB of squeezing).

\begin{table*}[ht!]
\centering
\begin{tabularx}{\textwidth}{
   >{\centering\arraybackslash\hsize=0.5\hsize\linewidth=\hsize}X 
   >{\centering\arraybackslash\hsize=0.75\hsize\linewidth=\hsize}X 
   >{\centering\arraybackslash\hsize=0.75\hsize\linewidth=\hsize}X 
   >{\centering\arraybackslash\hsize=2\hsize\linewidth=\hsize}X   
   }
\hline
\rowcolor[HTML]{05c0d9}[\dimexpr\tabcolsep+0.1pt\relax]
\textcolor{white}{Year}&\textcolor{white}{Reference}  &  \textcolor{white}{Squeezing threshold}& \textcolor{white}{Protocol / Scheme  } \\ \hline
2014&Menicucci et al.\cite{menicucci2014fault}&$ 20.5$ dB &Concatenated GKP codes\\
\rowcolor[HTML]{a6dce3}[\dimexpr\tabcolsep+0.05pt\relax]
2017&Fukui et al.\cite{Fukui2017}&$9$ dB& Hybrid quantum error correction \\
2018&Fukui et al.\cite{fukui2018high}&$<10$ dB& Postselection measurement  \\ 
\rowcolor[HTML]{a6dce3}[\dimexpr\tabcolsep+0.05pt\relax]
2021&Larsen et al.\cite{larsen2021fault}&$12.7$ dB&Surface GKP code\\
2021&Tzitrin et al.\cite{Tzitrin2021_FTQC}&$10.1$ dB&Topological protected MBQC\cite{fukui2020temporal}\\
\rowcolor[HTML]{a6dce3}[\dimexpr\tabcolsep+0.05pt\relax]
2022&Noh et al.\cite{Noh2022}&9.9 dB&Surface-GKP Code\\
2025& \O stergaard et al.\cite{ostergaard2025}&9.75 dB&Octo-Rail Lattice, Surface-GKP Code\\
\rowcolor[HTML]{a6dce3}[\dimexpr\tabcolsep+0.1pt\relax]
2025&Aghaee Rad et al.\cite{Rad2025}&9.75 dB&Heralded non-Gaussian resource generation\\\hline
\end{tabularx}
\caption{Quantum error correction protocols with GKP encoding and squeezing thresholds.}
\label{FTQC}
\end{table*}
In addition to the imperfections that originate from the finite squeezing of the initial states in GKP encoding, errors naturally occur during information processing due to thermal noise, photon loss, dephasing, or imprecise control operations. We assume that these errors introduce random displacements in the phase space of the encoded oscillator. Mathematically, they are described by a Gaussian displacement channel \cite{Holevo2001, Fukui_2022} acting on a quantum state $\hat{\rho}$ as follows:
\begin{equation}
    \mathcal{N}_\xi(\hat{\rho}) = \int \frac{d^2\alpha}{\pi \xi^2} e^{-\frac{|\alpha|^2}{\xi^2}} \hat{D}(\alpha) \hat{\rho}\hat{D}^\dagger(\alpha),
\end{equation}
where $\alpha=\frac{1}{\sqrt{2}}(\xi_q+i\xi_p)$ is a complex displacement in phase space, and $\xi^2$ characterizes the variance of the noise. This channel increases the variance such that $\Delta^2 \rightarrow \Delta^2+\xi^2$ and corresponds to independent Gaussian-distributed shifts applied to the position and momentum quadratures $\hat{q}$ and $\hat{p}$, respectively. Importantly, the GKP code is specifically designed to correct these small displacement errors, provided the magnitude of the noise stays below a critical threshold \cite{menicucci2014fault}. Displacements smaller than $\sqrt{\pi}/2$ in either quadrature can be uniquely identified modulo $\sqrt{\pi}$, and corrected by applying an opposite shift to bring the state back to the closest logical point on the GKP lattice. Practically, this is accomplished via syndrome extraction using Steane \cite{steane_error_1996} or Knill approaches \cite{Knill2005}, wherein the qubit mode is coupled to an ancilla GKP state (prepared in either the $\ket{0}_{\text{GKP}}$ or $\ket{+}_{\text{GKP}}$ state) followed by a homodyne measurement of the ancilla \cite{wang2019quantumerrorcorrectiongkp,Tzitrin2020,Fukui_2022}. Another primary source of error stems from the finite amount of squeezing in the physical GKP state, which leads to a non-zero probability of measuring the incorrect logical qubit value. If a single comb line of the GKP state has a width $\Delta$, the incorrect qubit value will be read with a probability given by \cite{Fukui2017}:
\begin{equation}
    P_{error}(\Delta)=1-\int_{\frac{-\sqrt{\pi}}{2}}^{\frac{\sqrt{\pi}}{2}}dx\frac{1}{\sqrt{2\pi\Delta^2}}e^{-x^2/2\Delta^2}.
\end{equation}

Logical errors can be significantly reduced by concatenating GKP qubits with higher-level encodings, such as Steane’s seven-qubit code \cite{steane_error_1996}, Knill's code \cite{Knill_2005}, or the surface code \cite{Noh2020}. In general, for the overall computation to remain fault-tolerant, the error probability of a given operation, $P_{\text{error}}$, must remain below a specific fault-tolerance threshold, $P_{\text{FT}}$. The exact value of $P_{\text{FT}}$ is determined by the underlying noise model and the chosen error correction architecture. In 2014, Menicucci\cite{menicucci2014fault} demonstrated that to achieve an error probability below $P_{\text{FT}}$, the minimum necessary squeezing level is given by $\Delta^2_{\text{dB}} > -10\log_{10}(2\Delta^2)$. For example, assuming a fault-tolerance threshold\cite{aharonov1999faulttolerantquantumcomputationconstant, Knill_1998} of $P_{\text{FT}} = 10^{-6}$  and utilizing a concatenated Steane code \cite{steane_error_1996}, the variance must satisfy $\Delta^2 < 4.44 \times 10^{-3}$. This corresponds to a highly demanding experimental squeezing level of $\Delta^2_{\text{dB}} > 20.5$ dB. Following Menicucci's seminal work \cite{menicucci2014fault}, various other error correction protocols have been proposed with the explicit goal of reducing this strict squeezing threshold; these protocols are summarized in Table~\ref{FTQC}.

\begin{figure*}[!htb] 
\centering
    \includegraphics[width=\textwidth]{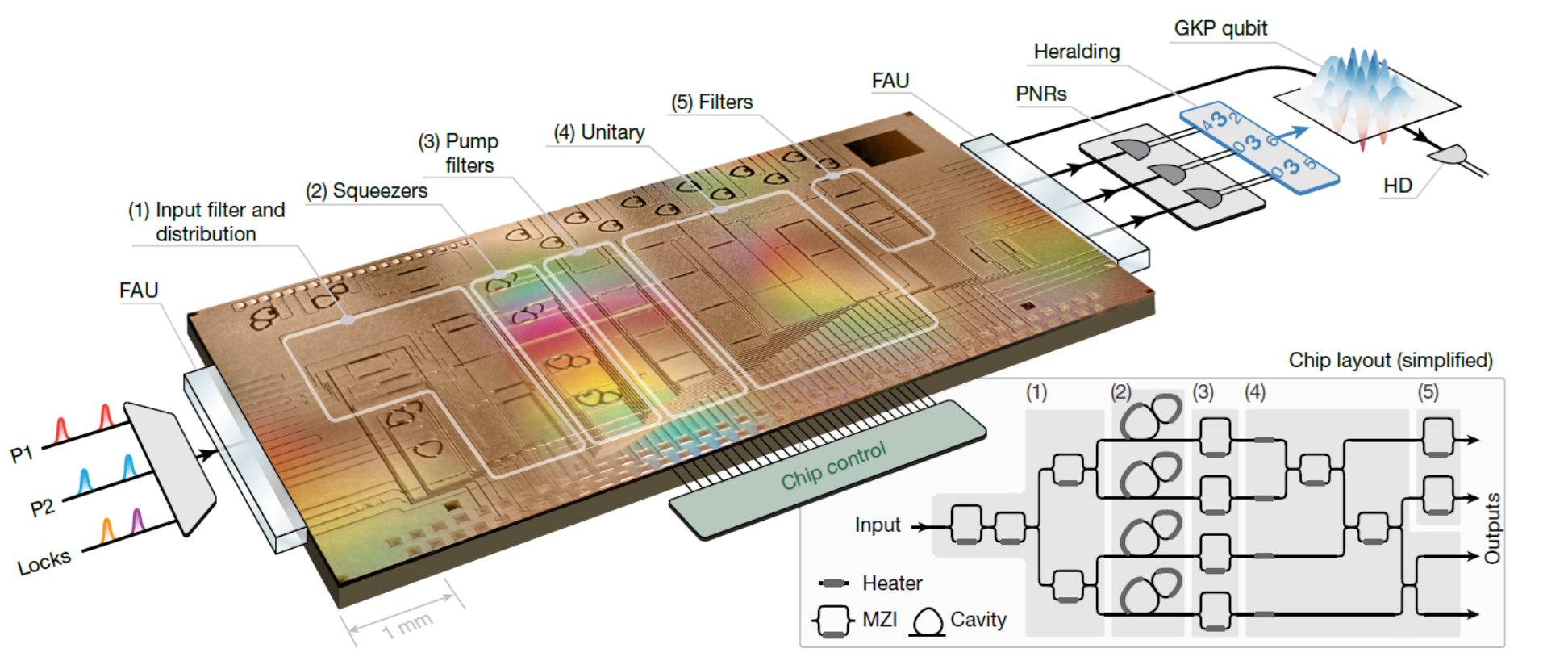}
    \caption{\textbf{Generation of GKP state on chip.} The chip is fabricated on a low-loss silicon nitride (SiN) waveguide platform and integrates necessary elements such as filters (1,3,5), squeezers (2) based on micro-ring multiresonator, and interferometers (4). The single-mode squeezed states of light are generated by degenerate Spontaneous Four-Wave Mixing (SFWM) using two pulsed pumps (P1 and P2) and a control light (locks) to stabilize the micro-ring resonators and the optical phase. The squeezed states are then separated from the pump light by optical filters (3), before being entangled by a programmable linear interferometer (4). 3 of the optical modes are sent to PNR detectors, heralding the production of a GKP qubit state in the remaining optical mode. (FAU) fibre array units. Reproduced from Larsen et al., Nature 642, 587 (2025); licensed under a Creative Commons Attribution (CC BY) license.\cite{Larsen2025Integrated}}
    \label{GKP-integrated}
\end{figure*}

Experimentally, GKP states were first generated in trapped ions \cite{fluhmann2019encoding, matsos2024robust} and superconducting circuits \cite{campagne2020quantum}. However, translating these results to optical systems remains a challenge due to the strict requirements for high squeezing levels and precise control over photon-number distributions. To overcome these hurdles, a variety of techniques are actively under development. These include light-ensemble interactions \cite{Motes2017}, cavity QED systems \cite{Hastrup2022}, cross-Kerr nonlinearities \cite{Pirandola_2004, Fukui_Crosskerr_2022}, breeding protocols \cite{Vasconcelos:10, Weigand_breeding2018, konno2024logical}, and post-selection via Gaussian Boson Sampling (GBS) devices \cite{Su2019, Tzitrin2020, Larsen2025Integrated}.

Recently, Konno \textit{et al.} \cite{konno2024logical} demonstrated the generation of GKP-like non-Gaussian states in a propagating optical system by interfering Schrödinger-cat states followed by homodyne measurements. Although this experiment successfully observed hallmark features—such as non-classicality, non-Gaussianity, and the signature "trident" peaks in the quadrature distribution—the prepared states did not yet exhibit the strict finite squeezing and well-resolved grid structure required of an approximate GKP code suitable for fault-tolerant error correction. Nevertheless, this work represents a crucial proof-of-principle realization of propagating-light GKP-inspired states. Pushing the boundary further, Larsen \textit{et al.} \cite{Larsen2025Integrated} recently presented the first integrated photonic chip source of GKP qubits, realized on a customized multilayer silicon nitride platform with ultra-low optical loss (see Fig.~\ref{GKP-integrated}). The chip integrates four ring resonators acting as squeezed vacuum sources \cite{Dutt2015}, coupled to a linear optical unitary circuit. By employing high-efficiency photon-number-resolving (PNR) detection on three of the four output modes, they successfully heralded the generation of a GKP state in the remaining fourth mode. Conditioned on specific multi-photon detection patterns, the system generated GKP states exhibiting clear lattice structures, at least a $3 \times 3$ grid of Wigner function negativity, and four well-resolved peaks in both quadratures. With further reductions in optical loss, this scalable device possesses the capacity to produce GKP states of sufficient quality for universal fault tolerance.

\section{Non-universal models}
Universal optical quantum computing aims to implement any quantum algorithm through a complete set of operations, but the technological requirements to make it scalable and fault-tolerant are challenging, which positions current quantum processors in the so-called Noisy Intermediate Scale Quantum (NISQ) regime. 
Nonetheless, such an intermediate regime encourages the investigation of non-universal models for quantum computation. These restricted models, though not capable of arbitrary quantum operations, can offer improvements in terms of demonstrating quantum advantage \cite{Harrow2017}, as well as increasing the performance of classical algorithms when working in synergy with classical processors, or be employed in quantum machine learning tasks. 

The non-universal models we discuss here belong to all the aforementioned classes. We first discuss Boson Sampling (BS) and its main variant, Gaussian Boson Sampling (GBS), which allowed for the first quantum advantage demonstration with photons \cite{zhong2020quantum, zhong2021phase, madsen2022, Deng2023}. Then, we discuss the most recent advances in the field of variational quantum algorithms and quantum machine learning based on photonic platforms. 
\begin{figure*}[t]
    \centering
    \includegraphics[width=\textwidth]{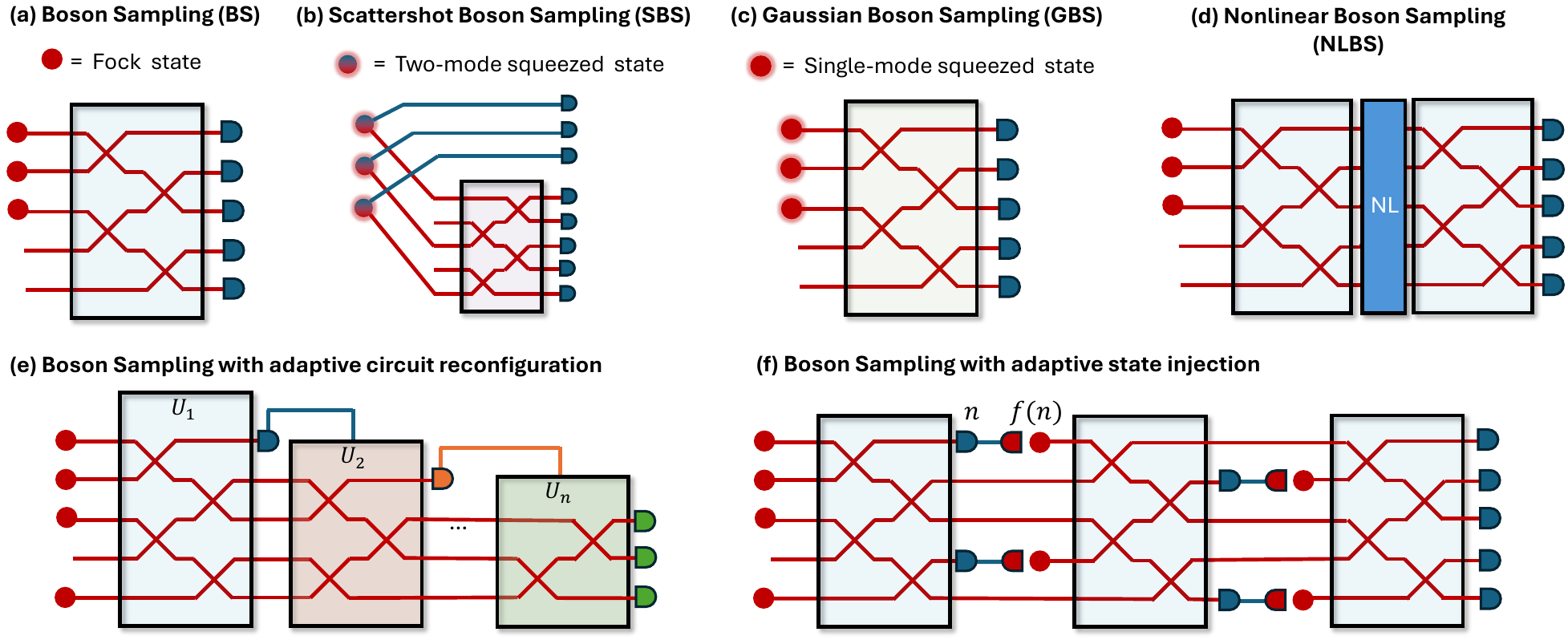}
    \caption{\textbf{Photonic Boson Sampling and its variants. (a)} Boson Sampling \cite{aaronson2011computational} consists of sending $n$ identical single photons through a random linear-optical transformation over $m>n$ spatial modes. With a large number of photons and modes, the output distribution is hard to sample for classical hardware. \textbf{(b)} In the Scattershot Boson Sampling scheme \cite{Lund2014, Bentivegna2015}, the input state is generated by using $k>n$ heralded two-mode squeezed state sources. In this way, one can enhance the probability of having $n$-photon events and, therefore, shorten the time of a Boson Sampling experiment when using SPDC single-photon sources. Moreover, this scheme allows to investigate different input states and input photon numbers in a single experiment. \textbf{(c)} Gaussian Boson Sampling \cite{Hamilton2017} employs $k$ sources of single-mode squeezed vacuum states that are sent through a random linear-optical transformation over $m$ spatial modes. Single-photon detectors are employed to record the output distribution. \textbf{(d)} Nonlinear Boson Sampling \cite{spagnolo2023nonlinear} proposes the introduction of a nonlinear transformation in between linear-optical transformations. The nonlinearity introduced increases the simulation complexity of the problem at a lower number of photons and modes involved, thus aiming at increasing the expressivity of Boson Sampling and bringing quantum advantage closer in Boson Sampling experiments. \textbf{(e)} The expressivity of Boson Sampling can also be enhanced by introducing adaptive measurements \cite{chabaud2021quantum} or state injections \cite{Monbroussou2024}. In the first case, the linear-optical circuit is reconfigured upon detection of a photon in a given mode. \textbf{(f)} In the second case, instead, the linear-optical transformation is unchanged, while the detection of $n$ in a given mode causes the injection of $f(n)$ photons in the circuit.}
    \label{fig:boson_sampling_machines}
\end{figure*}

\begin{figure*}[t]
    \centering
    \includegraphics[width=\textwidth]{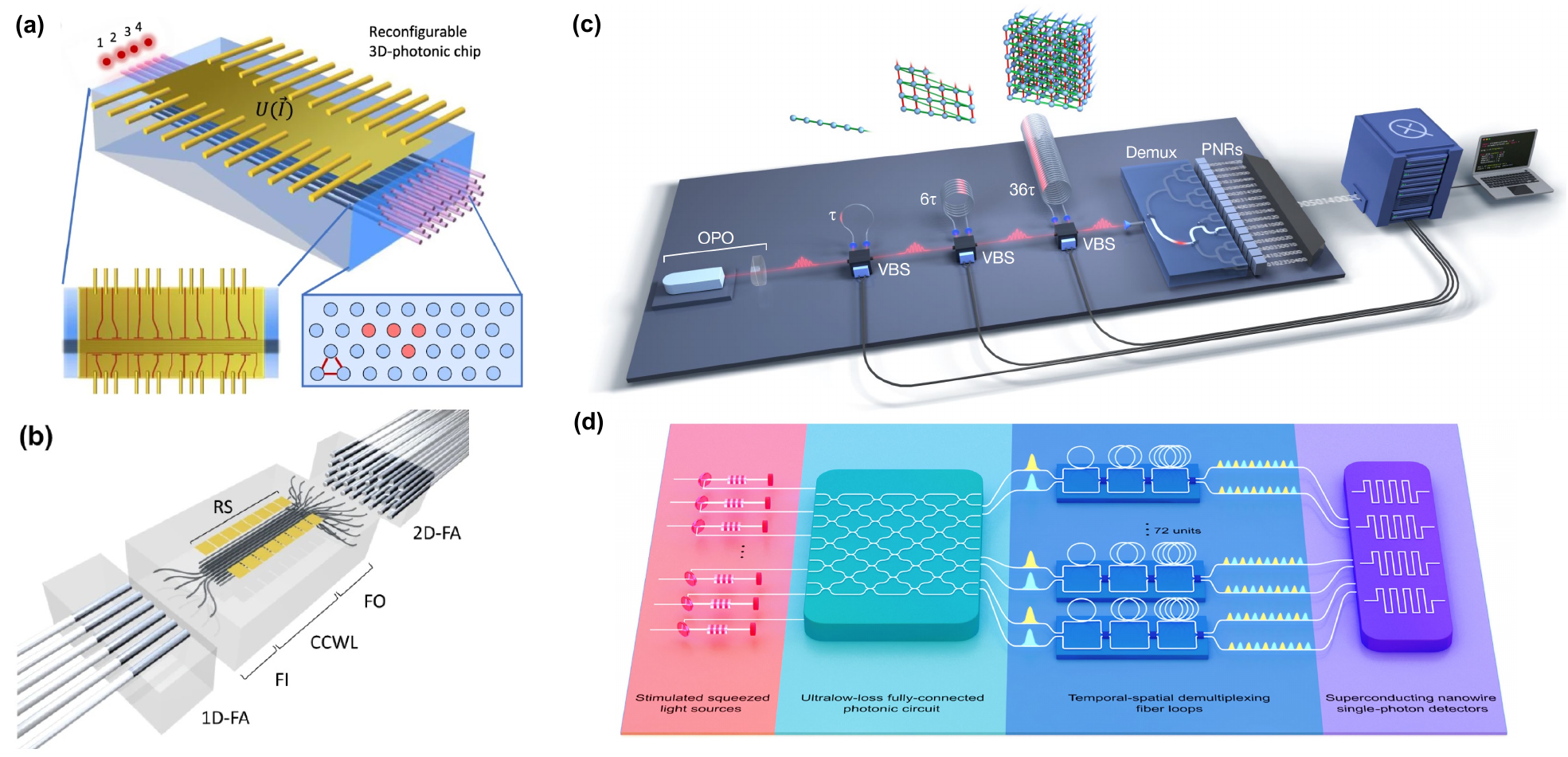}
    \caption{\textbf{Setups for the implementation of Boson Sampling and Gaussian Boson Sampling. (a)}  Reconfigurable 3D integrated photonic chip fabricated through femtosecond laser writing \cite{Crespi2013}. The device consists of 32 modes in a triangular lattice, as shown in the bottom right inset. The applied unitary can be reconfigured through 16 resistors on the top of the device. The bottom left inset shows the electrical circuits to monitor the resistors. Reproduced from Hoch et al., npj Quantum Information 8 (2022); licensed under a Creative Commons Attribution (CC BY) license.\cite{Hoch2022}  \textbf{(b)} One- and two-dimensional fiber arrays (1D-/2D-FA) connect the fan-in (FI) and fan-out (FO) stages for single photon input and output coupling in the device (CCWL = continuously-coupled waveguide lattice). Reproduced from Hoch et al., npj Quantum Information 8 (2022); licensed under a Creative Commons Attribution (CC BY) license.\cite{Hoch2022}  \textbf{(c)}  A pulsed optical parametric oscillator (OPO) generates a train of single-mode squeezed states, which pass through three reconfigurable loop-based interferometers. Each loop consists of a variable BS (VBS) employing programmable phase shifters and fiber delay lines. After interference, the output Gaussian state is partially demultiplexed (Demux) using a 1-to-16 binary switch and measured through photon-number-resolving detectors (PNRs). Reproduced from Madsen et al., Nature 606, 75–81 (2022); licensed under a Creative Commons Attribution (CC BY) license.\cite{madsen2022} \textbf{(d)} The GBS machine called \textit{Jiuzhang 3.0} involves 25 two-mode squeezed state photon sources coupled to a 144-mode interferometer. The photons are directed through 72 fiber loop modules for temporal-spatial demultiplexing. Final detection is performed through 144 superconducting nanowire single-photon detectors in a pseudo-PNR scheme. Reproduced from Deng et al., Phys. Rev. Lett. 131,150601 (2023); re-use license number RNP/25/OCT/097128.}
    \label{fig:exp_setups_bs}
\end{figure*}

\subsection{Boson Sampling machines}\label{BSmachines}
Boson Sampling and Gaussian Boson Sampling describe two computational machines which owe their interest to the fact that they are one of the most promising architectures for demonstrating quantum advantage with photonic platforms. After giving their mathematical description, we provide an overview of the experimental implementations performed to date. For a more detailed description of the problem, we suggest the following dedicated review works\cite{Lund2017, Brod_rev}.
Understanding Boson Sampling machines will also be useful for sections \ref{VQAsec} and \ref{QMLsec}, dedicated to variational quantum algorithms and quantum machine learning, as many of such tasks rely on the computational power arising from multi-photon interference.

\subsubsection{Boson Sampling}
Boson Sampling\cite{aaronson2011computational} was originally proposed to demonstrate that classical machines cannot simulate efficiently some problems that are easily solved through quantum machines. The photonic formulation of the problem, i.e., where the bosons involved are photons, consists of sampling from the output probability distribution of $n$ indistinguishable photons after their evolution through an $m-$mode linear interferometer, described by the unitary operator $\hat{U}$.

To estimate the transition from a given input state $\ket{I} = \ket{n_1, \cdots, n_m}$, where $n_i$ is the number of photons in each optical mode such that $\sum_{i=1}^m n_i = n $, to a given output state $\ket{O} $, requires the computation of the permanent of complex matrices, a problem that belongs to the $\#P-$hard complexity class  \cite{aaronson2011computational,valiant1979complexity}. 
More specifically, Ref.~[\onlinecite{aaronson2011computational}] showed that this sampling task is unfeasible for classical hardware if the following conditions are satisfied: the input state has at most one photon per mode, the unitary operator is randomly chosen according to the Haar measure, and the number of modes and photons are such that $m \gg n $ and $n > 50\div70$ photons.
Hence, such a mathematical problem, while being exponentially hard for classical computers, is naturally implemented by the described physical setup.

In formulas, the transition amplitude from $\ket{I} $ to $\ket{O} $ reads:

\begin{equation}
    A_{\hat{U}} (\ket{I} \rightarrow \ket{O}) = \bra{I} 
    \hat{U}
    \ket{O} = \frac{\text{Perm}(\hat{U}_{I,O})}{\sqrt{\prod_{k,j=1}^m i_k! o_j!}}
\label{eq:lin_boson_sampling}
\end{equation}

where $\text{Perm}(\cdot)$ indicates the permanent operation of a matrix, while $\{i_k \}$ and $\{o_j \}$ indicate respectively the occupation of the input and the output states.
The $n \times n$ matrix $\hat{U}_{I,O}$ is obtained by selecting rows and columns of the unitary operator $\hat{U}$, according to the numbers $\{i_k \}$ and $\{o_j \}$ respectively. 
In practice, the permanent operation corresponds to a sum over the possible ways to arrange the $n$ photons in the input state $\ket{I}$ into the output configuration that corresponds to the detection of the $n$ photons in the state $\ket{O}$.

A conceptual illustration of a Boson Sampling experiment is depicted in Fig.~\ref{fig:boson_sampling_machines}a. A number of photons enter a linear-optical circuit, 
that can be realized through several technologies using different encodings in the photons' degrees of freedom. One decomposition for the circuit is through a mesh of beam splitters and phase shifters, which allows for the implementation of any arbitrary $\hat{U}$ according to the universal schemes of Refs. [\onlinecite{Reck_1994, Clements2016, Bell2021}]. This approach has been investigated extensively thanks to the opportunity to realize these circuits in integrated photonic devices based on various fabrication technologies \cite{Wang2019, Giordani2023}. The first proof-of-concept experimental demonstrations were reported in integrated chips since 2013 \cite{Crespi2013, Tillmann2013, Broome2013, Spring2013}. An exhaustive timeline of the various implementations can be found in Table \ref{tab:boson_sampling}.

The most recent experiments explored alternative encoding and circuit decompositions to cover a broader range of modes (see Fig.~\ref{fig:exp_setups_bs}). One possibility is to consider continuously coupled integrated devices, in which the evolution is encoded in a compact lattice of waveguides, as shown in Refs.~[\onlinecite{Hoch2022, Yang2025}]. Time-bin encoded interferometers, realized through optical modulators and multiple passages through the same beam-splitters thanks to fiber-loop delays, are another architectures investigated for high-dimensional BS implementations \cite{He2017, Carosini2024}. Finally, the largest BS demonstrations have adopted evolutions in path-encoded circuits in micro-optic elements, such as prisms \cite{Wang2017,Wang2019}. Another challenge in scaling the size of the Boson Samplers and reaching the quantum advantage regime regards the efficiency of single-photon generation. Most of the seminal experiments employed probabilistic sources based on spontaneous parametric down-conversion. More efficient processes to obtain the input states required by the BS paradigm, which, we remind, is a Fock state with $n$ photons in $m$ modes, are offered by solid-state quantum emitters, like quantum dots (see Sec.~\ref{sec:DCGM_generalwork}). These sources interfaced with the previously mentioned optical circuits have allowed for BS demonstrations which involved from 5 to 14 photons \cite{Wang2017, Wang2019, maring2024versatile, Carosini2024, Rodari2025}.

In parallel to the experimental investigations, further interest has grown in finding the simulatability criteria of a noisy BS and, consequently, the threshold to reach the quantum advantage regime. Classical sampling algorithms have been developed \cite{clifford2017classical, Clifford2024}, some of them leveraging Markov Chain Monte Carlo methods \cite{Neville2017} and approximation of the BS output distribution via $k$-order marginal probabilities \cite{Renema_2018_classical}, i.e. distributions that reproduce the same photon-number correlations between the output modes up to a given order $k$ \cite{renema2020marginal}.
Indeed, the presence of experimental imperfections makes BS easier to approximate with classical sampling algorithms.
In particular, photons' partial distinguishability \cite{Renema_2018_classical} and photon losses \cite{Oszmaniec_2018,GarciaPatron2019simulatingboson, Brod2020classicalsimulation} are the main sources of noise that undermine the computational complexity of Boson Sampling. Other certification approaches regarded hypothesis tests that aimed to exclude that the collected samples could be drawn from classical simulatable models, such as the uniform and distinguishable particle samplers. In this context, several \emph{validation} algorithms have been formulated. They range from Bayesian tests \cite{aaronson2013bosonsamplingfaruniform,Spagnolo2014}, methods based on the properties of indistinguishable photons interference \cite{Tichy14, Crespi2016, Menssen17, Brod19,Pont22, Seron2023, Rodari2025}, of two-mode photon number correlation functions \cite{GBSVal3,Giordani18_correlators} and of binned probabilities \cite{Seron2024, CorreaAnguita2025}, to machine learning-based algorithms \cite{Agresti19}.
\begin{table*}[th!]
\centering
\begin{tabularx}{\textwidth}{ 
   >{\centering\arraybackslash\hsize=.5\hsize\linewidth=\hsize}X
   >{\centering\arraybackslash\hsize=1.5\hsize\linewidth=\hsize}X
   >{\centering\arraybackslash}X 
   >{\centering\arraybackslash}X
   >{\centering\arraybackslash}X
    >{\centering\arraybackslash}X}
\hline
\rowcolor[HTML]{F8A102}[\dimexpr\tabcolsep+0.1pt\relax]
\textcolor{white}{\textbf{Year}} & \textcolor{white}{\textbf{Reference}} & \textcolor{white}{\textbf{Photons}} & \textcolor{white}{\textbf{Modes}} &\textcolor{white}{\textbf{Platform}} &\textcolor{white}{\textbf{Programmability}}\\ \hline
2013                                 & Broome et al. \cite{Broome2013}           &            2,3                          &   6           & integrated optics           & partial  \\ 
\rowcolor[HTML]{FFE6C8}[\dimexpr\tabcolsep+0.1pt\relax]
2013                                     & Crespi et al. \cite{Crespi2013}           &            3                            &   5             & integrated optics          & no \\
2013                                     & Spring et al. \cite{Spring2013}           &            3                          &   6    & integrated optics                   & no \\ 
\rowcolor[HTML]{FFE6C8}[\dimexpr\tabcolsep+0.1pt\relax]
2013                                     & Tillmann et al. \cite{Tillmann2013}       &            3                          &   5      & integrated optics                 & no \\ 
2014                                 & Carolan et al. \cite{Carolan2014}         &            3                            &   9         & integrated optics             & no \\
\rowcolor[HTML]{FFE6C8}[\dimexpr\tabcolsep+0.1pt\relax]
2014                                     & Spagnolo et al. \cite{Spagnolo2014}       &            3                            &   5, 7, 9, 13   & integrated optics         & no \\
2014                                     & Carolan et al. \cite{Carolan2015}         &            3                            &   6   & integrated optics                   & full \\ 
\rowcolor[HTML]{FFE6C8}[\dimexpr\tabcolsep+0.1pt\relax]
2017                                 & He et al. \cite{He2017}                   &            3,4                          &   6,8    & time-bin and fiber loops              & partial \\ 
2017                                     & Loredo et al. \cite{Loredo2017}           &            2,3                          &   6        & in-fiber and bulk optics             & partial \\ 
\rowcolor[HTML]{FFE6C8}[\dimexpr\tabcolsep+0.1pt\relax]
2017                                     & Wang et al. \cite{Wang2017}               &            3,4,5                        &   9       & micro-optics circuit              & no \\ 
2019                                 & Wang et al. \cite{Wang_20ph}                   &            14                          &   60    & micro-optics circuit              & no   \\   
\rowcolor[HTML]{FFE6C8}[\dimexpr\tabcolsep+0.1pt\relax]
2022                            & Hoch et al. \cite{Hoch2022}                   &            3,4                          &   32    & integrated optics         &  partial \\
2024                            & Maring et al. \cite{maring2024versatile}                   &            6                         &   12    & integrated optics          & full \\
\rowcolor[HTML]{FFE6C8}[\dimexpr\tabcolsep+0.1pt\relax]
2024                           & Carosini et al. \cite{Carosini2024}                   &            5,6                         &   10,12    & time-bin and fiber loops         & partial \\
\hline
\end{tabularx}
\caption{\textbf{Experimental implementations of photonic Boson Sampling.} Chronology of experiments that demonstrated Boson Sampling on different optical platforms. For each work, we report the number of photons and modes, the employed platform, and the degree of programmability of the linear-optical circuit used.}
\label{tab:boson_sampling}
\end{table*}

\subsubsection{Gaussian Boson Sampling}

The first experimental proof-of-concept Boson Sampling demonstrations highlighted challenges in scaling up the size of the system towards the regime of non-classical simulability. One important aspect regards the brightness and heralding efficiency of single-photon sources to satisfy the need for multiple, pure, and indistinguishable single-photon Fock states. The most recent Boson Sampling implementations replaced parametric sources with near-deterministic photon sources based on quantum dot technologies \cite{Wang2017, Loredo2017,  Wang_20ph, maring2024versatile, Carosini2024} that allowed for a significant increase in the number of photons \cite{Wang_20ph}. An alternative approach considers the problem of sampling from a generic source of quantum light, exploring Gaussian states (see Sec. \ref{sec:1way QC} for a precise definition). Such a choice results to be the most convenient way to pursue the realization of high-dimensional Boson Sampling instances with parametric sources. In this direction, the first Boson Sampling variant, the so-called Scattershot Boson Sampling (SBS) \cite{Lund2014, Bentivegna2015}, represented the seminal instance of a sampling task from a Gaussian state. In this case, the Gaussian resources are the two-mode squeezed vacuum states generated by parametric single-photon sources. More precisely, the SBS envisages $k$ sources, each generating pairs of photons distributed over pairs of modes. In this way, the detection of $n<k$ photons performed on half of the modes heralds the injection of likewise twin photons in the linear optics interferometer. The gain compared to the standard Boson Sampling is in the samples generation rate that derives from the exponential number of input configurations equal to $\begin{pmatrix}
    n\\
    k
\end{pmatrix}$ that can be heralded (see also Fig.~\ref{fig:boson_sampling_machines}b). The proof of the SBS computational complexity follows similar reasoning to that of the Boson Sampling. Further constraints on the complexity regard the need for highly efficient photon number resolving detection in the heralding channels, and to keep low the probability of multi-pair emission from the sources \cite{Lund2014}.

Starting from this first generalization, the concept of Gaussian Boson Sampling was formalized\cite{Hamilton2017}, intending to further increase the sample generation rate by relaxing the hypothesis to work with a fixed number of input photons and, at the same time, preserving the computational complexity of Boson Sampling.
In summary, Gaussian Boson Sampling (GBS) is a fully linear-optical platform that generates photon samples drawn from the output state distribution generated by Gaussian light
sources evolved in a multi-port optical interferometer, as depicted in Fig.~\ref{fig:boson_sampling_machines}c.
Such a more generic sampling task has been proven to be hard to simulate only for certain classes of
Gaussian states, such as the single-photon signal generated by indistinguishable sources of single-mode squeezed vacuum states (SMSV)\cite{Lund2014, Keshari15, Hamilton2017}. Here and after, we will refer to Gaussian boson samplers with SMSV states as GBS.

Let us firstly review the main concept that regulates the probability to detect an outcome pattern $\vec{n}= (n_1, n_2, \dots, n_m)$ from a set of $m$ Gaussian input states distributed over $m$ optical modes. Here, $n_i$ denotes the number of photons detected in the $i$-th output port, and $n = \sum_{i=1}^m n_i$ is the total photon number.
Starting from the $2m\times2m$ covariance matrix $\sigma$ of the Gaussian state $\rho_{in}$, the corresponding detection probability is given by:
\begin{equation}
\text{Pr}(\vec{n}) = \frac{1}{\vec{n}! \sqrt{|\sigma_Q|}} \left| \text{Haf}\,(A_{\vec{n}}) \right|^2 \,.
\label{eq:hafn_gbs}
\end{equation}
Here, $\vec{n}! = \prod_{k=1}^m n_k!$, and $\sigma_Q = \sigma+\frac{1}{2}\mathbb{I}_{2m}$ (where $\mathbb{I}_{2m}$ is the identity matrix). 
The matrix $A$ summarizes the transformation $U$ of the optical circuit and the input covariance. The term $\text{Haf}$ denotes the \emph{Hafnian}.
The definition of the sub-matrix $A_{\vec{n}}$ differs from the standard Boson Sampling paradigm. $A_{\vec{n}}$ is a symmetric matrix of dimension \textbf{$2n \times 2n$}. 
Following the definition in Ref.~[\onlinecite{Hamilton2017}], to construct $A_{\vec{n}}$, we define the vector $\mu$ of length $2n$ as follows. For every photon detected in mode $k$ (where $1 \le k \le m$), the pair of indices $\{k, k+m\}$ is included in $\mu$. 
The sub-matrix is then obtained by selecting the rows and columns of $A$ corresponding to the indices in $\mu$:
\[ (A_{\vec{n}})_{rs} = A_{\mu_r, \mu_s} \quad \text{for } 1 \le r, s \le 2n \,. \]
Consequently, the Hafnian of $A_{\vec{n}}$ corresponds to the summation over all perfect matching permutations of the $2n$ indices in the sub-matrix, i.e., the ways to partition the index set $\{1,\cdots ,2n\}$ into $n$ pairs such that each index appears only in one pair (see also Ref.~[\onlinecite{Caianiello1953}]).
\begin{table*}[]
\centering
\begin{tabularx}{\textwidth}{ 
   >{\centering\arraybackslash\hsize=.5\hsize\linewidth=\hsize}X
   >{\centering\arraybackslash\hsize=1.5\hsize\linewidth=\hsize}X
   >{\centering\arraybackslash}X 
   >{\centering\arraybackslash}X 
   >{\centering\arraybackslash}X
    >{\centering\arraybackslash}X
    >{\centering\arraybackslash}X}
\hline
\rowcolor[HTML]{FD6864}[\dimexpr\tabcolsep+0.1pt\relax]
\textcolor{white}{\textbf{Year}} & \textcolor{white}{\textbf{Reference}} &  \textcolor{white}{\textbf{Type}} &\textcolor{white}{\textbf{Photons}} & \textcolor{white}{\textbf{Modes}} &\textcolor{white}{\textbf{Platform}}& \textcolor{white}{\textbf{Programmability}}\\ \hline
2015                                & Bentivegna et al. \cite{Bentivegna2015}   &   
SBS&          3                            &   9,13         & integrated optics & no\\
\rowcolor[HTML]{FFEDEC}[\dimexpr\tabcolsep+0.1pt\relax]  
2019       &  Zhong et al. \cite{Zhong2019_gbs}                 &   GBS                 &         3,4,5                         &   12    & micro-optics & no   \\
2019       &  Paesani et al. \cite{Paesani2019}                 &   SBS, GBS                 &         4,8                         &   12    & integrated optics & no    \\   
\rowcolor[HTML]{FFEDEC}[\dimexpr\tabcolsep+0.1pt\relax]   
2020       &  Zhong et al. \cite{zhong2020quantum}                 &   GBS                 &         76                        &   100    & micro-optics & no    \\ 
2021       &  Arrazola et al. \cite{Arrazola2021}                 &   GBS                 &         6                       &   8    & integrated optics & full   \\ 
\rowcolor[HTML]{FFEDEC}[\dimexpr\tabcolsep+0.1pt\relax]   
2021     &  Zhong et al. \cite{zhong2021phase}                 &   GBS                 &         113                      &   144    & micro-optics & partial   \\ 
2022     &  Madsen et al. \cite{madsen2022}                 &   GBS                 &         219                      &   216    & fiber loops & partial   \\ 
 \rowcolor[HTML]{FFEDEC}[\dimexpr\tabcolsep+0.1pt\relax]   
2022  &   Sempere-Llagostera et al. \cite{SempereLlagostera2022}                 &   GBS                 &         4                    &   20    & fiber loops & no  \\ 
2022    &  Thekkadath et al. \cite{Thekkadath2022}                 &   dGBS                 &         4                      &   15   & integrated optics & no \\
\rowcolor[HTML]{FFEDEC}[\dimexpr\tabcolsep+0.1pt\relax]   
 2023    &  Yu et al. \cite{Yu2023}                 &   GBS                 &         4                      &   32   & fiber loops & full   \\
2023     &  Deng et al. \cite{Deng2023}                 &   GBS                 &         255                      &   144   & micro-optics & partial   \\
     \hline
\end{tabularx}
\caption{\textbf{Experimental implementations of photonic Gaussian Boson Sampling.}  Chronology of experiments that demonstrated Gaussian Boson Sampling on different optical platforms. For each work, we report the number of photons and modes, the employed platform, and the degree of programmability of the linear-optical circuit used. GBS = Gaussian Boson Sampling. SBS = Scattershot Boson Sampling. dGBS = displaced squeezed states Gaussian Boson Sampling.}
\label{tab:gaussian_boson_sampling}
\end{table*}
In practice, the hafnian is a more general expression of the permanent of a matrix $M$. In the standard BS, the permanent calculation derives from the fact that one has to consider all the possible ways to arrange the $n$ photons prepared in a fixed input configuration in the output modes. In the presence of Gaussian sources in the inputs, like SMVS states, it is required to sum also on the possible ways to generate the $n$ photons from the $m$ sources. This naive interpretation provides an intuition regarding the meaning of the hafnian as a generalization of the permanent. Formally, their relationships can be expressed through the following expression:
\begin{equation}
    \text{Perm}(M)=\text{Haf}\begin{pmatrix}
    0 & M\\
    M^t & 0
    \end{pmatrix}\,.
    \label{eq:permanent}
\end{equation}
Then, it follows that computing hafnians is hard at least as computing permanents. The problem belongs to the same computational class, the \#P-hard problems. 
Such an argument is the basis to define a classically-hard sampling algorithm, for indistinguishable SMSV states evolving in a linear optical interferometer with photon-counting measurements
\cite{Keshari15, Hamilton2017, DetailedstudyGBS}. 
The proof of GBS computational complexity involves photon-number-resolving detection at the output ports of the device. Later, the proof was extended to the case of threshold detectors \cite{Quesada18}, in which the hafnians are replaced by the \emph{Torontonian} function. It turns out that computing Torontonians, even in an approximate way, is hard, thus proving the computational complexity of GBS with threshold detectors. Recently, the class of hard-to-simulate Gaussian states has been enlarged to displaced squeezed states (dGBS) whose sampling task is regulated by loop-hafnian functions, as reported in Ref.~[\onlinecite{Thekkadath2022}]. 

The scheme of the GBS was extensively investigated in the experiments for the chance to demonstrate a quantum computational advantage with the current photonic technologies. From the seminal demonstration of SBS \cite{Bentivegna2015} and the first fully-on-chip demonstration of GBS \cite{Paesani2019}, the quantum advantage regime was claimed by 5 different experiments since 2020. Such demonstrations reached more than 100 optical modes for the interferometer size and high generation rates for photons samples with more than 100 detected photons (see Table \ref{tab:gaussian_boson_sampling}). All the mentioned large-scale implementations have been realized in bulk-optics apparatus via micro-optics interferometers (Jiuzhang 1.0 \cite{zhong2020quantum}, Jiuzhang 2.0 \cite{zhong2021phase, Deng2023_graph}, and Jiuzhang 3.0 \cite{ Deng2023} machines) or in fiber-loops and time-bin encoding (Borealis machine \cite{madsen2022}). Both platforms display a limited programmability compared to previous proof-of-concept on-chip realizations on fully-programmable devices, such as the one in Ref.~[\onlinecite{Arrazola2021}]. Nonetheless, they represent milestones for the photonic quantum community, unavoidable intermediate steps for the building of photonic quantum processors. Very recently, the machine Jiuzhang 4.0 \cite{liu2025robustq} has been released. It combines the spatial and temporal encoding adopted by the previous machines, demonstrating to enlarge the size of the system and a more robust evidence of quantum advantage via GBS.

Similarly to the standard BS scenario, questions arise whether a GBS device can be classically simulated.  
In these directions, classical simulation algorithms of the GBS sampling process have been developed \cite{clifford2017classical,Neville2017,Quesada_exact_simulation,Quesada_exact_simulation_speedup, oh2022classical,Bulmer_Markov_GBS, Drummond2022, popova2021cracking, cilluffo2023simulating, TensorNetworkGBS}, to investigate the thresholds to claim a genuine computational advantage for what concerns the size of the device, number of modes and detected photons, and the level of noise in the apparatus. Dedicated works indicate the principal sources of experimental imperfections that undermine the computational complexity of the problem \cite{Keshari_2016_conditions} in photon losses \cite{Oszmaniec_2018,GarciaPatron2019simulatingboson,Brod2020classicalsimulation, Qi_lossyGBS, ohlossy, liu2023complexity} and in photon distinguishability \cite{Renema_2018_classical,Moylett_2019, Renema_partial_2020, Shi2022}. Following these studies, effective benchmarking procedures had been introduced according to validation approaches similar to those formulated in the BS context. In the GBS scenario, the validation tests aim at discerning when samples are drawn from the class of Gaussian classical simulatable states, such as thermal light, coherent and squashed states, and distinguishable particles models. The various methods range from Bayesian tests \cite{Spagnolo2014,martinezcifuentes2022classical}, statistical properties of two-point correlation functions \cite{GBSVal3,GBS_correlators, Giordani18_correlators}, detector binning \cite{Bressanini2025}, properties of marginal probabilities \cite{Renema_partial_2020, renema2020marginal, villalonga2021efficient}, to approaches related to graph theory \cite{TairaGBS, Stanev2025}. 

\subsubsection{Perspectives and applications}
The interest in Boson Sampling and its variants is not limited only to the demonstration of quantum advantages. Other properties of the sampling process can be leveraged to find applications in the context of graph problems \cite{Bromley_2020, GBSGraphTheory1, GBSGraphTheory2, GBSGraphTheory3}, quantum simulations \cite{Huh2015_vibronic, Sparrow2018, Banchi_vibronic}, and quantum machine learning \cite{steinbrecher2019quantum, chabaud2021quantum, Sakurai2025}.

The first natural connection of the photonic sampling process is with graphs. Permanents, and more generally hafnians, have a precise meaning in graph theory since they count the number of perfect matchings in undirected graphs. More precisely, the permanent counts the perfect matching of bipartite graphs \cite{valiant1979complexity} and it is a special case of the hafnian, which, generally, counts the perfect matching in a generic graph \cite{Caianiello1953}. This means that, as long as the sub-matrices in Eqs.~\eqref{eq:lin_boson_sampling} and \eqref{eq:hafn_gbs} belong to the adjacency matrix of a graph, interesting applications can be found in graph-related problems. In particular, some relevant problems, such as finding the dense subgraphs, max-clique, and graph similarity, have been formulated in GBS-based algorithms \cite{Bromley_2020, GBSGraphTheory1, GBSGraphTheory2, GBSGraphTheory3}. First tests on quantum devices have been performed on small-scale integrated photonic devices \cite{Arrazola2021} and on larger dimensions in bulk optics through time-\cite{SempereLlagostera2022} and path-encoded interferometers \cite{zhong2021phase}.

In quantum simulation, the BS and GBS sampling processes have been investigated for simulating the vibronic spectra of molecules. Such a connection arises from the similar properties of the photons' Hamiltonian, which may include squeezing and displacement operators, and the one that governs the vibronic transitions of molecules. Theoretical studies in the field \cite{Huh2015_vibronic, Banchi_vibronic} have been investigated in experiments with fully programmable integrated circuits \cite{Sparrow2018, Arrazola2021} and in time-bin interferometers \cite{Yu2023}. 

Further applications of the BS paradigm can be found in the context of quantum machine learning. The various proposals adopt different approaches that can be divided in two categories. The first category pursues the fully linear approach and add classical feedback to train the circuit. This is the case, for example, of variational photonic algorithms. The second explores the inclusion of some nonlinear operations and adaptivity, leveraging the recently introduced BS variants, the nonlinear Boson Sampling\cite{spagnolo2023nonlinear, Nielsen2025} (NLBS) and the adaptive Boson Sampling\cite{chabaud2021quantum, hoch2025quantum, Monbroussou2024, monbroussou2025photonic}  (ABS, see Fig. \ref{fig:boson_sampling_machines}d-f). The aforementioned applications in quantum machine learning are analyzed in detail in section \ref{QMLsec}.

\subsection{Variational quantum algorithms} \label{VQAsec}

\begin{table*}[ht!]
\centering
\begin{tabularx}{\textwidth}{
   >{\centering\arraybackslash\hsize=.5\hsize\linewidth=\hsize}X 
   >{\centering\arraybackslash}X 
   >{\centering\arraybackslash}X   
   >{\centering\arraybackslash\hsize=1.5\hsize\linewidth=\hsize}X 
   >{\centering\arraybackslash}X  }
\hline
\rowcolor[HTML]{9698ED}[\dimexpr\tabcolsep+0.1pt\relax]
\textcolor{white}{\textbf{Year}} & \textcolor{white}{\textbf{Reference}}  & \textcolor{white}{\textbf{Task}}  & \textcolor{white}{\textbf{Platform}} & \textcolor{white}{\textbf{Encoding}} \\ \hline
\rowcolor[HTML]{EEEFFF}[\dimexpr\tabcolsep+0.05pt\relax]
2014 & Peruzzo et al. \cite{peruzzo2014variational}   & Quantum eigensolver            & Integrated interferometer  & Path         \\ 
2017 & Wang et al. \cite{Wang2017_ham}          & Hamiltonian Learning           & Silicon integrated interferometer  & Path         \\ 
\rowcolor[HTML]{EEEFFF}[\dimexpr\tabcolsep+0.05pt\relax]
2017 & Paesani et al. \cite{Paesani2017}          & Phase estimation           & Silicon integrated interferometer  & Path         \\ 
2018 & Santagati et al. \cite{Santagati2018}          & Quantum eigensolver            & Silicon integrated interferometer  & Path         \\
\rowcolor[HTML]{EEEFFF}
[\dimexpr\tabcolsep+0.05pt\relax]
2019 & Jašek et al. \cite{jaek2019experimental}        & Quantum cloning                & Bulk optics                & Polarization \\ 
[\dimexpr\tabcolsep+0.05pt\relax]
2020 & Carolan et al. \cite{Carolan2020}              & Quantum unsampling             & Silicon integrated interferometer  & Path         \\ 
\rowcolor[HTML]{EEEFFF}
[\dimexpr\tabcolsep+0.05pt\relax]
2022 & Lee et al. \cite{Lee2022}                      & Quantum eigensolver            & Bulk optics                & Polarization/path \\
[\dimexpr\tabcolsep+0.05pt\relax]
2023 & Ding et al. \cite{Ding2023}                    & Quantum classifier             & Bulk optics        & Polarization         \\ 
\rowcolor[HTML]{EEEFFF}
[\dimexpr\tabcolsep+0.05pt\relax] 
2024     & Borzenkova et al. \cite{Borzenkova2024}        & Quantum eigensolver            & Glass integrated interferometer  & Path         \\
[\dimexpr\tabcolsep+0.05pt\relax]
2024     & Cimini et al. \cite{cimini2024variational}     & Multiparameter estimation      & Glass integrated interferometer  & Path         \\ 
\rowcolor[HTML]{EEEFFF}
[\dimexpr\tabcolsep+0.05pt\relax] 
2024     & Hoch et al. \cite{hoch2024variational}         & Quantum cloning                & Glass integrated interferometer  & Path         \\
[\dimexpr\tabcolsep+0.05pt\relax] 
2024     & Maring et al. \cite{maring2024versatile}       & Quantum eigensolver            & SiN integrated interferometer  & Path         \\      
\rowcolor[HTML]{EEEFFF}
[\dimexpr\tabcolsep+0.05pt\relax] 
2025     & Hoch et al. \cite{Hoch2025} & Quantum eigensolver, UNOT gate & Glass integrated interferometer  & Path \\        
[\dimexpr\tabcolsep+0.05pt\relax] 
2025    & Baldazzi et al. \cite{Baldazzi2025} & Quantum eigensolver, integers factorization & Silicon integrated interferometer  & Path         \\  \hline
\end{tabularx}
\caption{\textbf{Experimental implementations of photonic variational quantum algorithms.} Chronology of experiments that demonstrated variational quantum algorithms on a quantum-optical platform. For each work, we report the performed task, the kind of platform, and the qubit encoding.}
\label{tab:VQAs}
\end{table*}

The variational approach to quantum algorithms is among the most promising applications of quantum processors \cite{cerezo2021variational}. Originally proposed to work within the NISQ constraints, variational quantum algorithms (VQAs) are also compatible with universal quantum computing \cite{biamonte2021universal}, and they have been widely investigated for two main reasons. First, they employ a hybrid quantum-classical method that takes advantage of the features of both quantum and classical hardware, and, second, their structure is flexible and can be applied to a variety of computational tasks, such as Hamiltonian eigenvalue estimation \cite{peruzzo2014variational,abrams1999quantum,aspuru2005simulated,higgott2019variational,jones2019variational,nakanishi2019subspace,garcia2018addressing,cerezo2022variational,wang2019accelerated,cao2024accelerated}, quantum metrology \cite{cimini2024variational,polino2020photonic,giovannetti2011advances,beckey2022variational,paris2004quantum,yang2022variational,liu2025variational,yang2021hybrid,kaubruegger2019variational,kaubruegger2021quantum,koczor2020variational}, dynamical quantum simulation \cite{mclachlan1964variational,yao2021adaptive,zhang2023low,heya2019subspace,cirstoiu2020variational,commeau2020variational}, quantum machine learning tasks \cite{hoch2024variational,Hoch2025,jaek2019experimental,Coyle2022,Baldazzi2025}, and quantum error correction \cite{johnson2017qvector,xu2021variational}.

In such a context, quantum-optical platforms are promising for the realization of VQAs. Indeed, photons are naturally decoherence-free and allow for the realization of single-qubit gates with high fidelity \cite{Peters2003}.
Furthermore, they naturally embody \textit{flying qubits} for distributed tasks \cite{cacciapuoti2019quantum}.

In the following, we first give an overview of the variational approach, and then review existing realizations on quantum-optical platforms. \\

The first step when casting a VQA involves defining a cost function $C(\theta)$, which depends on a set of continuous or discrete parameters $\theta$, and that encodes the objective to be minimized or maximized, thereby representing the solution landscape of the problem.
In several cases, one must also choose a suitable training dataset.

\begin{figure*}[ht!]
    \centering
    \includegraphics[width=\textwidth]{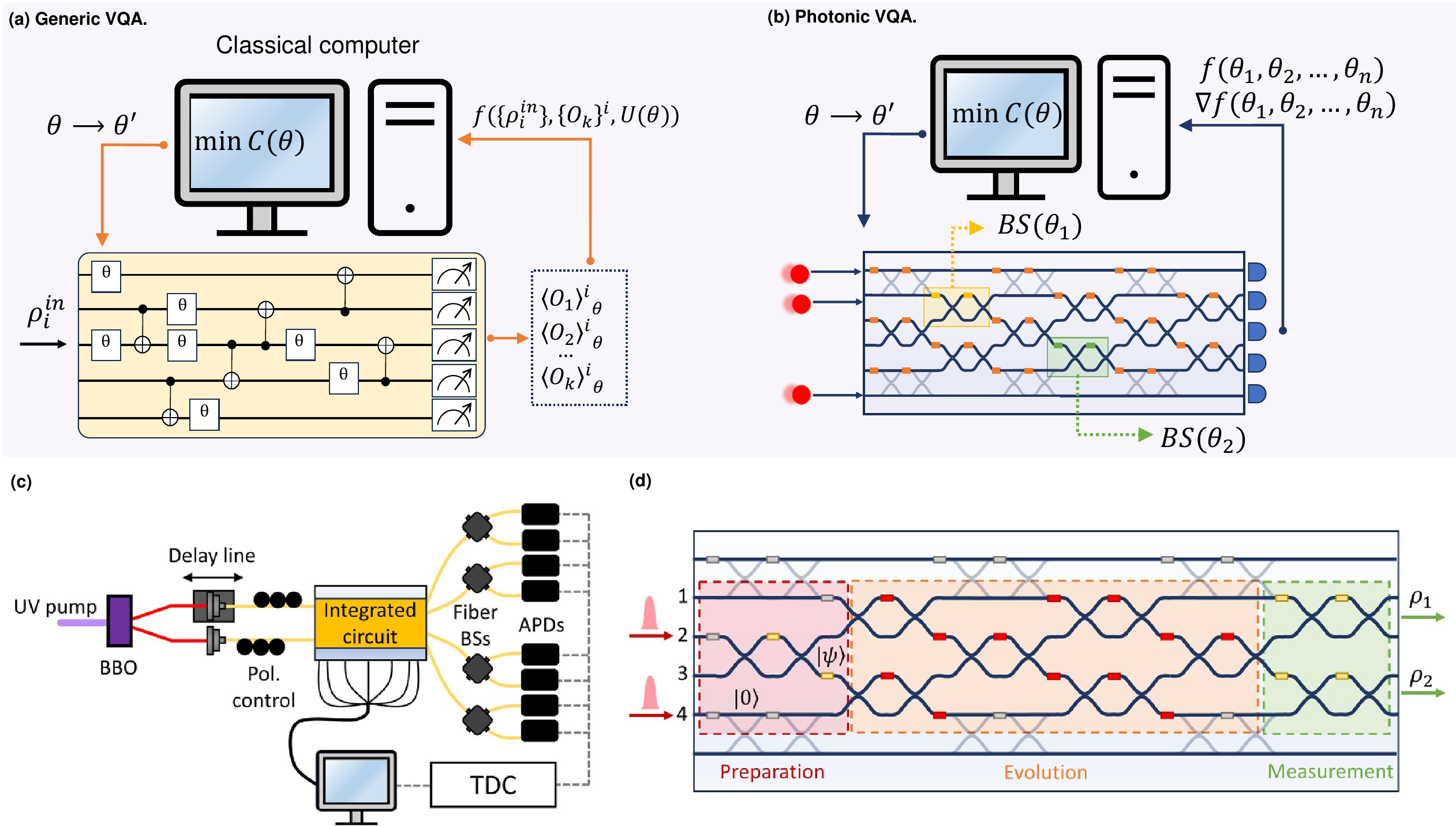}
    \caption{\textbf{Photonic realization of VQAs. (a) Generic VQA.} In a variational quantum algorithm, a parameterized quantum circuit gets as input a quantum state chosen from a training dataset $\rho^{in}_{i}$ and indexed $i$, that undergoes a unitary evolution depending on the parameters $\theta$. The set of observables $\{ \langle O_k \rangle \}^i$ measured at the output of the circuit is used by a classical computer to evaluate the cost function $C(\theta)$ and to update the circuit parameters $\theta$. \textbf{(b) Photonic VQA.} In a typical photonic realization of a VQA, qubits are encoded in the spatial or polarization modes of single photons. After generation, these are coupled into a photonic processor where they undergo a unitary transformation defined by the reflectivities of a mesh of (Polarizing) Beam Splitters (BS or PBS). The reflectivity of a single BS can be tuned by introducing suitable phase shifts between different spatial modes, while that of a PBS can be tuned by rotating the polarization of the input photon. At the output of the circuit, all modes are measured with single-photon detectors, and the pattern of detected events is used to estimate the cost function at a given algorithm iteration. 
    \textbf{(c)} Photon pairs are generated using spontaneous parametric down-conversion and then coupled into a six-mode universal photonic integrated processor fabricated through femtosecond laser writing \cite{Crespi2013}. Photon indistinguishability is optimized through polarization controllers and fiber delay lines. Single photons are measured through a pseudo-PNR detection scheme made up of fiber BSs and avalanche photodiodes. \textbf{(d)} The applied unitary is divided into three regions: preparation (red), trainable evolution (orange), and measurement (green). The fidelity between clones is estimated through a projective measurement of the clone qubit onto the target state. Panels (c) and (d) are reproduced from Hoch et al., Optica Quantum 3, 351–359 (2025); reproduced under the Optica Open Access Publishing Agreement.\cite{hoch2024variational}}
    \label{fig:variational_exps}
\end{figure*}

The parameters $\theta$ are optimized within hybrid quantum–classical iterations, where the evaluation of quantum operators is used in a classical optimization routine, as shown in Fig.~\ref{fig:variational_exps}a.
More specifically, a quantum computer is used to estimate the cost function $C(\theta)$ or its derivatives, while a classical optimizer is trained to find the best set of parameters $\theta^*$ according to the following update rule:
\begin{equation}
    \theta^* = \text{arg} \min_{\theta} C(\theta)
\label{eq:variational_optimization}
\end{equation}

A generic cost function can be expressed as:
\begin{equation}
    C(\theta) = f(\{ \rho_i \}, \{ O_k \}^i, U(\theta))
\label{eq:cost}
\end{equation}
where $f$ is a function depending on: (i) the unitary $U(\theta)$, corresponding to the parameterized quantum circuit, (ii) the input states $\rho_i$, chosen from the training set, and (iii) a set of observables $\{O_k\}^i$.
One can identify some desirable criteria when choosing the cost function of a VQA, such as\cite{cerezo2021variational}:

\begin{itemize}
    \item \textit{faithfulness}: the minimum of the cost function must correspond to the solution of the chosen problem;
    \item \textit{quantum-enhancement}: it must be more efficient to evaluate the cost function or its derivatives on NISQ quantum hardware, than on classical hardware;
    \item \textit{trainability}: it should be possible to efficiently optimize the parameters $\theta$, through gradient-based or gradient-free techniques.
\end{itemize}

Another crucial component of a VQA is the \textit{ansatz}, i.e., the form of the parameterized quantum circuit used in the optimization. Its structure determines the nature of the parameters $\theta$ and, consequently, how they can be optimized to minimize the cost function.
The design of an ansatz often depends on the specific addressed task (\textit{problem-inspired} ansatzes), or there can be more general architectures known as \textit{problem-agnostic} ansatzes, when only little information is known about the problem at hand.
As an example, the so-called \textit{hardware-efficient ansatz} can be chosen to limit the circuit depth on the available quantum hardware.

After defining the cost function and ansatz, the next step involves training the parameters $\theta$ by solving the optimization problem in Eq.~\eqref{eq:variational_optimization}.
The most common method is to compute the gradient of the cost function, which is usually chosen as a differentiable function.
However, since computing gradients on physical platforms is a non-trivial task, gradient-free classical optimizers can be used, such as Nelder-Mead \cite{powell1973search}, though they might converge to non-stationary points \cite{mckinnon1998convergence}.
Alternatively, the so-called parameter shift rule \cite{Mitarai2018,Schuld2019,Wierichs2022} can be adapted to several platforms and used to approximate the gradient of a function. In the next paragraph, we focus on recent theoretical and experimental investigations of a photonic version of this tool.

However, it is worth noting that serious issues may arise when training a VQA, and that might nullify a potential quantum advantage for these algorithms. 
Indeed, VQAs can lead to the phenomenon of Barren Plateaus \cite{McClean2018}. Specifically, this phenomenon causes vanishing gradients of the cost function, thus leading to an almost flat landscape.
Consequently, exponential precision would be required to identify an optimization direction, regardless of whether gradient-based or gradient-free optimization methods are used. Possible ways to mitigate this issue have been explored, such as a careful choice of the parameter initialization or specific ansatz strategies.
For a detailed summary of the most widely employed ansatzes in VQAs and possible solutions to mitigate Barren Plateaus, we refer to the review in Ref.~[\onlinecite{cerezo2021variational}].\\

The VQA framework requires the application of a classical optimizer to train the parameters of a quantum circuit. As aforementioned, many optimizers use gradient-based methods that estimate the derivatives of the cost function. The calculation is achieved via the finite difference numerical method, in which exact derivatives are substituted by evaluating the cost function at very nearby points.
However, such an approximation of the gradient becomes impractical on noisy quantum devices, as the choice of close values of the circuit's parameter settings can lead to increased noise sensitivity and, consequently, lower accuracy.

A common solution is to use gradient-free optimization methods, which have been successfully adopted on photonic platforms \cite{peruzzo2014variational,Santagati2018,maring2024versatile,hoch2024variational,Paesani2017,jaek2019experimental}. Alternatively, a gradient-based strategy that can be adopted even in the circuit-based quantum computing paradigm is the \textit{parameter shift rule}, which offers an exact, noise-resilient method to compute derivatives. The rule is based on the properties of the probabilities to measure a given outcome in quantum circuits, which are trigonometric functions of the 
variational parameters. In such a scenario, the exact derivative of the function at a point $x_0$ can be computed by evaluating the function in a finite set of translated points $x_0 + x_k$ \cite{Schuld2019, Wierichs2022}. Recently, efforts have been put into developing a photonic-native parameter shift rule \cite{cimini2024variational,facelli2024,pappalardo2025photonic, Hoch2025}. In Ref.~[\onlinecite{cimini2024variational}], a first generalized version of the rule was applied to a photonic-based multiparameter estimation task. 
In Refs.~[\onlinecite{Hoch2025},\onlinecite{pappalardo2025photonic}, \onlinecite{facelli2024}], a parametric shift rule tailored to the calculation of gradients via a photonic hardware has been fully formalized, and then demonstrated in an experimental variational quantum eigensolver and Universal-NOT gate tasks in Ref. [\onlinecite{Hoch2025}].\\

\subsubsection{Experimental implementations}

We now focus on the implementation of VQAs on photonic platforms. 
A typical realization is illustrated in Fig.~\ref{fig:variational_exps}b. Photonic VQAs strongly rely on the Boson Sampling paradigm, described in Section~\ref{BSmachines}. In a VQA, single photons are fed to an arbitrary complex interferometer whose unitary description at each algorithm iteration depends on the values of the parameters $\theta$.
Depending on the quantum state encoding, such parameters can be tuned through phase shifters (spatial encoding) or waveplates (polarization encoding).
At the output of the processor, photons are measured through single-photon detectors, and the resulting string is post-processed in a classical computer to estimate the cost function or its gradient. Finally, the updated values of the parameters are computed and applied to the processor for the next iteration.

In Table~\ref{tab:VQAs}, we summarize the tasks that have been addressed in existing demonstrations, the platforms, and the qubit encoding employed.
As one can see, in recent years photonic VQAs are drawing an increasing interest for a variety of tasks, and integrated photonic technologies have played a key role\cite{Giordani2023,Wang2019}, since they allow for the insertion of a large number of phase shifters in compact devices. In what follows, we review the tasks that have been most commonly addressed in the reported experiments.\\

\textit{Variational quantum eigensolver} (VQE) was first realized on a photonic platform in the work reported in Ref.~[\onlinecite{peruzzo2014variational}]. 
A similar task was also implemented in other experiments for different goals, such as approximating eigenvalues for both ground and excited states\cite{Santagati2018}, error mitigation in a VQE\cite{Lee2022,Borzenkova2024}, factorizing integer numbers\cite{Baldazzi2025}, demonstrating efficient ways to compute gradients for the cost function\cite{Hoch2025,pappalardo2025photonic}, and testing a prototypical cloud-accessible single-photon-based quantum computing platform\cite{maring2024versatile}.
The VQE task aims at finding the ground-state energy level of a given Hamiltonian. For this algorithm, the cost function is therefore defined as $C(\theta) = \braket{\psi(\theta)|H|\psi(\theta)}$, where the state is defined as $\ket{\psi(\theta)} = U(\theta) \ket{\psi_0}$, for an ansatz $U(\theta)$ and an initial state $\ket{\psi_0}$.
To encode such problems, the Hamiltonian $H$ can be represented as a linear combination of Pauli operators $\sigma_k$ as $H= \sum_k c_k \sigma_k~(c_k \in \mathbb{R})$.
In this way, the cost function can be efficiently estimated through a quantum computer.\\

\begin{figure*}[htb]
    \centering
    \includegraphics[width=0.85\textwidth]{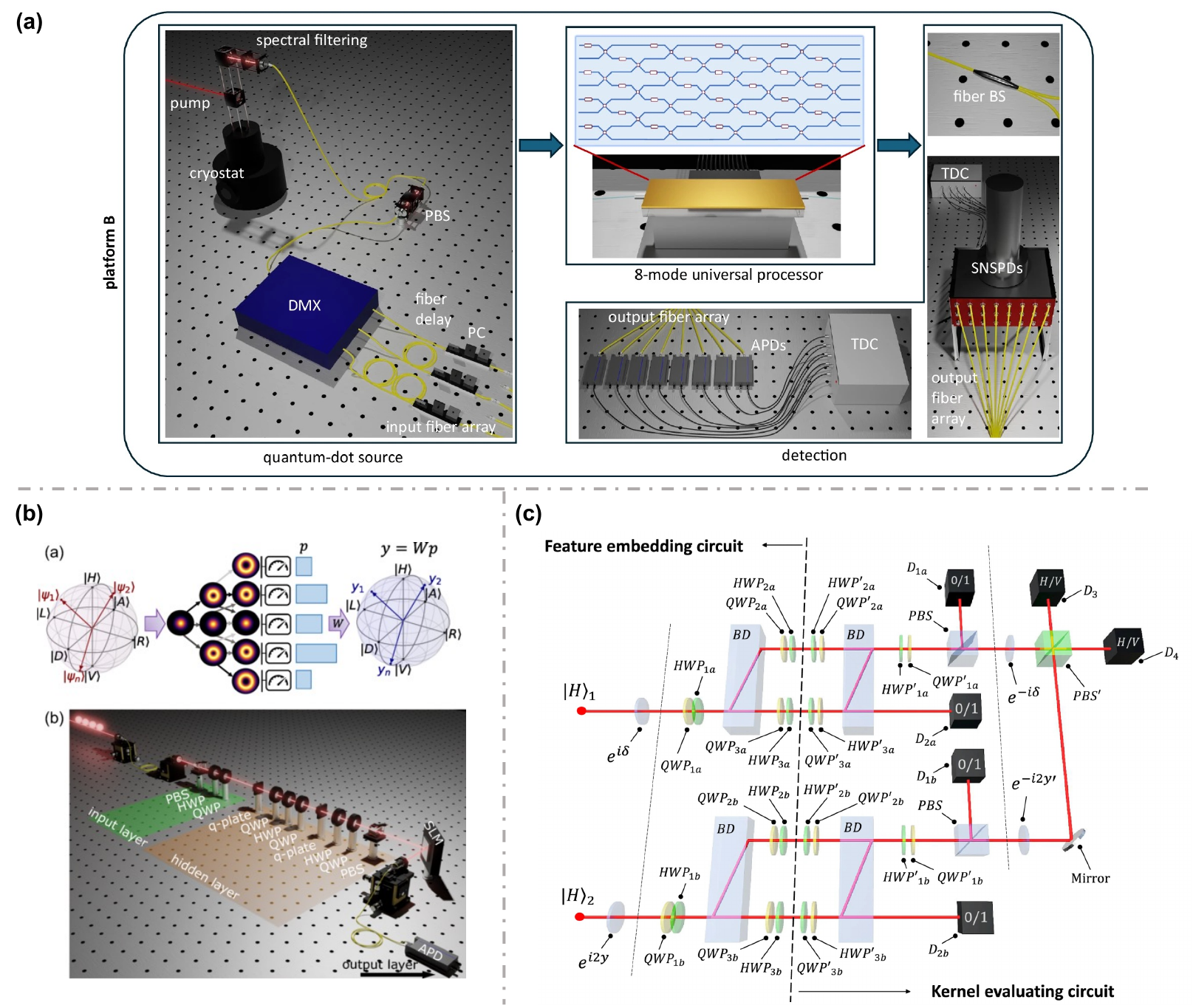}
    \caption{\textbf{Different platforms for photonic QML. (a)} The experiment shown involves three photons generated via a semiconductor quantum dot-based single-photon source. The temporal train of photons passes through a temporal-to-spatial demultiplexer (DMX), and the photons are synchronized over three different spatial modes. Photon indistinguishability is optimized through fiber delays and polarization controllers (PCs). Then, they enter an 8-mode integrated interferometer and are measured through superconducting single-photon detectors (SNSPDs) or avalanche photodiodes (APDs) and a time-to-digital converter module (TDC). Reproduced and modified from Hoch et al., Nat. Commun. 16 (2025); licensed under a Creative Commons Attribution (CC BY) license.\cite{hoch2025quantum} \textbf{(b)} (Top) Initial quantum states $\ket{\psi_1},\ket{\psi_2},...,\ket{\psi_n}$ are polarization-encoded and evolve through a reservoir dynamics employing the orbital angular momentum (OAM) encoding. The machine is trained to reconstruct a set of target labels $y_1,y_2,...,y_n$ by performing projective measurements in the OAM computational basis. (Bottom) Experimental setup: single photons are generated via spontaneous parametric down-conversion and prepared in a specific polarization state. The reservoir dynamics is implemented through half- and quarter-waveplates (HWP, QWP) and q-plates (QPs). Measurements in the OAM computational basis are performed through a spatial light modulator (SLM). Reproduced from Suprano et al., Phys. Rev. Lett. 132 (2024); re-use license number RNP/25/OCT/097129. \cite{suprano2024experimental} \textbf{(c)}  The bulk optic setup employs the polarization and path encoding to carry out kernel-based quantum machine learning classification tasks. The experimental setup consists of PBSs, beam displacers (BSs), QWPs, HWPs, and single photon detectors. Reproduced from Bartkiewicz et al., Scientific Reports 10 (2020); licensed under a Creative Commons Attribution (CC BY) license. \cite{Bartkiewicz2020}}
    \label{fig:QML_experiments}
\end{figure*}

\begin{table*}[htb]
\centering
\begin{tabularx}{\textwidth}{
   >{\centering\arraybackslash\hsize=.5\hsize\linewidth=\hsize}X 
   >{\centering\arraybackslash}X 
   >{\centering\arraybackslash\hsize=1.5\hsize\linewidth=\hsize}X 
   >{\centering\arraybackslash}X   
   >{\centering\arraybackslash}X  
   >{\centering\arraybackslash}X 
   }
\hline
\rowcolor[HTML]{3166FF}[\dimexpr\tabcolsep+0.1pt\relax]
\textcolor{white}{\textbf{Year}} & \textcolor{white}{\textbf{Reference}}  & \textcolor{white}{\textbf{Task}}  & \textcolor{white}{\textbf{Platform}}  & \textcolor{white}{\textbf{Encoding}} & \textcolor{white}{\textbf{Comput. photons}}  \\ \hline
2019 & Yu et al. \cite{Yu2019}                            & Quantum state reconstruction                    & Bulk optics        & Polarization               & 2 \\ 
\rowcolor[HTML]{ECF4FF}[\dimexpr\tabcolsep+0.05pt\relax]
2020 & Bartkiewicz et al. \cite{Bartkiewicz2020}          & Kernel method for classification task           & Bulk optics        & Polarization/path               & 2 \\
2021 & Saggio et al. \cite{Saggio2021}                    & Reinforcement learning                          & Integrated optics  & Path                       & 1 \\ 
\rowcolor[HTML]{ECF4FF}[\dimexpr\tabcolsep+0.05pt\relax]
2024 & Suprano et al. \cite{suprano2024experimental}      & Quantum extreme learning  machine                      & Bulk optics        & OAM   & 1 \\ 
2024     & Anai et al. \cite{Anai2024}                        & Kernel method for classification task           & Bulk optics, CV    & Field quadratures          &  \\ 
\rowcolor[HTML]{ECF4FF}[\dimexpr\tabcolsep+0.05pt\relax]
2025 & Hoch et al. \cite{hoch2025quantum}                 & Kernel method for classification task           & Integrated optics  & Path                       & 3 \\ 
2025 & Monbroussou et al. \cite{monbroussou2025photonic}  & Quantum convolutional neural network            & Integrated optics  & Path                       & 2 \\ 
\rowcolor[HTML]{ECF4FF}[\dimexpr\tabcolsep+0.05pt\relax]
2025     & Yin et al. \cite{Yin2025}                          & Kernel method for classification task           & Integrated optics  & Path &2\\
2025     & Zia et al. \cite{Zia2025}                          & Quantum reservoir computing for entanglement witness           & Bulk optics  & OAM 
& 2 \\ \hline

\end{tabularx}
\caption{\textbf{Photonic demonstrations of Quantum Machine Learning.} Chronology of experiments that demonstrated quantum machine learning protocols on a quantum-optical platform. For each reference, we report the performed machine learning task, the platform used, the chosen qubit encoding, and the number of computational photons involved. SVM = Support Vector Machine; CV = continuous-variable; OAM = orbital angular momentum.}
\label{tab:QML}
\end{table*}

\textit{Variational quantum cloning} aims at generating $N$ approximate copies of a given quantum state, starting from $M$ exact copies, e.g., $M \rightarrow N$ optimal cloning, where $M < N$. Though it is well-established that an unknown quantum state cannot be cloned with perfect fidelity\cite{Wootters1982,Park1970}, optimal strategies can be identified to realize imperfect copies of universal qubits or for specific classes of states \cite{Bru1998}. It is then reasonable to explore whether a VQA can be trained to achieve such optimal cloning fidelities, \textit{e.g.}, to analyze the cryptographic consequences of such a functionality\cite{Coyle2022}. To this aim, a parameterized quantum circuit is fed with $M$ copies of a quantum state $\ket{\psi}$ and $N-M$ ancillary dummy qubits, \textit{e.g.}, in the state $\ket{0}$. The optimal parameters will then be the ones that maximize a given figure of merit that quantifies the cloning quality. For instance, in a $1 \rightarrow 2$ cloning task, at a given step of the optimization, the algorithm will apply the transformation $U(\theta)\ket{\psi}_1\ket{0}_2 = \ket{\phi'_\theta}_1\ket{\phi'_\theta}_2$, where the state $\ket{\phi'_\theta}$ depends on the set of the current parameters $\theta$. To symmetrically maximize the cloning quality between the final state $\ket{\phi'_{\theta^*}}$ and the target state $\ket{\psi}$, a convenient choice of the cost function is the following:
\begin{equation}
\begin{split}
&C^{M \rightarrow N} (\theta) =\\ &\mathbb{E}_{\ket{\psi} \in \mathcal{S}} \left[ \sum_{i=1}^N(1 - F_i(\theta))^2  + \sum_{i<j}^N(F_i(\theta) - F_j(\theta))^2 \right]
\end{split}
\label{eq:cloning_cost}
\end{equation}
where $\mathcal{S}$ is the class of states to be cloned.
In the previous expression, the variable $F_i(\theta)$ indicates the fidelity between the target state and the $i-th$ ancillary qubit.
The first two terms of the cost function in Eq.~\eqref{eq:cloning_cost}  push the parameterized circuit to maximize the fidelity of each clone with the target state, while the third term forces the algorithm to output copies having the same fidelities,\textit{ i.e.}, \textit{symmetric} copies.
Such a cost function can also be modified to address different optimal cloning tasks, \textit{i.e.}, asymmetric cloning \cite{Coyle2022}.

The first experiment addressing a $1 \rightarrow 2$ optimal quantum cloning is reported in Ref.~[\onlinecite{jaek2019experimental}]. It employed bulk optics and polarization-encoding to train a two-parameter circuit to clone quantum states lying on the equator of the Bloch sphere\cite{Bru2000}.
A more recent work\cite{hoch2024variational} leveraged a photonic integrated processor with 12 trainable parameters to perform optimal quantum cloning of different classes of states.

Other experimental implementations of different variational tasks have involved quantum unsampling \cite{Carolan2020}, quantum classifiers \cite{Ding2023}, and multiparameter estimation for quantum metrology \cite{cimini2024variational}. Fig.~\ref{fig:variational_exps}c-d report the experimental setup used in Ref.~[\onlinecite{hoch2024variational}] to implement variational quantum cloning.

\subsection{Quantum machine learning}\label{QMLsec}

Quantum machine learning (QML) represents a wide class of different algorithms, all having the same goal: using a quantum processor to learn patterns, in classical or quantum data, more efficiently than classical machines. 
The debate about whether QML can in general provide a quantum speedup with respect to its classical counterpart is still open, and, while some quantum algorithms for machine learning, such as Fourier transforms, yield an exponential quantum advantage, for others such as finding eigenvectors and eigenvalues, and solving linear equations\cite{Harrow2009,Wiebe2012,Childs2017}, a general answer has not been given yet\cite{Biamonte2017}.
Nonetheless, several QML tasks have been theoretically proposed to address different problems, such as quantum autoencoders for data compression\cite{Romero2017}, quantum neural networks\cite{Schuld2014}, quantum convolutional neural networks\cite{cong2019quantum,monbroussou2025subspace}, quantum reservoir computing\cite{garcia2023scalable}, and quantum systems characterization\cite{Huang2022}. 
Moreover, many photonic experiments have been carried out to show that even NISQ platforms can solve QML tasks of varying difficulty. We review some of them in this section, and we provide a comprehensive list in Table~\ref{tab:QML}.
As shown in the Table, depending on the degree of freedom employed to encode quantum information, bulk or integrated setups have been used in the literature. In the following, we start from some existing implementations to briefly discuss both approaches.\\

Although Boson Sampling is not universal, recent demonstrations have been given that this paradigm can represent a starting point to develop QML algorithms that take advantage of multiphoton interference to carry out useful tasks. Indeed, a recent experiment demonstrated a quantum kernel estimation for a binary classification on a Boson Sampling platform \cite{Yin2025}. The scheme was shown to outperform state-of-the-art classical kernel methods due to the presence of bosonic interference.

Moreover, recent theoretical works investigated the potential of equipping Boson Sampling platforms with non-linear \cite{spagnolo2023nonlinear,steinbrecher2019quantum} or adaptive elements \cite{chabaud2021quantum,Monbroussou2024}. Such elements can enhance the expressivity of the Boson Sampling paradigm and, hence, enlarge their potential applications to useful computational tasks\cite{spagnolo2023nonlinear, monbroussou2025subspace, chabaud2021quantum}. Specifically, in Ref.~[\onlinecite{chabaud2021quantum}], the authors presented supervised learning algorithms in which a Boson Sampling platform, equipped with adaptive circuit reconfiguration, is used to perform computational subroutines such as probability estimation or kernel estimation via quantum states overlap.
The adaptive circuit reconfiguration occurs upon detection of some photons in given modes at a specific location in the circuit, such that the undetected photons undergo a unitary that is reconfigured accordingly to the detection pattern and outcomes. 
A scheme of this technique is sketched in Fig.~\ref{fig:boson_sampling_machines}e.
Such a theoretical proposal was firstly demonstrated experimentally in Ref.~[\onlinecite{hoch2025quantum}], with the platform reported in Fig.~\ref{fig:QML_experiments}a.
In this work, the authors leverage two different hybrid photonic platforms to carry out one-dimensional and two-dimensional classification tasks, based on a well-known kernel-based method called Support Vector Machine \cite{Cortes1995}.

An alternative scheme was proposed in Ref.~[\onlinecite{Monbroussou2024}], where the adaptivity is moved from circuit reconfiguration to photonic state generation. This approach is depicted in Fig.~\ref{fig:boson_sampling_machines}f.
In detail, in this paradigm, photon measurements are performed at given locations inside the circuit, but, unlike the previous architecture, such measurements cause the re-injection of a number of photons in the circuit, which depends on the detection pattern.
This tool was experimentally tested in Ref.~[\onlinecite{monbroussou2025photonic}], where the authors propose a quantum convolutional neural network (QCNN) modular architecture based on adaptive photon injection and on the QCNN structure formalized in Ref.~[\onlinecite{monbroussou2025subspace}]. The architecture is tailored to a binary image classification task and implemented on a hybrid photonic platform featuring two integrated photonic processors of different sizes.

Bulk setups have been used to address QML tasks based on the polarization or OAM degrees of freedom of photons.
In Fig.~\ref{fig:QML_experiments}b-c, we show two bulk optics setups that employ, respectively, these two degrees of freedom. 
The experimental setup in Fig.~\ref{fig:QML_experiments}b is taken from Ref.~[\onlinecite{suprano2024experimental}] and is used to implement a photonic quantum extreme learning machine \cite{Innocenti2023, Sakurai2025} to reconstruct the properties of unknown quantum states. 
This algorithm builds upon the concept of reservoir dynamics \cite{garcia2023scalable,Fujii2017,Mujal2021}, a computational framework based on a fixed nonlinear layer to efficiently extract features from input data.
To implement this dynamics experimentally, the authors employ a coined quantum walk encoded in the polarization and OAM of single photons \cite{Innocenti2017,Giordani2019}. This is achieved using polarizing optical elements and $q$-plates \cite{Marrucci2006,Marrucci2011}, which are devices that convert the spin angular momentum of circularly polarized light into OAM.
In this way, the dimension of the considered Hilbert space is increased from 2 to $d$. 
An experimental application of OAM quantum walks to quantum reservoir computing has been reported in Ref.~[\onlinecite{Zia2025}].
Recently, other experimental implementations of photonic extreme learning machines have been reported \cite{joly2025_QELM, gong2025_GBS_QELM, cimini2025_GBS_QELM}. These demonstrations encoded the high-dimensional reservoir dynamics in the spatial modes of multimode fibers \cite{joly2025_QELM} or in large-scale GBS setups, exploiting hybrid time-spatial modes \cite{gong2025_GBS_QELM} and frequency modes \cite{cimini2025_GBS_QELM}.

The experimental setup in Fig.~\ref{fig:QML_experiments}c, instead, is taken from Ref.~[\onlinecite{Bartkiewicz2020}], and addresses a kernel-based classification task through a bulk setup based on the polarization and path degrees of freedom of single photons.

As a further demonstration of QML fundamentals through quantum optics, we mention the experimental implementation of neuromorphic computing based on quantum memristors.
A quantum memristor is the quantum counterpart of a classical memristor, i.e., a variable resistor that exhibits memory properties \cite{Chua1971}.
Analogously to classical memristors, quantum memristors introduce memory effects into quantum circuits, representing promising devices for applications to quantum reservoir computing schemes \cite{Lamata2024,garcia2023scalable,Fujii2017,Mujal2021}.
Optical quantum memristors have been implemented with single photons in different degrees of freedom and platforms, such as dual-rail encoding and integrated optics \cite{spagnolo2022experimental} and single-rail encoding and bulk optics \cite{dimicco2025quantum,spagnolo2022experimental}. Moreover, they have been recently applied to non-linear functions and time series prediction \cite{selimovic2025experimental}.

\begin{figure*}[th!]
    \centering
    \includegraphics[width=\linewidth]{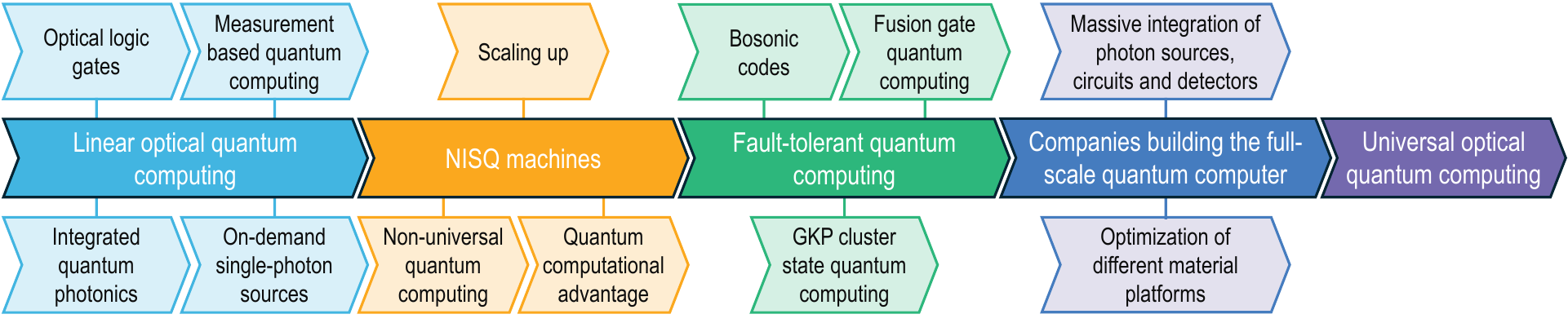}
    \caption{Evolution and future perspectives of optical quantum computing.}
    \label{fig:conslusions}
\end{figure*}
\section{Conclusions and Perspectives}
Since the realization of the first optical CNOT gate in 2003 \cite{OBrien2003Demonstration}, photonic quantum computing has evolved from fundamental proof-of-principle demonstrations into a leading candidate for large-scale, fault-tolerant quantum computation as schematically shown in Fig.~\ref{fig:conslusions}. The unique advantages of the optical platform—namely, room-temperature operation and inherent protection against environmental decoherence—make photons exceptionally robust carriers of quantum information. On one hand, this exact lack of interaction is a double-edged sword: while it preserves quantum coherence, it also makes building a quantum computer challenging by hindering the direct realization of deterministic two-qubit entangling gates. On the other hand, the photonic architecture is fully appropriate for modular approaches. This modularity also implies the capability to obtain a very large degree of connectivity between the different information carriers.

The challenge in the implementation of interactions between different photons has led to a strong effort in  the continuous-variable (CV) approach, as quadrature entanglement and two-qumode operations can be generated deterministically using linear optics and squeezing. However, this encoding paradigm still requires nonlinearity at the single-photon level to implement the non-Gaussian operations which are necessary for universal quantum computation. 

Realizing a large-scale, fault-tolerant quantum computer, whether approached via discrete-variable architectures (requiring deterministic single-photon logic) or continuous-variable encodings (necessitating complex cubic phase gates or GKP magic state distillation), must circumvent the limitation of the weak photon-photon interaction for the implementation of deterministic two qubit gates or the generation of non-Gaussian resources. In 2001 it was proposed to use linear optics and single photon detection with ancillary photons\cite{knill2001scheme} to engineer these functionalities, but these schemes introduce significant overhead and a heightened sensitivity to optical loss. To overcome these challenges, the field is advancing along two parallel trajectories: (1) devising architectures that are less resource-intensive and more resilient to optical loss, and (2) enhancing the performance of fundamental hardware components, such as single-photon sources, detectors, and reconfigurable manipulation circuits.

A prime example of this evolution is the paradigm shift away from the traditional gate-based model—which mimics classical computer architectures—toward Measurement-Based Quantum Computing (MBQC) \cite{Raussendorf2001}. Because MBQC bypasses the need for sequential, deterministic two-qubit gates, it can offer better scalability, provided that massive entangled cluster states can be successfully generated. This scalability is practically realized by multiplexing across time, spatial modes, frequency, or a combination thereof.  

The CV approach currently has recently reported the successful generation of large-scale \cite{yoshikawa2016invited, yokoyama2013ultra} and two-dimensional cluster states \cite{asavanant2019generation, larsen2019deterministic}. Concurrently, Xanadu is advancing the CV approach with a roadmap focused on fully integrated, fault-tolerant CV-MBQC, relying on on-chip squeezed light sources, time-domain multiplexing, and GKP error-correcting codes \cite{Bourassa2021blueprintscalable}. In particular their recent partnership with Hyperlight\cite{Xanadu_news} - a leader in TFLN manufacturing - signals an industrial interest in adopting this material platform for next-generation scalable architectures. TFLN offers two key properties for scalable architectures: high optical nonlinearity for efficient squeezed vacuum generation, and ultra-fast electro-optical reconfigurability for dynamic photon manipulation. 

Discrete-variable protocols have theoretically demonstrated that the probabilistic nature of optical entanglement can be decisively overcome utilizing topological architectures and percolation theory \cite{Rudoplh2017}. An advantage of this framework is its inherent tolerance to a significant fraction of photon loss, a threshold that has been progressively improved in recent theoretical proposals\cite{Segovia2015,Bartolucci2023}. PsiQuantum is pushing toward utility-scale, fault-tolerant DV systems by leveraging Tier-1 semiconductor foundries \cite{Alexander2025}. Their fusion-based architecture aims to mass-produce millions of silicon photonic components, effectively bridging the gap between quantum information theory and commercial CMOS manufacturing. 
 
Concurrently, alternative hybrid approaches have emerged, accompanied by initial experimental demonstrations. Notably, Quandela is developing a scalable quantum computing architecture based on spin-photon entanglement \cite{Wein2024minimizing,quandela_architecture}. By encoding quantum information within the spin states of solid-state quantum emitters and entangling them with flying photons, this platform enables the deterministic generation of robust graph states \cite{Huet2025Deterministic} that can be efficiently manipulated and distributed over macroscopic distances. This hybrid scheme significantly enhances network connectivity and supports high-fidelity operations by mitigating decoherence and errors. Furthermore, integrating stationary spin qubits with photonic carriers facilitates seamless communication and distributed computation, laying crucial groundwork for future quantum networks \cite{Wein2024minimizing,quandela_architecture}

Realizing a useful, fault-tolerant quantum computer will ultimately require on the order of one million physical qubits, the vast majority of which will be dedicated to error correction. Whether this milestone will be achieved first via a CV cluster state leveraging GKP encoding, a DV protocol utilizing fusion gates and percolation theory, or an entirely novel hybrid approach, remains an open question. However, the convergence of theoretical ingenuity and industrial-scale photonics engineering suggests that the realization of a universal optical quantum computer is no longer a question of fundamental physics, but one of time and scalable manufacturing. Today, comprehensive roadmaps chart viable pathways toward fully fault-tolerant quantum computation \cite{Bourassa2021blueprintscalable, Rudoplh2017, larsen2021fault}.

In parallel with the pursuit of fault-tolerant universal architectures, non-universal photonic devices can play an important role in this NISQ era. Driven initially by the theoretical framework of boson sampling, these specialized processors provided the field with its first experimental claims of quantum computational advantage \cite{zhong2020quantum, madsen2022}. Moving beyond sampling tasks, current efforts are focused on leveraging these platforms for heuristic applications, such as variational quantum eigensolvers for quantum chemistry \cite{Wang2017_ham,Santagati2018, Hoch2025}, quantum simulations \cite{Huh2015_vibronic, Sparrow2018, Arrazola2021,Nielsen2025} and quantum machine learning \cite{chabaud2021quantum, hoch2025quantum, monbroussou2025photonic, Yin2025}. While these non-universal architectures operate without full error correction, placing inherent constraints on circuit depth, this trade-off is a strategic choice that allows for the testing of quantum utilities in the near term. These specialized schemes circumvent the massive hardware requirements of fault tolerance, allowing them to be tailored for robustness and functionality within their operational limits, even in the presence of noise. Rather than being mere precursors, these protocols represent a promising pathway to task-specific quantum algorithms. They allow for the immediate benchmarking of quantum photonic hardware and provide a viable near-term framework for solving practical problems before the advent of fully fault-tolerant, large-scale machines. 

\begin{acknowledgments}
H.H., L.L., M.B., A.Q., and M.L. acknowledge support from INFN, Italy through the CSN5-UNIDET project. H.H. and M.L.  acknowledge support from INFN, Italy through the CSN5-SQUEEZE project. M.B. and M.L. acknowledges support from the PRIN MUR 2022 project 2022R7RSHL (Integrated hybrid quantum detector). T.G., B.P., and F.S. acknowledge support from the ERC Advanced Grant QU-BOSS (QUantum advantage via nonlinear BOSon Sampling, grant agreement no. 884676), the European Union’s Horizon Europe research and innovation program under EPIQUE Project (Grant Agreement No. 101135288), and from the PNRR MUR project PE0000023-NQSTI (National Quantum Science and Technology Institute, Spoke 4). L.L., and M.B. acknowledge support from the PNRR MUR project PE0000023-NQSTI (National Quantum Science and Technology Institute, Spoke 4). The authors have no conflicts to disclose.
\end{acknowledgments}

\section*{Data Availability Statement}
Data sharing is not applicable to this article as no new data were created or analyzed in this study. 

\section*{REFERENCES}
\bibliography{Bibliography}
\end{document}